\documentclass[times,twocolumn]{aastex62}

\usepackage{enumitem}
\usepackage{amsmath}
\usepackage{multirow}
\usepackage{cases}
\usepackage{graphicx}
\usepackage{subfigure}
\usepackage{natbib}
\usepackage{color}
\usepackage{tabularx}
\usepackage{bm}
\usepackage{threeparttable}
\usepackage{gensymb}
\usepackage{longtable}

\newcommand{\HL}[1]{\textcolor{black}{#1}}

\providecommand{\noprint}[1]{}

\DeclareGraphicsExtensions{.pdf,.png,.jpg}

\shorttitle{KMTNet Bulge Asteroids}
\shortauthors{Huang et al.}

\begin{document}
\title{{\large Measuring Asteroid Rotation Periods Using the KMTNet Bulge Survey Data}}

\correspondingauthor{Hongjing Yang, Bin Li}
\email{hongjing.yang@qq.com, binli@pmo.ac.cn}

\author[0009-0001-8695-0896]{Haitao Huang}
\affiliation{Department of Astronomy, Westlake University, Hangzhou 310030, Zhejiang Province, China}

\author[0000-0003-0626-8465]{Hongjing Yang}
\affiliation{Department of Astronomy, Westlake University, Hangzhou 310030, Zhejiang Province, China}
\affiliation{Westlake Institute for Advanced Study, Hangzhou 310030, Zhejiang Province, China}

\author[0000-0003-0043-3925]{Chung-Uk Lee}
\affiliation{Korea Astronomy and Space Science Institute, Daejeon 34055, Republic of Korea}

\author{Bin Li}
\affiliation{CAS Key Laboratory of Planetary Sciences, Purple Mountain Observatory, Chinese Academy of Sciences, Nanjing 210008, China}
\affiliation{University of Science and Technology of China, Hefei 230026, China}

\author{Qiyue Qian}
\affiliation{Department of Astronomy, Tsinghua University, Beijing 100084, China}

\author{Jun Tian}
\affiliation{CAS Key Laboratory of Planetary Sciences, Purple Mountain Observatory, Chinese Academy of Sciences, Nanjing 210008, China}

\author[0000-0002-4503-9705]{Tianjun Gan}
\affil{Instituto de Astrof\'isica de Canarias (IAC), V\'ia L\'actea s/n, E-38205 La Laguna, Tenerife, Spain}
\affil{Dept. Astrof\'sica, Universidad de La Laguna (ULL), E-38206 La Laguna, Tenerife, Spain}
\affiliation{Department of Astronomy, Westlake University, Hangzhou 310030, Zhejiang Province, China}

\author[0000-0001-8317-2788]{Shude Mao}
\affiliation{Department of Astronomy, Westlake University, Hangzhou 310030, Zhejiang Province, China}


\author[0000-0001-6000-3463]{Weicheng Zang}
\affiliation{Department of Astronomy, Westlake University, Hangzhou 310030, Zhejiang Province, China}





\author{Dong-Jin Kim}
\affiliation{Korea Astronomy and Space Science Institute, Daejeon 34055, Republic of Korea}

\begin{abstract}

Since 2015, the Korea Microlensing Telescope Network (KMTNet) has conducted high-cadence, near-continuous observations of the Galactic bulge for nearly nine months each year from three sites in Chile, South Africa, and Australia, and its wide field of view provides a unique opportunity to extract asteroid lightcurves from archival survey data. In this work, we performed photometric measurements of bright asteroids ($V< 20$~mag) identified within a one-square-degree field during the 2018 KMTNet bulge season. We derived reliable rotation periods for 96 asteroids, including 84 objects without previously published lightcurves. The reliable spin-rates of the asteroids in our sample are broadly consistent with those reported in the Asteroid Lightcurve Database. This archival mining approach can be readily extended to a much larger KMTNet footprint, and the existing $\sim12$~deg$^2$ high-cadence KMTNet dataset has the potential to yield reliable rotation periods (U $\geq 2+$) for more than 5,500 asteroids, substantially expanding the current database of asteroid rotational properties.

\end{abstract}

\section{Introduction}\label{sec:intro}

As of 16 May 2026, more than 1.54 million asteroids have been discovered according to the Jet Propulsion Laboratory (JPL) database\footnote{\url{https://ssd.jpl.nasa.gov/}}. However, based on the most comprehensive photometric database of asteroids, the Asteroid Lightcurve Database\footnote{\url{https://www.minorplanet.info/php/lcdb.php}.} (LCDB; \citealt{2009Icarus-Warner-LCDB}), \HL{only 32,852 asteroids have lightcurve parameters published} as of October 2023, corresponding to less than 2.2\% of the currently known asteroid population \citep{2023MNoRA-Xu}. The asteroid lightcurves provide fundamental information about several physical properties of asteroids; for example, the rotation period can be measured from lightcurves \citep{1989Icarus-Harris-rotation,2020AJ-Yeh-rotation}, the general shape can be estimated from the lightcurve amplitude \citep{2000Icarus-Pravec-shape,2003Icarus-Lacerda-shape,2022AA-Tian-shape}, the detailed shape model can be obtained from lightcurve inversion \citep{1992bAA-Kaasalainen-3D,1992aAA-kaasalainen-3D,2018Icarus-Durech-3D}, and the albedo can be roughly inferred from the phase–curve relation \citep{1989AsteroidII-Bowell-Albedo}. In addition, statistical distributions of asteroid spin-rates and pole orientations are essential for understanding how rotational states are influenced by multiple mechanisms, including collisional evolution, tidal interactions, internal structure, \HL{and the Yarkovsky--O'Keefe--Radzievskii--Paddack (YORP) effect \citep{2015AsteroidIV-Vokrouhlicky-YORP,2008Icarus-Pravec-YORP,2000Icarus-Pravec-shape}.}

A fraction of the asteroid lightcurve parameters compiled in the LCDB were derived from targeted ground-based photometric observations using small- and medium-sized telescopes. Such efforts include the Small Main-Belt Asteroid Lightcurve Survey, which derived lightcurves for 32 small main-belt asteroids using the 1.8-m Perkins, 1.3-m McGraw–Hill, and 2.4-m Hiltner telescopes \citep{1992Icarus-Binzel-SMALS}, the Thousand Asteroid Light Curve Survey (TALCS), which measured lightcurves for 828 asteroids using the Canada–France–Hawaii Telescope \citep{2009Icarus-Masiero-TALCS}, and the EURONEAR Lightcurve Survey, which reported lightcurves for 101 near-Earth asteroids based on coordinated observations from 11 telescopes \citep{2017EMaP-vaduvescu-EURONEAR}. With high cadence and high photometric precision, such observations enable the robust determination of asteroid rotation periods. However, because they typically focus on individual targets, only a limited number of asteroids can be monitored during a single observing night \citep{2011Icarus-Warner-targeted}.

In recent years, photometric data from several large time-domain surveys, originally not designed for asteroid studies, have increasingly been exploited for asteroid research and have significantly enlarged the lightcurve sample. Successful examples include the K2 mission of the Kepler Space Telescope, which obtained lightcurves for 608 main-belt asteroids and 10 Centaurs \citep{2018AJ-Molnar-K2,2020Icarus-Marton-K2}, the Transiting Exoplanet Survey Satellite (TESS), which produced lightcurves for 17,189 asteroids \citep{2020AJ-Pal-TESS,2019AJ-McNeill-TESS}, the Palomar Transient Factory (PTF), which derived reliable rotation periods for more than 8,300 asteroids \citep{2015AJ-Waszczak-PTF}, the MOA-II microlensing survey, which derived 26 asteroid rotation periods \citep{2022MNoRAS-Cordwell-MOAII}, the Vera C. Rubin Observatory Legacy Survey of Space and Time (LSST), which derived lightcurves, rotation periods, and colors for approximately 2,000 asteroids \citep{2026AJ-Greenstreet-LSST}, and the European Space Agency Gaia mission, which provided unique spin-state solutions for more than 8,600 asteroids \citep{2023AA-Durech-GAIA}. \HL{Together, these surveys have contributed tens of thousands of asteroid lightcurves and/or rotational solutions.} Nevertheless, each of these surveys has its own observational limitations. For instance, although TESS provides high-cadence photometry, its relatively shallow limiting magnitude restricts the accessible asteroid population. Conversely, deep wide-field surveys such as the LSST will reach much fainter objects, but their nominal survey cadence is not generally optimized for detailed lightcurve characterization.

The Korea Microlensing Telescope Network (KMTNet) is a wide-field time-domain survey primarily designed for gravitational microlensing observations toward the Galactic bulge \citep{KMT2016}. The network consists of three identical 1.6-m telescopes, each equipped with a 4 deg$^2$ camera, located in Chile, South Africa, and Australia, enabling near-continuous monitoring of the same fields. \HL{With its typical deep limiting magnitude of approximately V $\sim$21~mag \citep{2023LPICo-Lee-LimitingMag} }and high observational cadence (as short as 15 min for prime fields), KMTNet bulge observations are well suited for detecting and characterizing asteroid lightcurves.

In this work, we analyze the data from a one-square-degree KMTNet field during the 2018 bulge season and extract asteroid lightcurves from the difference images produced by \cite{2025PoASoP-Qian-FFPs} using a ``trailed'' PSF photometry (see Sect.~\ref{3.2photometry} for more details). While asteroid lightcurve photometry has previously been performed using KMTNet data obtained during the non-bulge observing seasons \citep{2017AJ-Erasmus-KMT,2018AJ-Erasmus-KMT}, this work presents the first systematic extraction of asteroid lightcurves from KMTNet bulge observations, where the crowding background present additional challenges for asteroid photometry.

The outline of the paper is as follows. Details of the telescope and our observations can be found in Section~\ref{sec:obs}; the methodology of target selection, photometry, and \HL{lightcurve fitting} is described in Section~\ref{sec:method}; the resulting asteroid period measurements are given in Section~\ref{sec:result}; and in Section~\ref{sec:dis} we summarize the results and discuss the expected yields from the KMTNet prime fields as well as future surveys. 

\section{Observations}\label{sec:obs}

KMTNet is a wide-field telescope network operated by the Korea Astronomy and Space Science Institute (KASI). The network comprises three identical 1.6-m telescopes equipped with a mosaic CCD camera consisting of four $9K \times 9K$ chips, providing a total field of view of $2.0 \times 2.0$~deg$^2$ with a pixel scale of $\sim0.4''$. The three telescopes are located at the Cerro Tololo Inter-American Observatory (CTIO) in Chile, the South African Astronomical Observatory (SAAO) in South Africa, and the Siding Spring Observatory (SSO) in Australia. Owing to their longitudinal distribution across similar southern latitudes ($\sim$−30$^\circ$), the network enables near-continuous (24-hour) monitoring of targets in the Southern Hemisphere. KMTNet was designed primarily for microlensing surveys toward the Galactic bulge, which are conducted annually from mid-February to mid-October.

The KMTNet Galactic bulge survey began full operations in 2016, covering a total area of 96~deg$^2$. Within this region, about 12~deg$^2$ are designated as ``prime'' fields, observed with a cadence of 4~hr$^{-1}$ (i.e., one exposure every 15 minutes). The remaining fields are observed with cadences ranging from 1~hr$^{-1}$ to 0.2~hr$^{-1}$. Observations are primarily obtained in the Cousins $I$ band, with $\sim$9\% of the images taken in the $V$ band to enable color measurements. The exposure times of the $I$-band and $V$-band images are 60~s and 90~s, respectively.
Although this observing strategy was originally optimized for detecting microlensing exoplanets, it is also well suited for asteroid lightcurve studies. For typical main-belt asteroids with apparent motions of $\sim$0.5~arcsec/min, the crossing time for a 1$\times$1~deg$^2$ field is $\sim$5 days. Consequently, a typical asteroid located in the prime field can be observed over $\sim 10$ days with up to $\sim 10^3$ exposures, making the bulge survey data a valuable resource for asteroid lightcurve measurements.

In this work, we utilize one-year $I$-band images from a $\sim$1~deg$^2$ field centered at ($\alpha$, $\delta$) = (269.22$^\circ$, $-$29.58$^\circ$) to explore the potential of KMTNet bulge data for asteroid lightcurve analysis (hereafter the target field). This area corresponds to the N CCD chip of the 02/42 bulge fields.

\section{Methodology}\label{sec:method}

\subsection{Target Selection}

To identify asteroids within the given KMTNet field during the 2018 observing season, we first query the Minor Planet Center (MPC) database\footnote{\url{https://minorplanetcenter.net/iau/MPCORB.html}.} for orbital elements. We exclude approximately 300,000 objects with poorly constrained orbital solutions, defined operationally as those with fewer than 30 reported observations in the MPC database, leaving about 1.2 million asteroids for further analysis. Ephemerides are then computed for these remaining objects at the times of each KMTNet exposure. For each exposure, we select asteroids whose predicted positions fall within the target field. 
For each selected asteroid, we record its pixel coordinates, on-sky position, orbital parameters, frame identifier, and the MPC-predicted $V$-band magnitude, phase angle, heliocentric distance, and geocentric distance.
To ensure robust photometric measurements, we restrict our sample to asteroids with a sufficiently bright mean apparent magnitude, $V < 20~mag$. 
In total, 4,288 asteroids are expected to traverse the field during the season, of which 376 satisfy the magnitude criterion and are retained in the final sample. The final sample is dominated by main-belt asteroids, with typical ephemeris uncertainties of $\sim$0.1~arcsec.

\begin{figure*}
    \centering
    \includegraphics[width=0.8\linewidth]{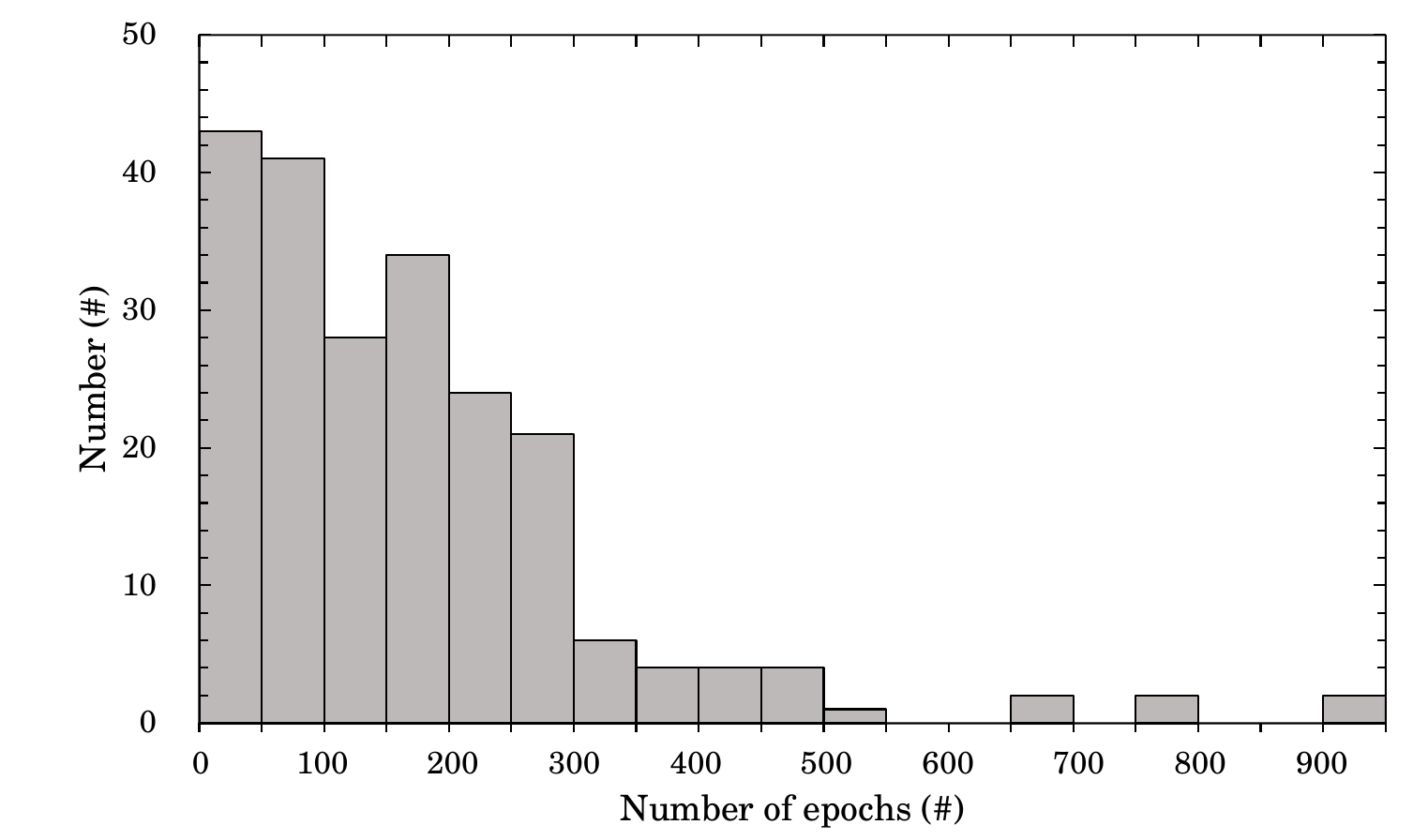}
    \caption{Distribution of Asteroid Observation Counts in the Pilot Dataset.}
    \label{fig:Observation count}
\end{figure*}

We perform photometry for all 376 asteroids. The distribution of their number of epochs is shown in Figure~\ref{fig:Observation count}, with most asteroids having fewer than 300 observations.

\subsection{Photometry}\label{3.2photometry}

\begin{figure*}
    \centering
    \includegraphics[width=0.8\linewidth]{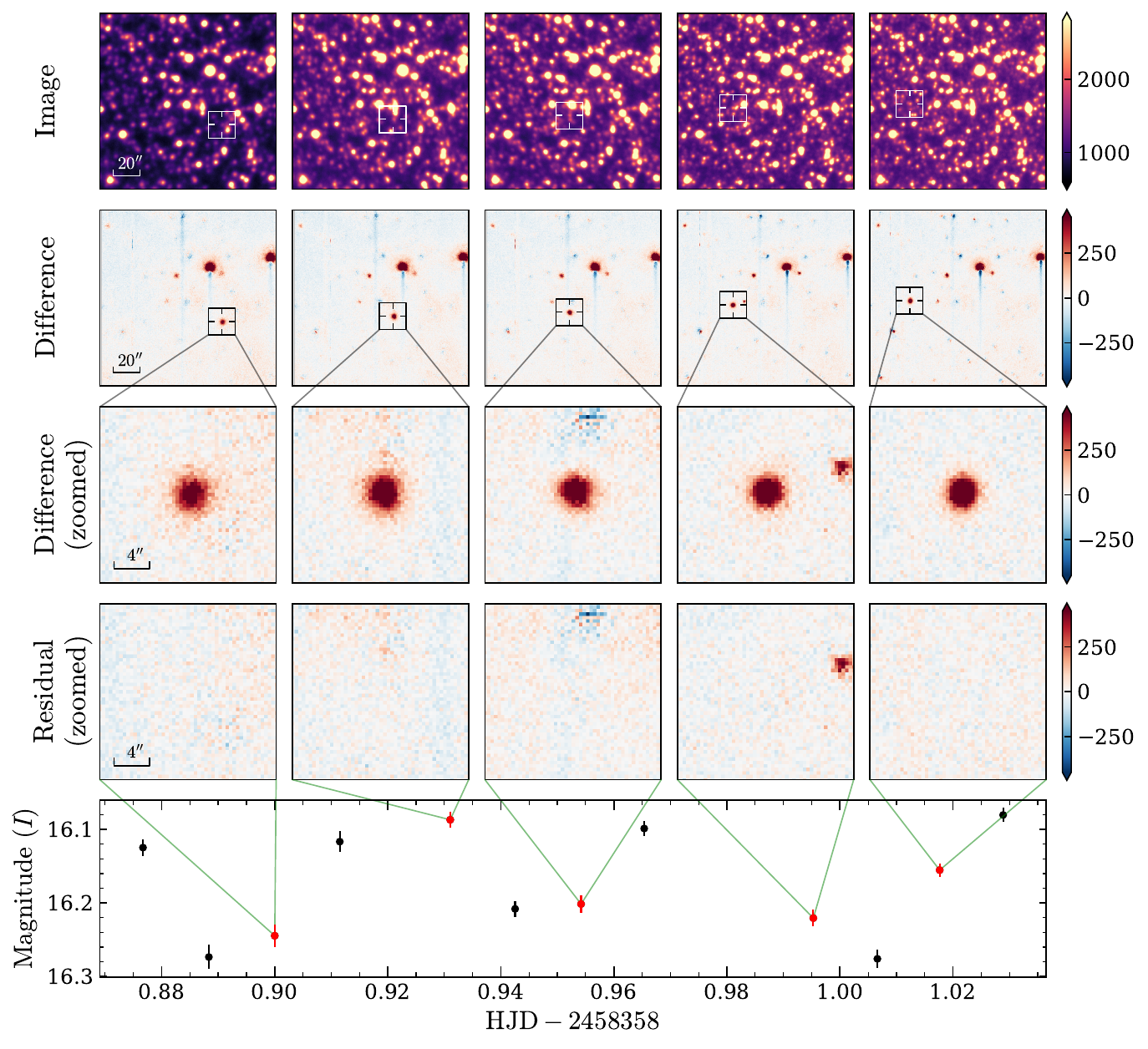}
    \caption{Illustration of the difference image and the trailed PSF photometry for asteroid (4362) Carlisle. Rows from top to bottom are: original KMTNet SSO images, difference images, zoomed cutouts of the difference images, residuals after subtracting the asteroid's trailed PSF, and the resulting lightcurve, respectively. Columns from left to right show five different epochs. The asteroid (4362) Carlisle is marked in boxes in the top two rows. Fluxes are in instrumental units.}
    \label{fig:photometry}
\end{figure*}

We perform photometric measurements based on the difference images generated by \citet{2025PoASoP-Qian-FFPs}. 
\HL{A difference image is obtained by subtracting a suitably convolved reference image from each science exposure. Specifically, the reference image for each field is constructed by stacking multiple high-quality exposures with $3\sigma$ clipping. Because asteroids move across exposures, their signals are removed during stacking, leaving only stationary sources. The reference image is then convolved with a spatially varying kernel to match the PSF and sky background of the science exposure before subtraction \citep{DanDIA, pysis}. Static sources are thus largely removed, while moving objects (asteroids) and transient signals remain. Residual stars in the difference images (e.g., in Figure~\ref{fig:photometry}) are primarily saturated stars and variable stars.}

\HL{Asteroids move significantly during the 60~s exposure of KMTNet, causing their images to appear as trails rather than point-like sources. For a typical main-belt asteroid, the on-sky motion is approximately 0.5~arcsec per 60~s exposure. Given the KMTNet pixel scale of 0.4~arcsec~pix$^{-1}$, this corresponds to a typical displacement of more than one pixel per exposure, indicating the necessity of a trailed PSF treatment.} 

To accurately extract fluxes for moving asteroids, we adopt a trailed PSF photometry approach \HL{on the difference images}, extending the method described in \citet{2024MNoRAS-Yang-KMT} by replacing the stationary PSF model with a trailed PSF model that accounts for the asteroid's apparent motion during each exposure.

We model the two-dimensional intensity distribution of each exposure of each asteroid as a stationary PSF convolved with a linear kernel representing constant-velocity motion during the exposure. The stationary PSF for each image is derived from the reference image as part of the difference image pipeline \citep{2025PoASoP-Qian-FFPs}.
The asteroid's position $(x, y)$ is defined as its centroid at the midpoint of the exposure, and the linear motion kernel is constructed accordingly to represent the trail centered on this position. The trailed PSF model is then fitted to the difference image cutout for each exposure. The model includes parameters for the asteroid's position $(x, y)$, flux $f$, and local (difference) background $b$. We minimize the $\chi^2$
\begin{equation}
    \chi^2(x,y,f,b)=\sum_{i,j}\frac{(\mathcal{D}_{i,j}^2-f\cdot \mathcal{P}_{i,j}(x,y)-b)^2}{\sigma_{i,j}^2},
\end{equation}
to measure $(x,y,f,b)$, where $\mathcal{D}_{i,j}$ is the difference image pixel value at pixel $(i, j)$, $\mathcal{P}_{i,j}(x,y)$ is the pixelated trailed PSF model, and $\sigma_{i,j}$ is the pixel-wise uncertainty.

In principle, the measured flux $f$ corresponds to the flux in the difference image. However, for an asteroid that is present in the science image but absent from the reference image (due to its motion), its entire signal appears in the difference image. Therefore, $f$ directly represents the asteroid's total flux.

To construct the trailed PSF model for each exposure, we first compute the asteroid's instantaneous velocity from its orbital ephemeris. This initial velocity is used to build the linear motion kernel. We then perform the trailed PSF fitting for each exposure to obtain the position $(x, y)$ and flux $f$. Using the fitted positions across multiple exposures within the same observational night, we refine the velocity estimate, update the motion kernel, and repeat the fitting process. This iterative procedure continues until the velocity converges, ensuring that the trailed PSF model accurately represents the asteroid's motion.

Uncertainties in the fitted parameters are estimated following the method of \citet{2024MNoRAS-Yang-KMT}. For each measurement, we also derive a photometric quality indicator $\sigma_{\rm phot}$ (defined in \citealt{2024MNoRAS-Yang-KMT, 2025AJ_Yang_KMT2}), which quantifies the goodness-of-fit and helps identify problematic measurements, such as those affected by saturated stars, bad columns, or other image artifacts. 

Finally, the instrumental fluxes are converted to semi-calibrated magnitudes using a photometric zero point of $m_0=28.1$, determined from previous KMTNet observations.
\begin{equation}\label{equ-m}
    m = m_0 - 2.5 \log_{10}(F),
\end{equation}
where $F$ is the instrumental flux. This conversion does not include detailed corrections (e.g., cross-matching to standard reference stars and measureing the actual zero point). 
However, this is sufficient because the primary goal of this study is to identify asteroid rotation periods, which are not sensitive to detailed photometric calibration.

As an example, Figure~\ref{fig:photometry} illustrates the photometry procedure for asteroid (4362) Carlisle, showing from top to bottom: the original KMTNet SSO images, difference images, zoomed cutouts of the difference images, residuals after subtracting the best-fit trailed PSF model, and the resulting lightcurve across five epochs. The clean residuals demonstrate the effectiveness of the trailed PSF photometry.

\subsection{Lightcurve Analysis}

\subsubsection{Orbital Effect Correction}
For each asteroid, to remove systematic brightness variations caused by changes in heliocentric distance, geocentric distance, and phase angle during the orbital motion, we apply distance and phase-angle corrections to each individual observation. After these corrections, the remaining magnitude variations---hereafter referred to as the ``reduced magnitude''---primarily reflect the rotational modulation of the asteroid.

The magnitude of an asteroid can be described as
\begin{equation}\label{equ-V}
    V = H + \delta + 5\log_{10}(r\Delta) - 2.5\log_{10}(\phi(\alpha)),
\end{equation}
here, $V$ corresponds to the apparent magnitude $m$ defined in Eq.(\ref{equ-m}), where $H$ is the absolute magnitude, $r$ is the Sun-asteroid distance in A.U., $\Delta$ is the Earth-asteroid distance in A.U., $\delta$ describes the rotational and shape variation, and $\phi(\alpha)$ is the phase function, where $\alpha$ represents the phase angle, the angle between the Sun, the asteroid, and the Earth.

We model the phase function using the Lumme-Bowell model \citep{1989AsteroidII-Bowell-Albedo}, which is defined as

\begin{equation}
\phi = (1 - \mathrm{G})\phi_1 + \mathrm{G}\phi_2,
\end{equation}

\begin{equation}
\phi_1 = \exp\left(-3.33 \tan^{0.63}\left(\frac{\alpha}{2}\right)\right),
\end{equation}

\begin{equation}
\phi_2 = \exp\left(-1.87 \tan^{1.22}\left(\frac{\alpha}{2}\right)\right),
\end{equation}
where the phase parameter G is fixed at 0.15.

\subsubsection{Rotation Period Search}

Before we proceed the period analysis, we first exclude poor photometric measurements. Data with negative flux, large ($>3.4''$) or unphysically small seeing ($<0.4''$), high sky background ($>5000$\,ADU), large positional uncertainty ($\geq 2.5$\,pix or $1.0''$ in any direction), or bad image-subtraction quality (quality indicator $\sigma_{\rm subt}>1.7$; \citealp{2024MNoRAS-Yang-KMT}) are excluded.

To determine the asteroid rotation period, we adopt a multi-stage procedure combining grid search, downhill refinement, and Markov Chain Monte Carlo (MCMC). The lightcurve is modeled using a second-order Fourier series with coefficients $a_0$, $B_1$, $B_2$, $C_1$, and $C_2$ \citep{1989Icarus-Harris-rotation}, which provides sufficient flexibility to reproduce the typical double-peaked morphology of asteroid rotational modulation. The model is expressed as
\begin{equation}
\begin{split}
    M_j = a_0 + & \sum_{k=1,2}^{N_k} B_k \sin\left[\frac{2\pi k}{P}(t_j - t_0)\right]\\
    + &\sum_{k=1,2}^{N_k} C_k \cos\left[\frac{2\pi k}{P}(t_j - t_0)\right],
\end{split}
\end{equation}
where $M_j$ is the $I$-band reduced magnitude (i.e., $H + \delta$ as defined in Eq.~(\ref{equ-V})) measured at epoch $t_j$; $t_j$ denotes the observation time of the $j$th data point; $a_0$ represents the mean magnitude level of the lightcurve, approximately corresponding to the absolute magnitude $H$; $B_k$ and $C_k$ are the Fourier coefficients; $P$ is the rotation period; and $t_0$ is an arbitrary reference epoch. \HL{The present analysis assumes that the observed lightcurve is dominated by the rotational modulation of a single asteroid. Potential binary asteroid signatures, such as mutual events or additional periodic components, are not explicitly investigated in this work and will be addressed in a future study.}

We first perform a period grid search over the range of 0.48–100 h with a uniform step of 0.01 h. The lower bound is chosen to be close to the Nyquist limit set by the typical cadence of $\sim$15~min, below which period determinations may be affected by aliasing, corresponding to a spin frequency of 50 cycles per day. For each grid, the Fourier coefficients are determined via weighted least-squares fitting, and the goodness-of-fit is quantified using the $\chi^2$ statistic, yielding a discrete $\chi^2(P)$ distribution. To improve robustness against outliers, the fitting procedure is iterated three times; after each of the first two iterations, data points deviating by more than $2\sigma$ from the model are rejected, and the fit is repeated on the remaining cleaned dataset.

Candidate periods are identified not only from the global minimum of the $\chi^2(P)$ distribution, but also from local minima satisfying $\Delta \chi^2 < 100$ relative to the global minimum. If no such solutions exist, the period corresponding to the second smallest $\chi^2$ is adopted as a candidate. This selection accounts for the presence of aliasing (stroboscopic) effects caused by quasi-periodic sampling, which can produce multiple competing minima in $\chi^2$ space. Retaining near-degenerate solutions prevents premature selection of an alias period at this stage \citep{2020AJ-Pal-TESS}. Even though KMTNet's near-continuous coverage reduces diurnal-gap aliasing, we retain multiple candidate periods for each lightcurve at this stage to ensure robust period determination and avoid prematurely adopting an alias solution. \HL{For objects whose $\chi^2$ distributions continued to decrease toward the upper boundary of the initial search range, the period search was manually extended beyond 100 h to identify additional candidate minima.}

All candidate periods are subsequently refined using the Nelder--Mead simplex algorithm, jointly optimizing the period and Fourier coefficients to minimize $\chi^2$. The final best-fit period is chosen as the solution with the lowest $\chi^2$ after refinement. The uncertainty of the derived period is then estimated using a Markov Chain Monte Carlo (MCMC) approach, sampling the joint posterior distribution of the period and Fourier coefficients. The reported period value and its uncertainty are the median and the (16\%, 84\%) percentiles of the marginalized posterior distribution.

Because asteroid lightcurves are generally expected to exhibit a double-peaked structure at the physical rotation frequency, we further examine the phase-folded light curves of candidate solutions via visual inspection. The adopted period produces a coherent and smooth double-peaked morphology with minimal scatter, whereas alternative (alias) solutions often yield less consistent phase-folded structures. If only a single peak is present, the solution at half the corresponding frequency is additionally evaluated by manually adjusting the trial period and re-examining the folded lightcurve, followed by the same optimization and uncertainty estimation procedure described above.

The final phase-folded lightcurve of each asteroid is then visually assigned a quality code ``U''. According to the criteria of the LCDB, U = 3, 2, and 1 correspond to highly reliable, ambiguous, and low-reliability solutions, respectively. \HL{The updated system} also allows `$+$' and `$-$' subdivisions, e.g., $2+$ or $3−$, to refine the assessments even more \citep{2009Icarus-Warner-LCDB}. The resulting distribution of quality codes is as follows: 7, 29, 60, 40, 48, 74, and 118 asteroids are classified as U = 3, $3-$, $2+$, 2, $2-$, $1+$, and 1, respectively.

\section{Asteroid Rotation Periods}\label{sec:result}

\subsection{Rotation Period and Quality}

In total, 96 reliable rotation periods (i.e., U $\geq$ 2+) are obtained, which are listed in Table~\ref{tab1:list}, where only the first ten entries are shown. The complete table is provided in Appendix~\ref{appendix:table}. In addition to these reliable solutions, we also report less reliable rotation periods (i.e., U$ < 2+$) for 123 asteroids in Appendix~\ref{appendix:table}, excluding asteroids for which no period could be determined due to insufficient data points.

\begin{table*}
\caption{List of reliable rotation periods (first 10 entries). The complete table is available in Appendix~\ref{appendix:table}.}
\label{tab1:list}
\centering
\small

\begin{threeparttable}

\begin{tabular*}{\textwidth}{@{\extracolsep{\fill}}lccccccc}
\hline\hline
Object & Period & Quality & $H$\tnote{*} & Magnitude        & Amplitude & \HL{Epochs} & Phase Angle\tnote{*} \\
       & (h)    &         & (mag)        & (\textit{I}-band) & (mag)    &       & ($^\circ$) \\
\hline

(1077) Campanula   & $3.8378 \pm 0.0008$  & $2+$  & 12.36 & 15.407 & 0.4375 & 48 & 26.4939 \\
(1152) Pawona      & $3.4348 \pm 0.0004$  & $2+$  & 11.17 & 14.873 & 0.2216 & 83 & 24.4671 \\
(1187) Afra        & $13.8938 \pm 0.0119$ & $2+$  & 11.62 & 16.150 & 0.4563 & 47 & 21.0534 \\
(1324) Knysna      & $2.5503 \pm 0.0007$  & $2+$  & 13.20 & 15.548 & 0.1308 & 87 & 30.2324 \\
(2294) Andronikov  & $3.1534 \pm 0.0001$  & $3-$  & 11.91 & 14.925 & 0.4518 & 189 & 14.1802 \\
(2438) Oleshko     & $3.2222 \pm 0.0003$  & $2+$  & 13.02 & 15.660 & 0.2007 & 142 & 24.7712 \\
(3180) Morgan      & $2.4481 \pm 0.0004$  & $3-$  & 14.21 & 17.809 & 0.2003 & 200 & 22.0880 \\
(3314) Beals       & $5.4650 \pm 0.0039$  & $2+$  & 13.11 & 17.038 & 1.3315 & 24 & 23.7157 \\
(4362) Carlisle    & $2.6320 \pm 0.0002$  & $2+$  & 13.18 & 16.178 & 0.1878 & 170 & 24.5133 \\
(5116) Korsor      & $4.5033 \pm 0.0004$  & $3 $  & 12.06 & 16.349 & 0.4812 & 114 & 2.2259 \\

\hline
\end{tabular*}

\begin{tablenotes}
\item[*] Obtained from the MPCORB database.
\end{tablenotes}

\end{threeparttable}
\end{table*}

\begin{figure*}
    \centering
    \includegraphics[width=0.7\linewidth]{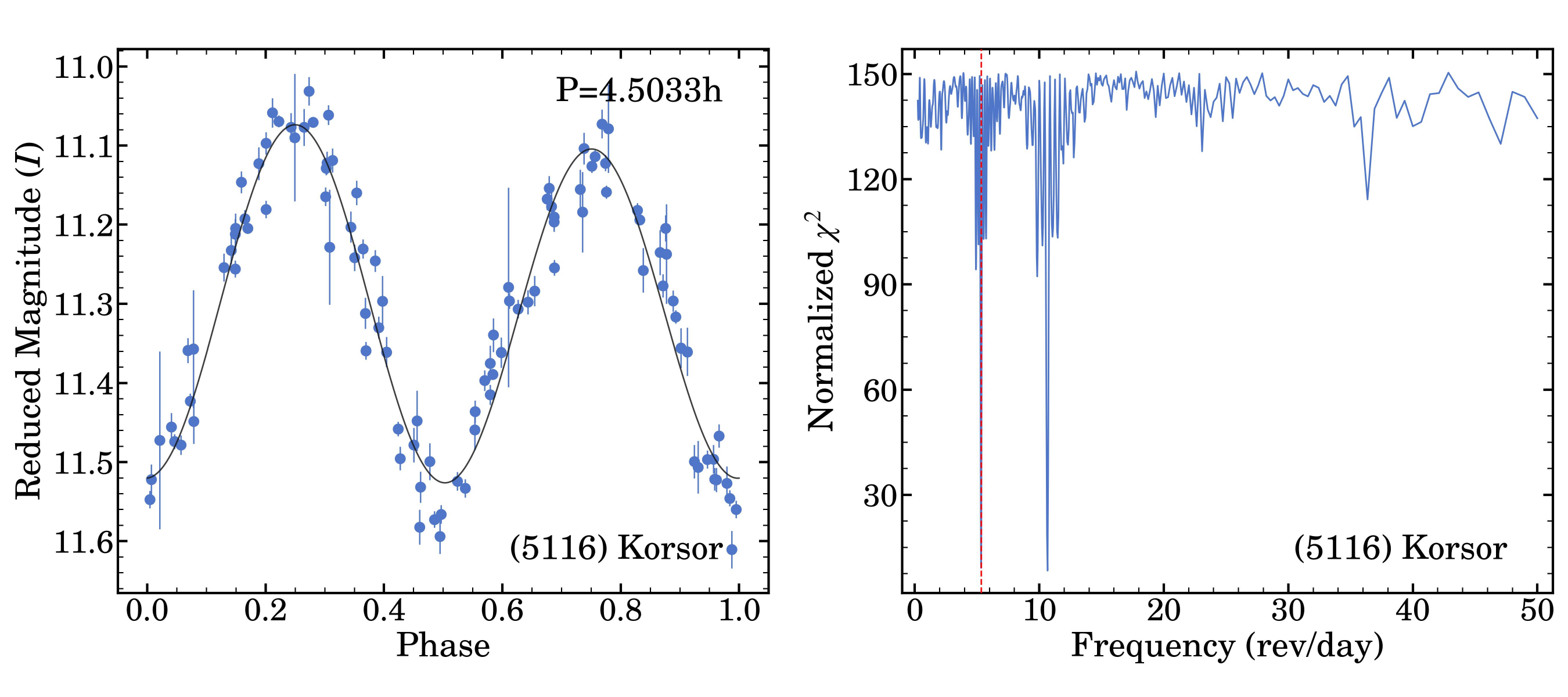}
    \caption{Folded lightcurve (left) and the normalized $\chi^2$ distribution as a function of frequency (right) for asteroid with rotation-period quality code U = 3. Only the first panels are shown here to illustrate the format; the remaining panels are provided in \HL{Appendix~\ref{appendix:figure}}. }
    \label{fig:U3}
\end{figure*}

\HL{Figure~\ref{fig:U3} shows an example of a folded lightcurve and normalised $\chi^2$ distribution of the lowest numbered asteroid for which we found the period with U = 3.} The folded lightcurves and normalized $\chi^2$ distribution of the remaining U = 3 objects, as well as those with U = 3− and U = 2+, are available in Appendix~\ref{appendix:figure}.

\subsection{Comparison with LCDB and Individual Case}

\begin{table*}
\caption{Comparison of the Rotation Period for the 12 Objects having known LCDB measurements.}
\label{tab2:comparison}
\centering
\small
\begin{tabular}{lccc}
\hline\hline
Object & Quality & LCDB period & Our derived period \\
       & (this work) & (h)         & (h)            \\
\hline
(1077) Campanula  & $2+$  & 3.8509  & $3.8378 \pm 0.0008$ \\
(1152) Pawona     & $2+$  & 3.4154  & $3.4348 \pm 0.0004$ \\
(1187) Afra       & $2+$  & 14.0701 & $13.8938 \pm 0.0119$ \\
(1324) Knysna     & $2+$  & 2.5538  & $2.5503 \pm 0.0007$ \\
(2294) Andronikov & $3-$  & 3.1529  & $3.1534 \pm 0.0001$ \\
(2438) Oleshko    & $2+$  & 3.2270  & $3.2222 \pm 0.0003$ \\
(3180) Morgan     & $3-$  & 2.4477  & $2.4481 \pm 0.0004$ \\
(3314) Beals      & $2+$  & 5.4616  & $5.4650 \pm 0.0039$ \\
(4362) Carlisle   & $2+$  & 2.6329  & $2.6320 \pm 0.0002$ \\
(5116) Korsor     & $3 $  & 4.5030  & $4.5033 \pm 0.0004$ \\
(6399) Harada     & $3 $  & 11.0200 & $10.9662 \pm 0.0002$ \\
(7284) 1989~VW    & $2+$  & 26.4500 & $26.8283 \pm 0.0549$ \\
\hline
\end{tabular}
\end{table*}

Among the 96 reliable rotation periods, \HL{12 of them also have rotation periods published} with U $\geq$ 2+ in the LCDB. Table~\ref{tab2:comparison} shows a comparison of rotation periods for the 12 objects. 

Most of the derived rotation periods are consistent with previously published results. One notable exception is asteroid (15699) Lyytinen. In both datasets, the rotation period of this object is assigned a quality code of U = 2−, indicating a tentative but not fully reliable solution.

\begin{figure*}
    \centering
    \includegraphics[width=1\linewidth]{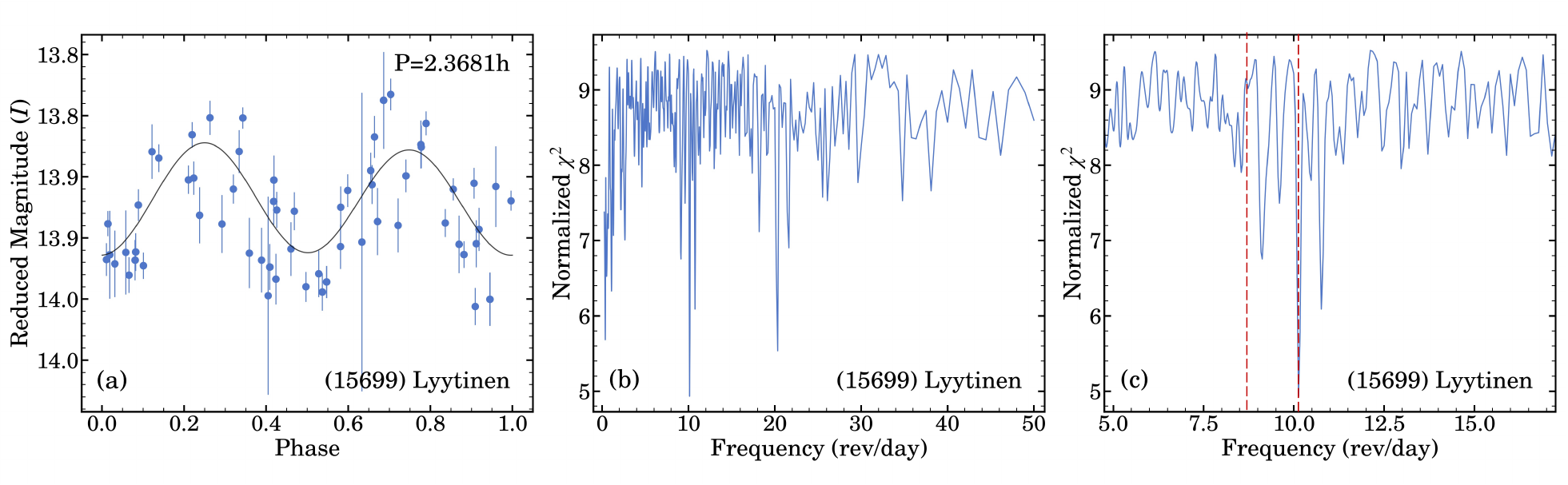}
    \caption{Phase-folded lightcurve and the normalized $\chi^2$ distribution as a function of frequency for asteroid (15699) Lyytinen. Panel (a) shows the folded lightcurve, (b) presents the normalized $\chi^2$ distribution as a function of frequency, and (c) is a zoom-in of panel (b) around the best-fit frequency. The two dashed lines indicate the LCDB (left) frequency and our measured period (right).}
    \label{fig:15699}
\end{figure*}

In our data, the folded lightcurve of (15699) Lyytinen is characterized by a relatively small number of photometric measurements, and the data are noticeably scattered. This limited sampling is primarily due to the short time span during which the asteroid traverses the field of view, resulting in fewer observational epochs than ideal. The corresponding $\chi^2$–frequency distribution exhibits multiple local minima, leading to several competing candidate periods and making it difficult to uniquely identify a preferred solution. Based on these limitations, we conservatively assign a quality code of U = $2−$ to this object. Among the candidate solutions, the formally best-fit period is 2.3681~h, which is somewhat shorter than the previously reported LCDB value of 2.7818~h, and the associated uncertainty remains non-negligible (Figure~\ref{fig:15699}).

The similarly low quality rating reported in the LCDB suggests that the ambiguity in the rotation period determination is not unique to our dataset, but may reflect an intrinsic difficulty in characterizing the rotational properties of this object. One possible explanation is that the asteroid exhibits a complex or non-principal axis rotation state (i.e., tumbling), which can produce irregular or non-periodic lightcurve variations that are not well described by a single-period model \citep{1985Icarus-Harris-Tumbling,2005Icarus-Pravec-Tumbling}. Alternatively, the apparent ambiguity may arise from observational limitations, such as sparse temporal coverage or unfavorable viewing geometries, which can obscure or distort the underlying periodic signal \citep{1999Icarus-Harris-Equa}.

Given the current data quality, it is not possible to unambiguously distinguish between these scenarios. However, the persistent difficulty in obtaining a stable and high-quality period solution for (15699) Lyytinen makes it a valuable target for follow-up observations. A denser and longer lightcurve coverage would be essential to confirm its rotational state and to determine whether its behavior is indeed indicative of more complex rotational dynamics.

In our sample, most ambiguous period determinations arise from insufficient photometric coverage of asteroids (typically fewer than 50 epochs), primarily due to limited observational epochs, with a minor contribution from data gaps caused by adverse weather conditions. When adequate coverage is available, our measured rotation periods are generally robust and reliable.

\subsection{Statistical Properties}

\begin{figure*}
    \centering
    \includegraphics[width=0.7\linewidth]{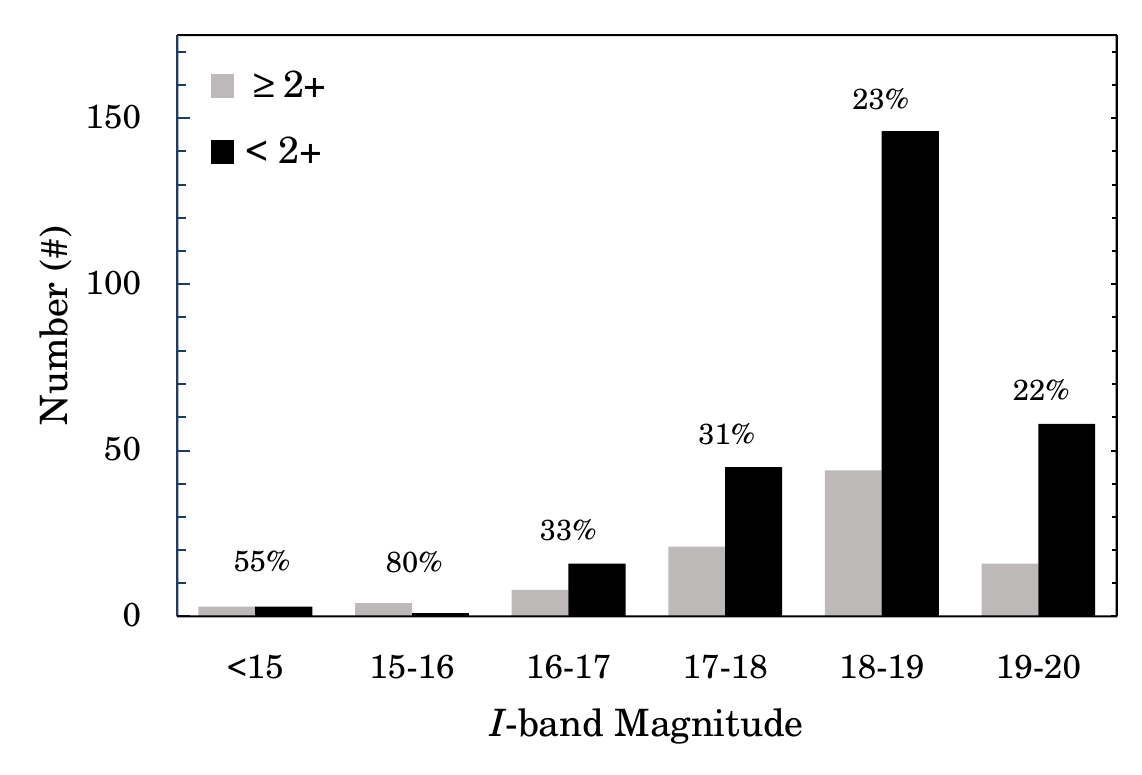}
    \caption{Asteroid number distribution of $I$-band magnitude, grouped in quality codes. The gray and black colors represent the rotation periods of U $\geq$ 2+ and U < 2+ groups, respectively. The fraction listed above each magnitude bin indicates the fraction of U $\geq$ 2+ objects.}
    \label{fig:stability}
\end{figure*}

\begin{figure*}
    \centering
    \includegraphics[width=0.9\linewidth]{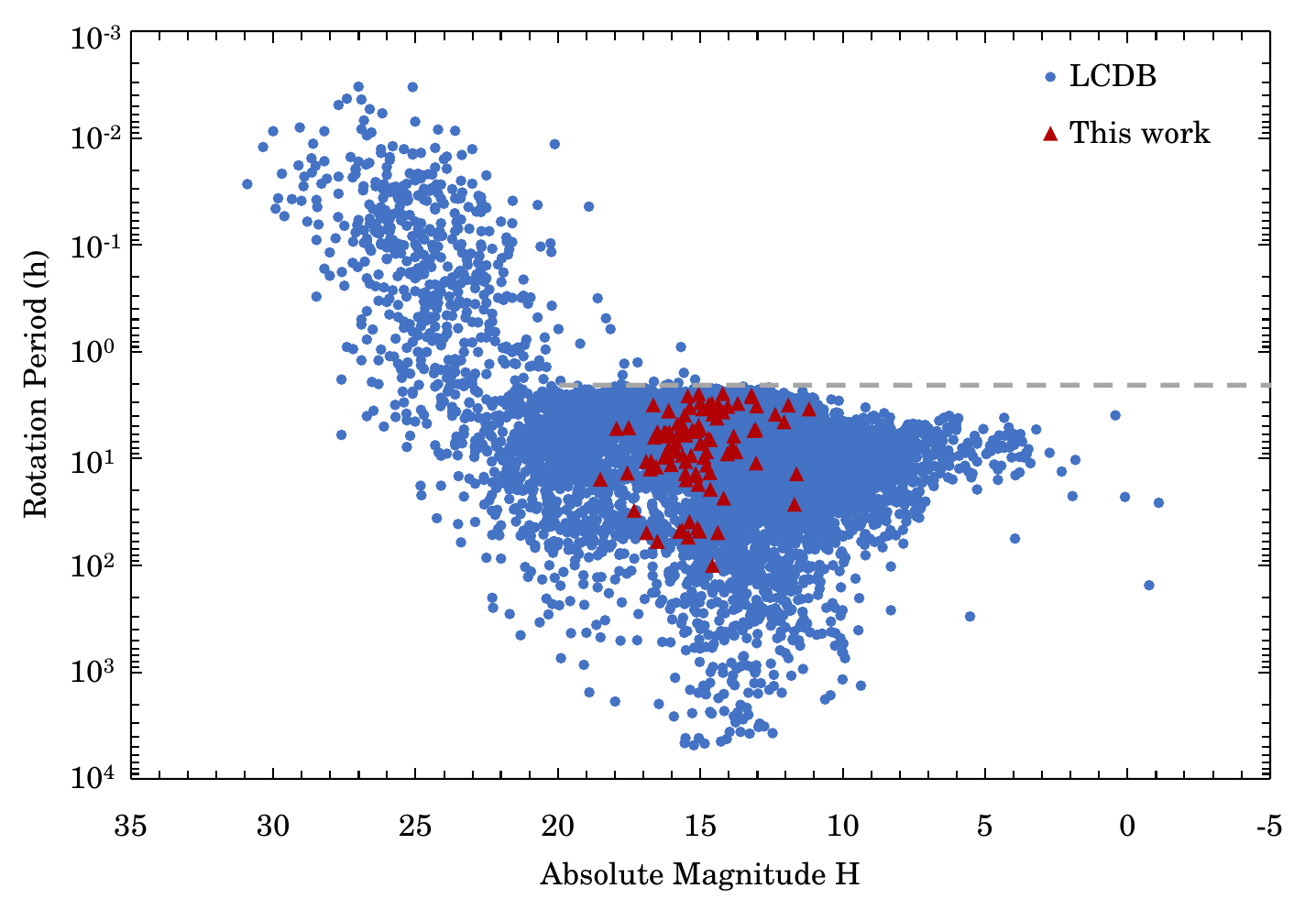}
    \caption{The magnitude–period two-dimensional distribution of asteroids with reliable rotation periods derived in this work, compared with those from the LCDB-BASIC \citep{2009Icarus-Warner-LCDB}. The gray dashed line indicates the asteroid spin barrier at $\sim$2.2~h.}
    \label{fig:distribution}
\end{figure*}

\begin{figure*}
    \centering
    \includegraphics[width=0.7\linewidth]{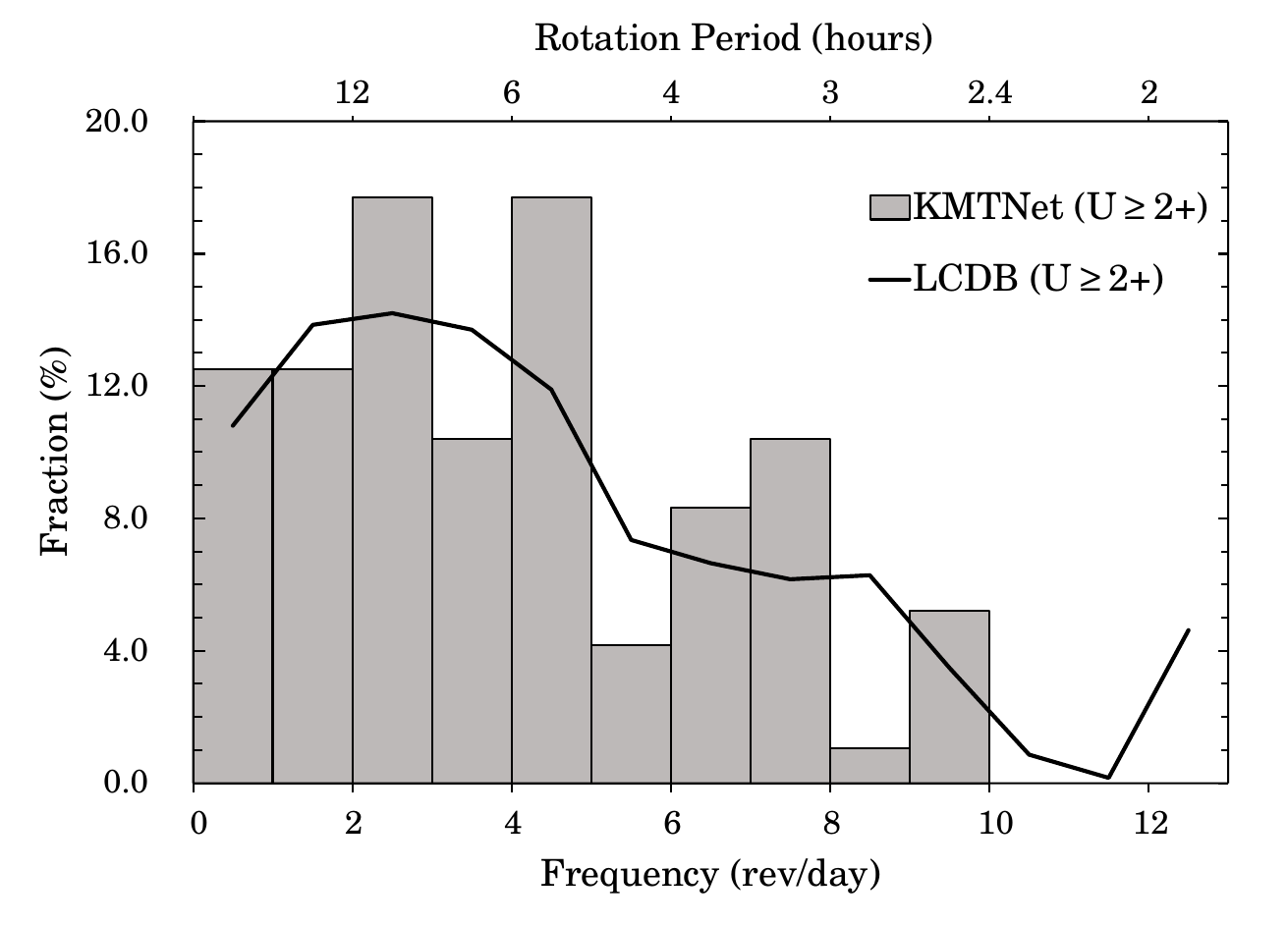}
    \caption{Density distribution of the asteroid spin frequency of U $\geq$ 2+ asteroids derived from this work compared to \HL{that of} the LCDB U $\geq$ 2+ asteroids. The top axis showing the corresponding rotation period.}
    \label{fig:Frequency}
\end{figure*}

Figure~\ref{fig:stability} shows the distribution of the quality code as a function of the magnitude. The total number of asteroids exhibits a unimodal distribution, with a clear maximum in the 18--19~mag interval, where both high-quality (U $\geq$ 2+) and lower-quality (U < $2+$) solutions reach their highest counts. Toward both brighter and fainter magnitudes, the number of objects decreases progressively. This behavior reflects the combined effects of the intrinsic asteroid brightness distribution and the observational selection function of our survey. 

The decline toward brighter magnitudes is primarily driven by the intrinsic scarcity of large (and thus bright) asteroids. In contrast, the decrease at the faint end is largely a consequence of the magnitude-limited sample selection. Although our sample is selected with a limiting magnitude of $V \leq 20$, the use of the mean instrumental $I$-band magnitude introduces a systematic offset between the selection band and the analysis band. As a result, objects with $I$-band magnitudes in the range of 19--20~mag are underrepresented compared to those in the 18--19~mag bin. 

The similar unimodal behavior observed in both the U $\geq$ 2+ and U < 2+ subsamples indicates that the quality classification does not introduce a strong additional bias as a function of magnitude within the well-sampled range.  Therefore, we suggest that our sample is broadly representative within the well-sampled magnitude range, providing a robust foundation for the subsequent statistical analysis of spin-rate properties.                                      
As shown in Figure~\ref{fig:distribution}, the reliable rotation periods derived from KMTNet observations span a wide range, from a few hours to over 100~h, with the shortest periods approaching the well-known spin barrier at $\sim$2.2~h. This limit is generally interpreted as a consequence of the structural and compositional properties of asteroids, representing the critical rotation period beyond which gravitationally bound rubble-pile bodies would undergo rotational disruption \citep{1984Icarus-Stanly-spin,2000Icarus-Pravec-shape}.

To further quantify the overall characteristics of this distribution, Figure~\ref{fig:Frequency} presents the spin-rate distribution of the asteroids in our sample and compares it with that reported in the LCDB. The primary horizontal axis shows the spin rate, while the secondary axis indicates the corresponding rotation period. The two distributions exhibit a generally consistent overall shape, with the majority of objects concentrated in the 1–5 rev/day range, broadly consistent with the well-established peak from previous studies. In terms of rotation period, most objects cluster between 6 and 12~h, and no objects with rotation periods shorter than $\sim$2.2~h are found in our sample.

The absence of very fast rotators in our sample is likely primarily due to the magnitude-limited selection ($V \leq 20$), which biases the sample toward brighter and therefore typically larger asteroids that are less likely to exhibit extremely short rotation periods. Nevertheless, the near-continuous temporal coverage enabled by the multi-site observations of KMTNet, together with its photometric depth, suggests that extending the analysis to fainter targets and broader sky coverage should improve sensitivity to short-period rotators, potentially enabling the detection of objects approaching or even below the spin barrier. This interpretation is also supported by Figure~\ref{fig:stability}, where the fraction of U $\geq$ 2+ solutions does not show a sharp decline toward fainter magnitudes, indicating that reliable period determinations remain achievable beyond the adopted sample limit.

\HL{We compared the period distributions of the reliable and less reliable samples. The latter have substantially longer periods (mean/median: 47.54/44.85~h) than the reliable sample (12.57/6.21~h), and also fewer observation epochs (134.0 vs. 176.4). We attribute this trend primarily to the limited field coverage (only one square degree): long-period asteroids may exit the field before sufficient rotational phase coverage is obtained, making their period determination less reliable. Furthermore, for those already classified as less reliable, the reported period values themselves may not be accurate. A future study over a larger area could help resolve this bias.}

\section{Discussion and Conclusion}\label{sec:dis}

\subsection{Estimated Yields from the KMTNet Prime-Field Archive}

To estimate the potential of the KMTNet archive for asteroid studies, we performed a systematic estimation of the number of asteroids traversing the $\sim12$~deg$^2$ KMTNet $\geq4$~hr$^{-1}$ cadence prime fields \citep{KMTNet2018EF} during the 2016–2025 period. 

For each year, we considered an eight-month observing window from February 20 to October 20, corresponding to the typical Galactic bulge season. Representative observing epochs were adopted for each of the three KMTNet sites to approximate the temporal sampling: 02:30 and 06:30 UT for Chile, 12:30 and 14:30 UT for Australia, and 18:30 and 22:30 UT for South Africa. The analysis is restricted to the prime fields \citep{KMTNet2018EF}. For consistency in the yield estimation and to reflect the total discovery potential of KMTNet, asteroids identified within the 1~deg$^2$ pilot field analyzed in this work (from the 2018 dataset) were not excluded from the calculation. 

Based on the adopted observational sampling, we find that a total of 160,616 asteroids pass through the KMTNet fields over the considered time span. Applying a magnitude cut of $V \leq 21$, consistent with the expected photometric limit of KMTNet, yields 31,535 asteroids that are potentially detectable. Among these, 1,564 objects traverse the field more than once during this 10-year period, including 1,442 asteroids observed twice and 122 observed three times over the ten-year baseline. 

In our sample, approximately 25\% of the asteroids yield well-determined rotation periods ($U \geq 2+$). However, this recovery fraction is expected to decrease toward fainter magnitudes due to the lower signal-to-noise ratio and reduced lightcurve quality. Adopting a conservative recovery fraction of 20\% for the full sample, we estimate that more than 6,300 asteroids could yield well-determined rotation periods. Among them, about \HL{5,500} would be new measurements. This would represent a substantial increase compared to the current sample size of 8,998 objects with known rotation periods of quality code U $\geq$ 2+ in the LCDB-BASIC (Figure~\ref{fig:distribution}), highlighting the considerable potential of the KMTNet archive for asteroid rotational studies.

In addition, asteroids that are observed over multiple seasons constitute a particularly valuable subsample. For these objects, the extended temporal baseline offers the opportunity to investigate potential variations in rotation period over timescales of several years, which may provide insights into rotational evolution processes such as the YORP effect or external perturbations.

\subsection{Future Surveys}

Future high-cadence time-domain surveys will further expand the potential of this approach. For example, the DECam Rogue Earths and Mars Survey (DREAMS) \citep{DREAMS_DR1} employs the 3-deg$^2$ DECam instrument on the 4-m Blanco telescope and is observing the Galactic bulge field at minute-level cadence with substantially deeper photometric limits. The minute-cadence sampling makes it possible to detect and characterize very fast and super-fast rotators with rotation periods shorter than $\sim$2.2~h, which are particularly important for constraining the internal structure and cohesion of small asteroids. 

In addition, compared to KMTNet, the $>1.5$~mag deeper limiting magnitude of DREAMS will enable the detection of significantly fainter objects, potentially extending asteroid lightcurve studies to $V \geq 22$. Given the steep increase in asteroid number counts toward fainter magnitudes, this capability may increase the accessible sample size by orders of magnitude, further enriching the statistical sample of asteroid rotational properties. 

Other large-scale surveys, such as the Vera C. Rubin Observatory Legacy Survey of Space and Time (LSST), will provide an important complementary capability. Although the nominal cadence of the main LSST survey is generally not optimized for detailed asteroid lightcurve characterization, recent high-cadence commissioning observations have demonstrated the strong potential of Rubin data for asteroid rotation studies \citep{2026AJ-Greenstreet-LSST}, particularly because of its substantially wider sky coverage and much larger asteroid sample compared with focused high-cadence surveys. Meanwhile, TESS offers stable high-cadence space-based photometry with nearly continuous temporal coverage, while KMTNet enables long-duration multi-site monitoring over consecutive nights. Together, these facilities provide highly complementary capabilities in cadence, depth, temporal coverage, and survey area.

In the future, combining datasets from multiple time-domain surveys will significantly improve the characterization of asteroid populations across a broad range of sizes and dynamical classes. The combination of high cadence, deep photometric limits, continuous monitoring, and ultra-wide sky coverage will not only enlarge the sample of asteroids with reliable rotational measurements, but also improve sensitivity to faint, fast-rotating, and previously unknown objects, thereby enabling more comprehensive investigations of asteroid physical properties, rotational evolution, and the broader small-body population of the Solar System.

\subsection{Summary}

In this work, we show that KMTNet Galactic bulge observations can be effectively exploited to extract reliable rotation periods of asteroids, even within extremely crowded stellar fields toward the Galactic center. Using archival data from the 2018 bulge season and focusing on a one-square-degree field centered at ($\alpha$, $\delta$) = (269.22$^\circ$, −29.58$^\circ$), we extract lightcurves for asteroids with $V < 20$ and derive 96 reliable rotation periods ($U \geq 2+$). 

Among these, 84 asteroids have no previously published lightcurve information, representing a substantial addition to the existing sample, while the remaining 12 objects with previously reported periods in the LCDB show good agreement with our measurements. 
These results demonstrate that small bodies traversing dense stellar regions can be robustly detected and their photometric variations accurately characterized despite the challenging background conditions, highlighting the untapped potential of KMTNet bulge observations for Solar System studies.

Although based on a single pilot field, a systematic calculation based on the full KMTNet survey cadence, sky coverage, and multi-year baseline indicates that the archival dataset has the potential to yield $\sim$5500 previously uncharacterized asteroids with reliable rotation periods, significantly expanding current rotation statistics. 


\acknowledgments

H.H., H.Y., Q.Q., T.G., and S.M. acknowledge support by the National Natural Science Foundation of China (Grant No. 12133005). H.Y. acknowledge support by the China Postdoctoral Science Foundation (No. 2024M762938).
The authors thank Dr. In-Gu Shin for helpful discussions regarding the KMTNet observations.
This research has made use of the KMTNet system operated by the Korea Astronomy and Space Science Institute (KASI) at three host sites of CTIO in Chile, SAAO in South Africa, and SSO in Australia. Data transfer from the host site to KASI was supported by the Korea Research Environment Open NETwork (KREONET). This research was supported by KASI under the R\&D program (project No. 2026-1-904-01) supervised by the Ministry of Science and ICT. 
The authors acknowledge the High-performance Computing center at Westlake University for providing computational and data storage resources that have contributed to the research results reported within this paper.

\bibliography{Huang.bib}

\appendix

\section{Asteroids List}\label{appendix:table}

This appendix provides the derived rotation periods and associated parameters for the asteroid sample. Table~\ref{appendix table3} lists the reliable rotation period determinations (U $\geq2+$), while Table~\ref{appendix table4} presents the less reliable rotation period determinations (U < 2+). Only asteroids with measurable rotation periods are included; objects for which no rotation period could be determined owing to insufficient data are excluded.

\begin{longtable}{lccccccc}
\label{appendix table3}\\
\caption{List of reliable rotation periods. $^{a}$ Derived from the MPCORB database.} \\
\hline\hline
Object & Period & Quality & $H^{a}$ & Magnitude & Amplitude & \HL{Epochs} & $Phase~Angle^{a}$ \\
       & (h)    &         & (mag) & (\textit{I}-band)  & (mag)   &     & ($^\circ$) \\
\hline
\endfirsthead

\caption{Continued.} \\
\hline\hline
Object & Period & Quality & $H^{a}$ & Magnitude & Amplitude & \HL{Epochs} & $Phase~Angle^{a}$ \\
       & (h)    &         & (mag) & (\textit{I}-band)  &  (mag)  &     & ($^\circ$) \\
\hline
\endhead

\hline
\endfoot

\hline\hline
\endlastfoot

(1077) Campanula   & $3.8378 \pm 0.0008$  & $2+$  & 12.36 & 15.407 & 0.4375 & 48 & 26.4939 \\
(1152) Pawona      & $3.4348 \pm 0.0004$  & $2+$  & 11.17 & 14.873 & 0.2216 & 83 & 24.4671 \\
(1187) Afra        & $13.8938 \pm 0.0119$ & $2+$  & 11.62 & 16.150 & 0.4563 & 47 & 21.0534 \\
(1324) Knysna      & $2.5503 \pm 0.0007$  & $2+$  & 13.20 & 15.548 & 0.1308 & 87 & 30.2324 \\
(2294) Andronikov  & $3.1534 \pm 0.0001$  & $3-$  & 11.91 & 14.925 & 0.4518 & 189 & 14.1802 \\
(2438) Oleshko     & $3.2222 \pm 0.0003$  & $2+$  & 13.02 & 15.660 & 0.2007 & 142 & 24.7712 \\
(3180) Morgan      & $2.4481 \pm 0.0004$  & $3-$  & 14.21 & 17.809 & 0.2003 & 200 & 22.0880 \\
(3314) Beals       & $5.4650 \pm 0.0039$  & $2+$  & 13.11 & 17.038 & 1.3315 & 24 & 23.7157 \\
(4362) Carlisle    & $2.6320 \pm 0.0002$  & $2+$  & 13.18 & 16.178 & 0.1878 & 170 & 24.5133 \\
(5116) Korsor      & $4.5033 \pm 0.0004$  & $3 $  & 12.06 & 16.349 & 0.4812 & 114 & 2.2259 \\
(6399) Harada      & $10.9662 \pm 0.0002$ & $3 $  & 13.04 & 16.226 & 1.2739 & 267 & 22.8387 \\
(6690) Messick     & $3.0308 \pm 0.0003$  & $3-$  & 13.68 & 14.568 & 0.2339 & 106 & 3.4887 \\
(7284) 1989 VW     & $26.8283 \pm 0.0549$ & $2+$  & 11.69 & 17.585 & 0.2260 & 191 & 6.0828 \\
(10229) 1997 WR3   & $3.2508 \pm 0.0004$  & $3-$  & 14.03 & 15.978 & 0.2766 & 166 & 11.4435 \\
(12015) 1996 WA    & $5.2515 \pm 0.0039$  & $2+$  & 15.00 & 17.755 & 0.1395 & 133 & 10.3298 \\
(13564) Kodomomiraikan & $2.9924 \pm 0.0007$ & $3-$ & 15.00 & 17.501 & 0.3484 & 162 & 2.6714 \\
(16673) 1994 BF1   & $3.1515 \pm 0.0008$  & $3-$  & 14.37 & 17.234 & 0.2178 & 162 & 2.4585 \\
(17039) Yeuseyenka & $9.0525 \pm 0.0118$  & $2+$  & 14.05 & 18.274 & 0.1753 & 126 & 21.2855 \\
(17234) 2000 EL11  & $6.6400 \pm 0.0002$  & $3-$  & 14.64 & 18.516 & 0.6868 & 332 & 11.6672 \\
(17262) Winokur    & $6.4775 \pm 0.0004$  & $3-$  & 14.71 & 18.286 & 0.2399 & 753 & 18.6021 \\
(20753) 2000 AW211 & $6.1301 \pm 0.0004$  & $3-$  & 13.83 & 17.599 & 1.0858 & 138 & 23.2769 \\
(20917) 5016 P-L   & $54.4984 \pm 0.1515$ & $2+$  & 15.42 & 17.602 & 0.3816 & 191 & 25.7187 \\
(23444) Kukucin    & $8.5928 \pm 0.0010$  & $3 $  & 13.73 & 16.229 & 0.4395 & 170 & 22.6733 \\
(23616) 1996 HY10  & $23.3896 \pm 0.0244$ & $2+$  & 14.18 & 18.693 & 0.2901 & 291 & 18.3271 \\
(27113) 1998 VY54  & $7.1788 \pm 0.0087$  & $3-$  & 15.00 & 18.458 & 0.7787 & 61 & 27.2459 \\
(28149) Arieldaniel& $48.1271 \pm 0.1530$ & $2+$  & 15.72 & 18.986 & 0.1814 & 316 & 19.6338 \\
(29377) 1996 GV18  & $8.6125 \pm 0.0012$  & $3-$  & 14.77 & 17.863 & 0.6702 & 323 & 13.4368 \\
(31929) 2000 GF79  & $3.9052 \pm 0.0062$  & $2+$  & 15.55 & 18.339 & 0.6843 & 31 & 33.8674 \\
(34843) 2001 SZ276 & $3.0965 \pm 0.0008$  & $3-$  & 14.69 & 17.730 & 0.3522 & 123 & 7.0559 \\
(36342) 2000 NX15  & $10.6395 \pm 0.0027$ & $3-$  & 15.50 & 19.487 & 0.6603 & 388 & 18.3337 \\
(39414) 3283 T-1   & $8.5067 \pm 0.0227$  & $2+$  & 15.85 & 19.128 & 0.2670 & 120 & 24.8223 \\
(43749) 1981 EG46  & $5.6297 \pm 0.0013$  & $2+$  & 16.07 & 18.721 & 0.1831 & 425 & 19.2054 \\
(45119) 1999 XA86  & $5.2865 \pm 0.0005$  & $3-$  & 13.05 & 16.681 & 0.4802 & 147 & 17.7351 \\
(45711) 2000 FD43  & $3.3310 \pm 0.0035$  & $2+$  & 14.53 & 18.892 & 0.3808 & 90 & 20.8770 \\
(46339) 2001 RU81  & $4.1786 \pm 0.0003$  & $3 $  & 14.41 & 17.503 & 0.6991 & 178 & 9.5006 \\
(48986) 1998 QJ47  & $4.9197 \pm 0.0007$  & $3-$  & 15.06 & 17.411 & 0.8307 & 81 & 3.0105 \\
(50739) Gracecook  & $7.7088 \pm 0.0026$  & $2+$  & 13.89 & 16.782 & 0.4636 & 91 & 24.3527 \\
(51905) 2001 QM51  & $19.4273 \pm 0.1946$ & $2+$  & 14.65 & 18.339 & 0.3014 & 60 & 7.8539 \\
(53834) 2000 ES179 & $17.4201 \pm 0.0213$ & $2+$  & 15.07 & 16.487 & 0.2494 & 156 & 4.9848 \\
(55585) 2002 PQ45  & $11.3697 \pm 0.0007$ & $3-$  & 16.02 & 17.847 & 0.7716 & 185 & 7.5894 \\
(58243) 1993 NG1   & $9.3101 \pm 0.0007$  & $2+$  & 15.32 & 18.615 & 0.2318 & 287 & 15.8864 \\
(58631) 1997 WE2   & $9.6653 \pm 0.0109$  & $2+$  & 14.87 & 17.019 & 0.8775 & 45 & 8.0982 \\
(59268) 1999 CU34  & $44.7391 \pm 0.0800$ & $2+$  & 15.10 & 18.512 & 0.2640 & 326 & 19.4154 \\
(68155) 2001 BM9   & $5.9969 \pm 0.0077$  & $2+$  & 15.51 & 18.741 & 0.3538 & 110 & 26.8556 \\
(70818) 1999 VJ77  & $5.3952 \pm 0.0027$  & $2+$  & 15.27 & 19.161 & 0.3914 & 219 & 12.5757 \\
(71628) 2000 EJ69  & $3.6411 \pm 0.0006$  & $2+$  & 14.20 & 19.553 & 0.2034 & 652 & 14.3664 \\
(74164) 1998 QL104 & $4.6043 \pm 0.0026$  & $2+$  & 15.65 & 18.642 & 0.3799 & 133 & 23.5719 \\
(74179) 1998 RZ23  & $5.4627 \pm 0.0042$  & $3 $  & 15.55 & 18.259 & 0.8135 & 60 & 30.2836 \\
(74455) 1999 CW28  & $3.8223 \pm 0.0010$  & $3-$  & 14.55 & 18.503 & 0.6268 & 190 & 2.1568 \\
(77143) 2001 EN3   & $49.0609 \pm 0.1328$ & $2+$  & 16.88 & 18.812 & 0.2190 & 555 & 18.5822 \\
(80017) 1999 GQ39  & $13.7048 \pm 0.0087$ & $2+$  & 15.16 & 17.809 & 0.3048 & 236 & 17.2110 \\
(81055) 2000 EJ65  & $39.0776 \pm 0.0024$ & $2+$  & 15.37 & 18.736 & 1.2282 & 110 & 25.5256 \\
(81146) 2000 EW142 & $13.7950 \pm 0.0513$ & $2+$  & 15.52 & 19.475 & 0.7600 & 53 & 26.5698 \\
(82123) 2001 FY77  & $46.8159 \pm 0.0094$ & $2+$  & 15.64 & 17.521 & 0.1841 & 228 & 16.9608 \\
(83278) 2001 RQ84  & $3.3259 \pm 0.0012$  & $3-$  & 15.36 & 17.914 & 0.2819 & 139 & 4.0799 \\
(84674) 2002 VT87  & $5.5964 \pm 0.0108$  & $2+$  & 15.54 & 19.219 & 0.2681 & 183 & 1.9749 \\
(84964) 2003 YF14  & $14.4092 \pm 0.0148$ & $2+$  & 15.16 & 18.785 & 0.2188 & 302 & 15.8500 \\
(87265) 2000 OL58  & $47.4877 \pm 0.2020$ & $2+$  & 15.03 & 18.929 & 0.3220 & 214 & 20.0696 \\
(89573) 2001 XD118 & $6.0378 \pm 0.0011$  & $2+$  & 15.78 & 18.628 & 0.7110 & 206 & 15.6900 \\
(94431) 2001 TZ41  & $4.7054 \pm 0.0024$  & $2+$  & 15.80 & 19.183 & 0.2722 & 229 & 16.4916 \\
(97595) 2000 EX37  & $49.1708 \pm 0.3245$ & $2+$  & 14.38 & 18.478 & 0.2023 & 281 & 10.5501 \\
(98355) 2000 SZ335 & $2.5911 \pm 0.0014$  & $2+$  & 15.43 & 17.968 & 0.2388 & 95 & 8.4357 \\
(100945) 1998 OE8  & $9.7430 \pm 0.0002$  & $3-$  & 16.23 & 18.463 & 0.9910 & 123 & 17.5123 \\
(101221) 1998 SB62 & $6.1311 \pm 0.0002$  & $2+$  & 16.20 & 18.629 & 0.5877 & 174 & 13.4155 \\
(103266) 2000 AT26 & $13.5153 \pm 0.0072$ & $3-$  & 14.66 & 18.430 & 0.7134 & 210 & 3.1524 \\
(105939) 2000 SY227& $9.1709 \pm 0.0004$  & $3 $  & 16.13 & 18.143 & 0.7499 & 182 & 10.4101 \\
(107095) 2001 AA30 & $7.9434 \pm 0.0262$  & $2+$  & 16.11 & 18.877 & 0.2249 & 111 & 3.4438 \\
(107806) 2001 FZ58 & $6.9973 \pm 0.0020$  & $3 $  & 15.85 & 17.312 & 0.6329 & 91 & 6.6548 \\
(108332) 2001 KP2  & $3.5687 \pm 0.0037$  & $2+$  & 16.10 & 18.884 & 1.1915 & 30 & 35.9095 \\
(116375) 2003 YU111& $11.4796 \pm 0.0084$ & $3-$  & 14.78 & 19.044 & 0.8060 & 223 & 4.3588 \\
(117092) 2004 NB7  & $5.5600 \pm 0.0002$  & $3-$  & 15.20 & 18.333 & 0.4220 & 228 & 11.6151 \\
(127443) 2002 PX69 & $3.0104 \pm 0.0005$  & $3-$  & 14.58 & 17.716 & 0.2525 & 255 & 9.9252 \\
(128072) 2003 OV9  & $2.4736 \pm 0.0012$  & $2+$  & 15.06 & 18.256 & 0.1525 & 224 & 6.0009 \\
(136881) 1998 FL133& $10.5715 \pm 0.0048$ & $3-$  & 16.90 & 18.237 & 0.7585 & 144 & 8.7321 \\
(141541) 2002 GL14 & $10.5601 \pm 0.0003$ & $2+$  & 16.72 & 18.952 & 0.9825 & 102 & 25.5147 \\
(148858) 2001 VH56 & $15.7087 \pm 0.0819$ & $2+$  & 15.49 & 18.775 & 0.2896 & 107 & 2.4846 \\
(149647) 2004 FC38 & $12.4895 \pm 0.0283$ & $2+$  & 16.75 & 18.767 & 0.3547 & 138 & 7.9643 \\
(201332) 2002 TV160& $3.4291 \pm 0.0017$  & $3-$  & 14.90 & 17.480 & 0.2505 & 62 & 2.8104 \\
(208317) 2001 MF3  & $6.2835 \pm 0.0082$  & $2+$  & 16.61 & 18.207 & 0.2616 & 147 & 8.7252 \\
(211013) 2001 YK33 & $5.6454 \pm 0.0045$  & $2+$  & 16.21 & 19.290 & 0.5907 & 86 & 2.4928 \\
(217434) 2005 SY95 & $5.9390 \pm 0.0030$  & $3-$  & 15.77 & 18.345 & 0.5383 & 138 & 10.6327 \\
(251421) 2008 AR81 & $11.8864 \pm 0.0245$ & $2+$  & 16.55 & 19.014 & 0.4151 & 136 & 7.2758 \\
(252478) 2001 UY44 & $3.5498 \pm 0.0001$  & $2+$  & 16.11 & 18.837 & 0.4019 & 176 & 10.2614 \\
(283926) 2004 NT8  & $13.6128 \pm 0.0078$ & $2+$  & 17.55 & 19.102 & 0.2815 & 360 & 23.0960 \\
(306902) 2001 TV171& $3.1264 \pm 0.0014$  & $2+$  & 16.65 & 18.394 & 0.2426 & 128 & 17.9808 \\
(307441) 2002 UQ28 & $100.3625 \pm 0.4320$& $2+$  & 14.57 & 19.264 & 0.6446 & 242 & 18.3822 \\
(307779) 2003 WK64 & $59.4284 \pm 0.5341$ & $2+$  & 16.50 & 19.309 & 0.9659 & 137 & 11.3891 \\
(367941) 2012 DZ12 & $5.7286 \pm 0.0035$  & $2+$  & 16.27 & 18.421 & 0.4193 & 104 & 3.3783 \\
(394438) 2007 PJ35 & $30.5142 \pm 0.1717$ & $2+$  & 17.32 & 18.433 & 0.2350 & 157 & 13.1460 \\
(400148) 2006 UT288& $5.5949 \pm 0.0048$  & $2+$  & 16.50 & 18.953 & 0.5920 & 66 & 5.4733 \\
(434102) 2002 JY100& $5.2161 \pm 0.0159$  & $3-$  & 17.93 & 18.609 & 0.4315 & 52 & 4.8421 \\
(639562) 2017 FD53 & $5.4108 \pm 0.0007$  & $3-$  & 15.13 & 18.944 & 0.4859 & 305 & 19.1602 \\
(661761) 2005 NT79 & $15.6609 \pm 0.0855$ & $2+$  & 18.51 & 19.054 & 0.6273 & 66 & 3.7807 \\
(666079) 2009 WM173& $9.3606 \pm 0.0099$  & $2+$  & 15.63 & 18.593 & 0.8881 & 84 & 7.0842 \\
(700192) 2000 ED211& $9.4238 \pm 0.0064$  & $3-$  & 16.24 & 19.225 & 0.5297 & 196 & 7.9677 \\
(742117) 2007 CB84 & $5.1335 \pm 0.0093$  & $2+$  & 17.51 & 19.398 & 0.8952 & 30 & 5.0937 \\

\end{longtable}

\clearpage

\begin{longtable}{lccccccc}
\caption{\HL{List of less reliable rotation periods.} $^{a}$ Derived from the MPCORB database.} \\
\hline\hline
Object & Period & Quality & $H^{a}$ & Magnitude & Amplitude & \HL{Epochs} & Phase \\
       & (h)    &         & (mag) & (\textit{I}-band)  & (mag)     &     & ($^\circ$) \\
\hline
\label{appendix table4}
\endfirsthead

\caption{Continued.} \\
\hline\hline
Object & Period & Quality & $H^{a}$ & Magnitude & Amplitude & \HL{Epochs} & Phase \\
       & (h)    &         & (mag) & (\textit{I}-band)  & (mag)     & (au)    & ($^\circ$) \\
\hline
\endhead

\hline
\endfoot

\hline\hline
\endlastfoot

(2472) Bradman & $56.2092 \pm 0.3750$ & $2-$ & 14.0 & 17.353 & 0.3283 & 77 & 25.3814 \\
(2899) Runrun Shaw & $23.7101 \pm 0.0198$ & $1+$ & 13.57 & 14.911 & 0.2438 & 183 & 9.6227 \\
(9764) Morgenstern & $12.4226 \pm 0.0040$ & $2-$ & 14.56 & 17.567 & 0.1373 & 171 & 5.7014 \\
\HL{(15699) Lyytinen} & $2.3682 \pm 0.0012$ & $2-$ & 14.62 & 17.670 & 0.1174 & 66 & 6.8504 \\
(18877) Stevendodds & $44.9685 \pm 0.0382$ & $2$ & 13.71 & 18.318 & 0.1152 & 381 & 15.0026 \\
(19734) 1999 XE175 & $47.8534 \pm 0.1026$ & $2$ & 14.2 & 18.946 & 0.1889 & 582 & 15.2399 \\
(27215) 1999 CK128 & $26.9112 \pm 0.1636$ & $2-$ & 13.99 & 17.96 & 0.3624 & 59 & 11.8447 \\
(28165) Bayanmashat & $13.7013 \pm 0.0164$ & $2-$ & 14.9 & 18.024 & 1.1257 & 64 & 28.9287 \\
(30818) 1990 RH2 & $46.1798 \pm 0.0951$ & $2$ & 15.31 & 18.914 & 0.2327 & 134 & 24.9217 \\
(31939) Thananon & $69.8331 \pm 0.0037$ & $2$ & 14.81 & 16.815 & 0.5612 & 140 & 20.2091 \\
(32357) 2000 QR124 & $51.6199 \pm 0.6893$ & $1+$ & 15.15 & 18.595 & 0.3553 & 73 & 25.0919 \\
(33062) 1997 VT2 & $36.5535 \pm 0.4157$ & $1+$ & 15.28 & 17.503 & 0.2532 & 69 & 8.1180 \\
(35523) 1998 FQ63 & $50.9474 \pm 0.6100$ & $1+$ & 14.18 & 17.817 & 0.2690 & 73 & 24.9356 \\
(36932) 2000 SK221 & $13.3921 \pm 0.0829$ & $2-$ & 14.27 & 19.267 & 0.7117 & 49 & 19.5345 \\
(37108) 2000 UG102 & $72.0379 \pm 0.5290$ & $1+$ & 15.12 & 17.785 & 0.1940 & 200 & 6.5183 \\
(39085) 2000 VW34 & $59.5706 \pm 0.8246$ & $1+$ & 14.3 & 18.426 & 0.4069 & 123 & 9.5215 \\
(42287) 2001 TE51 & $44.0706 \pm 0.2008$ & $2-$ & 15.66 & 17.423 & 0.2429 & 118 & 7.5268 \\
(43325) 2000 KY50 & $46.9862 \pm 0.2692$ & $2-$ & 14.06 & 16.457 & 0.3212 & 57 & 4.6746 \\
(43562) 2001 FE97 & $76.5544 \pm 0.4930$ & $2$ & 14.02 & 18.288 & 0.4266 & 142 & 14.9578 \\
(45394) 2000 AO132 & $39.3431 \pm 0.3160$ & $1+$ & 14.27 & 18.905 & 0.1613 & 133 & 18.9285 \\
(46980) 1998 SW156 & $18.2270 \pm 0.0009$ & $1+$ & 14.84 & 18.262 & 0.7326 & 98 & 24.2534 \\
(47664) 2000 CE54 & $180.0741 \pm 0.7711$ & $1+$ & 13.59 & 16.823 & 0.6819 & 175 & 3.1813 \\
(47690) 2000 CQ92 & $15.5726 \pm 0.0365$ & $1$ & 14.18 & 18.705 & 0.7098 & 82 & 9.3349 \\
(48867) 1998 HR67 & $13.6249 \pm 0.0892$ & $2$ & 14.99 & 17.5 & 0.2318 & 57 & 30.1097 \\
(50006) 2000 AY19 & $53.1549 \pm 0.6756$ & $1+$ & 15.15 & 18.564 & 0.2835 & 115 & 26.1825 \\
(50783) 2000 FE17 & $124.3644 \pm 0.5744$ & $2-$ & 14.9 & 18.595 & 0.7914 & 223 & 11.0326 \\
(51256) 2000 JB58 & $49.1065 \pm 0.3104$ & $2-$ & 14.52 & 17.594 & 0.1730 & 179 & 10.8052 \\
(54877) 2001 OU51 & $24.4071 \pm 0.2212$ & $2$ & 15.23 & 18.715 & 0.3978 & 65 & 26.4843 \\
(56126) 1999 CT31 & $14.3987 \pm 0.0023$ & $2$ & 15.21 & 17.702 & 0.1569 & 227 & 9.3990 \\
(56447) 2000 GR76 & $51.2280 \pm 0.3813$ & $2$ & 16.07 & 18.888 & 0.7592 & 125 & 16.5030 \\
(57171) 2001 QT20 & $50.8154 \pm 0.1933$ & $2$ & 15.46 & 18.107 & 0.2655 & 158 & 16.2304 \\
(58261) 1993 SD1 & $16.9603 \pm 0.0578$ & $1+$ & 15.0 & 17.231 & 0.1017 & 148 & 8.3756 \\
(58586) 1997 SG23 & $43.9946 \pm 0.1133$ & $2-$ & 15.62 & 18.831 & 0.6461 & 74 & 23.4172 \\
(59292) 1999 CN56 & $24.0643 \pm 0.0325$ & $2-$ & 15.09 & 18.414 & 0.1178 & 505 & 19.5274 \\
(62476) 2000 SH219 & $52.3341 \pm 0.7510$ & $2-$ & 14.17 & 19.048 & 0.6768 & 51 & 21.2614 \\
(66426) 1999 NW33 & $29.5877 \pm 0.2598$ & $2-$ & 15.33 & 18.639 & 0.2122 & 84 & 28.0022 \\
(66950) 1999 XQ11 & $21.9396 \pm 0.2319$ & $2-$ & 14.43 & 18.304 & 0.2556 & 101 & 7.7834 \\
(71588) 2000 DK68 & $23.0660 \pm 0.1926$ & $2-$ & 14.51 & 18.772 & 0.6749 & 73 & 3.7874 \\
(72747) 2001 FR121 & $4.7434 \pm 0.0005$ & $2$ & 14.17 & 18.279 & 0.1064 & 713 & 17.2128 \\
(72855) 2001 HX43 & $17.5550 \pm 0.1220$ & $2-$ & 14.77 & 18.952 & 0.6906 & 63 & 12.7518 \\
(76868) 2000 YC11 & $53.5829 \pm 0.5000$ & $1+$ & 15.46 & 17.81 & 0.2665 & 81 & 29.9274 \\
(77602) 2001 KZ26 & $67.3999 \pm 0.4939$ & $2-$ & 14.94 & 17.842 & 0.2192 & 160 & 21.2950 \\
(77684) 2001 NO & $208.5826 \pm 0.9087$ & $1+$ & 15.2 & 17.331 & 2.3970 & 117 & 2.9159 \\
(78743) 2002 TZ273 & $36.5630 \pm 0.6716$ & $2-$ & 14.43 & 18.145 & 0.3937 & 43 & 3.5201 \\
(79538) 1998 QN34 & $23.9277 \pm 0.0470$ & $2-$ & 15.65 & 17.756 & 0.2928 & 142 & 23.5240 \\
(80144) 1999 TY124 & $13.5813 \pm 0.0638$ & $2-$ & 16.56 & 18.816 & 0.2785 & 94 & 5.3624 \\
(86857) 2000 HW9 & $20.6596 \pm 0.1916$ & $1+$ & 15.86 & 17.691 & 0.1084 & 176 & 8.2107 \\
(86979) 2000 JW19 & $46.0559 \pm 0.6352$ & $1+$ & 15.58 & 18.532 & 0.2500 & 136 & 25.7173 \\
(88348) 2001 OT75 & $3.3026 \pm 0.0045$ & $2$ & 16.36 & 19.611 & 0.3100 & 139 & 24.6681 \\
(88607) 2001 QE296 & $46.0567 \pm 0.2387$ & $1+$ & 16.47 & 16.977 & 0.3040 & 74 & 8.9281 \\
(89019) 2001 TC90 & $36.9515 \pm 0.3346$ & $2-$ & 15.13 & 19.126 & 0.5542 & 92 & 23.3249 \\
(89691) 2001 YC70 & $51.9042 \pm 0.1057$ & $2$ & 16.07 & 18.92 & 0.2358 & 311 & 16.3178 \\
(89720) 2001 YQ122 & $84.1470 \pm 0.1498$ & $2-$ & 15.36 & 16.966 & 0.8038 & 175 & 9.1279 \\
(90614) 6646 P-L & $25.1947 \pm 0.1016$ & $1$ & 15.1 & 18.471 & 0.4017 & 156 & 4.7730 \\
(97617) 2000 EW104 & $48.9426 \pm 0.5383$ & $1+$ & 13.86 & 18.28 & 0.3785 & 53 & 21.8038 \\
(97968) 2000 QT149 & $460.3904 \pm 0.8444$ & $1+$ & 16.28 & 18.565 & 1.1542 & 156 & 13.4797 \\
(98436) 2000 UF42 & $41.0651 \pm 0.4535$ & $2-$ & 15.54 & 18.589 & 0.1273 & 219 & 9.9560 \\
(100728) Kamenice n Lipou & $93.3757 \pm 0.8197$ & $1+$ & 15.92 & 19.215 & 0.7432 & 118 & 5.0539 \\
(101187) 1998 SV13 & $31.1032 \pm 0.5827$ & $2-$ & 16.12 & 19.133 & 0.3628 & 54 & 5.6327 \\
(101486) 1998 WJ38 & $82.0681 \pm 0.8385$ & $1+$ & 15.23 & 19.29 & 0.4666 & 122 & 1.9485 \\
(101649) 1999 CO62 & $16.7503 \pm 0.1015$ & $2$ & 13.66 & 19.098 & 0.3991 & 66 & 16.3625 \\
(102237) 1999 TQ21 & $5.1312 \pm 0.0073$ & $2-$ & 15.01 & 19.503 & 0.2557 & 114 & 9.3657 \\
(103250) 2000 AY6 & $53.3729 \pm 0.7454$ & $1$ & 14.96 & 17.736 & 0.2066 & 86 & 5.9229 \\
(104270) 2000 EC149 & $56.2806 \pm 0.5053$ & $2-$ & 16.31 & 18.761 & 0.3996 & 166 & 2.5179 \\
(108042) 2001 FG155 & $50.6397 \pm 0.5122$ & $2-$ & 16.84 & 17.708 & 0.2196 & 83 & 7.4255 \\
(108149) 2001 HG2 & $21.9927 \pm 0.3784$ & $1+$ & 15.91 & 19.257 & 0.4407 & 57 & 28.1275 \\
(110835) 2001 UA65 & $83.3481 \pm 0.5373$ & $2-$ & 15.15 & 18.326 & 0.3320 & 163 & 8.1478 \\
(113697) 2002 TM121 & $47.6207 \pm 0.0372$ & $2-$ & 16.44 & 18.768 & 0.3576 & 174 & 12.4105 \\
(113901) 2002 TX274 & $8.5787 \pm 0.0439$ & $2-$ & 16.91 & 19.106 & 0.3682 & 72 & 16.8859 \\
(115599) 2003 UJ99 & $48.9481 \pm 0.7319$ & $1+$ & 15.11 & 17.146 & 0.2059 & 65 & 7.4849 \\
(119256) 2001 RW28 & $5.3012 \pm 0.0816$ & $1+$ & 15.89 & 18.759 & 0.1635 & 89 & 2.5528 \\
(124358) 2001 QA131 & $20.7772 \pm 0.3458$ & $1$ & 16.17 & 19.193 & 0.3409 & 63 & 25.6011 \\
(142389) 2002 SQ11 & $27.8556 \pm 0.5270$ & $2$ & 16.32 & 19.807 & 0.2154 & 121 & 23.1698 \\
(145499) 2006 BV33 & $59.6000 \pm 0.7891$ & $1$ & 15.38 & 18.178 & 0.2209 & 93 & 2.9077 \\
(146337) 2001 OW14 & $30.4319 \pm 0.4747$ & $2$ & 14.75 & 18.466 & 0.4706 & 67 & 25.7849 \\
(147455) 2004 BK20 & $28.7172 \pm 0.3066$ & $2-$ & 16.57 & 18.44 & 0.2121 & 145 & 2.7659 \\
(149456) 2003 DA13 & $149.6609 \pm 0.7822$ & $1+$ & 15.65 & 18.727 & 0.6662 & 176 & 9.3988 \\
(153030) 2000 PV1 & $62.2428 \pm 0.7463$ & $2-$ & 14.73 & 19.4 & 0.3223 & 121 & 7.2885 \\
(153370) 2001 QU7 & $100.3049 \pm 0.6155$ & $1+$ & 16.5 & 18.578 & 0.3416 & 153 & 13.3888 \\
(153401) 2001 QG119 & $64.6498 \pm 0.1424$ & $1+$ & 16.37 & 18.078 & 0.1519 & 262 & 8.7915 \\
(154315) 2002 VO20 & $10.7503 \pm 0.0585$ & $2-$ & 16.84 & 19.388 & 0.2310 & 108 & 5.9875 \\
(154669) 2004 FA127 & $47.8187 \pm 0.2481$ & $2$ & 17.11 & 18.028 & 0.2288 & 208 & 3.1640 \\
(171188) 2005 JR20 & $46.3577 \pm 0.7724$ & $1$ & 15.69 & 18.979 & 0.4701 & 86 & 24.4399 \\
(173817) 2001 SU305 & $47.7476 \pm 0.1773$ & $1+$ & 15.96 & 18.635 & 0.6862 & 188 & 7.7918 \\
(180213) 2003 UM8 & $44.0264 \pm 0.3784$ & $2$ & 17.03 & 18.643 & 0.2148 & 147 & 12.7505 \\
(187043) 2005 KE12 & $32.8428 \pm 0.1473$ & $2$ & 17.09 & 17.983 & 0.1566 & 141 & 29.1229 \\
(192583) 1999 AY9 & $69.2668 \pm 0.7715$ & $1+$ & 17.34 & 18.704 & 0.2476 & 89 & 10.2077 \\
(196021) 2002 ST2 & $35.6629 \pm 0.6409$ & $1+$ & 16.07 & 18.264 & 0.7812 & 67 & 17.8138 \\
(197627) 2004 KO & $66.1513 \pm 0.2014$ & $1+$ & 17.75 & 19.467 & 0.2510 & 151 & 16.2388 \\
(217474) 2005 WP113 & $48.5248 \pm 0.6076$ & $2-$ & 16.07 & 18.368 & 0.2684 & 76 & 11.2875 \\
(220436) 2003 UO283 & $49.8860 \pm 0.7651$ & $1+$ & 17.01 & 19.034 & 0.1780 & 157 & 2.6949 \\
(223998) 2005 EN324 & $47.6260 \pm 0.1627$ & $2$ & 17.68 & 18.825 & 0.2027 & 415 & 18.5846 \\
(226525) 2003 UF98 & $63.1282 \pm 0.6991$ & $1+$ & 15.01 & 18.526 & 0.4649 & 122 & 10.3969 \\
(228737) 2002 TB211 & $44.2943 \pm 0.4023$ & $2-$ & 16.79 & 19.226 & 0.3984 & 183 & 12.0934 \\
(242994) 2006 SX412 & $43.3321 \pm 0.6401$ & $1+$ & 15.74 & 17.707 & 0.2080 & 98 & 3.0360 \\
(246823) 2009 TO35 & $2.7887 \pm 0.0034$ & $2$ & 14.89 & 18.624 & 0.1364 & 161 & 8.1242 \\
(248294) 2005 LZ2 & $11.5238 \pm 0.0813$ & $2$ & 16.11 & 18.947 & 0.4985 & 67 & 25.4408 \\
(249205) 2008 DT39 & $50.8843 \pm 0.5408$ & $1$ & 16.19 & 19.475 & 0.5588 & 136 & 5.6187 \\
(254797) 2005 QO86 & $47.8117 \pm 0.6782$ & $1+$ & 16.07 & 18.556 & 0.2939 & 100 & 6.0095 \\
(259541) 2003 UW83 & $13.2512 \pm 0.0972$ & $1$ & 14.95 & 17.97 & 0.3462 & 60 & 5.0465 \\
(270596) 2002 NS15 & $46.0932 \pm 0.7192$ & $2$ & 15.52 & 18.867 & 0.1610 & 163 & 10.6732 \\
(272067) 2005 EK215 & $2.1530 \pm 0.0043$ & $2$ & 18.24 & 19.063 & 0.1422 & 55 & 3.8758 \\
(285800) 2000 XG19 & $39.5676 \pm 0.6672$ & $2-$ & 16.23 & 19.38 & 0.1962 & 129 & 2.1635 \\
(290837) 2005 WO14 & $49.2187 \pm 0.6318$ & $2$ & 16.64 & 19.243 & 0.3553 & 152 & 6.1133 \\
(301922) 1999 VH159 & $19.0213 \pm 0.0956$ & $1+$ & 16.06 & 18.856 & 0.5001 & 119 & 4.9312 \\
(306308) 2011 SL70 & $13.4124 \pm 0.0940$ & $2-$ & 17.57 & 19.317 & 0.2673 & 98 & 8.6754 \\
(307829) 2003 YB26 & $76.9895 \pm 0.7886$ & $1+$ & 17.03 & 18.972 & 1.1297 & 117 & 14.1466 \\
(307847) 2003 YN130 & $3.3822 \pm 0.0108$ & $2$ & 15.98 & 19.233 & 0.3338 & 61 & 6.8430 \\
(310997) 2003 WT141 & $16.9212 \pm 0.6872$ & $1+$ & 17.33 & 18.744 & 0.4069 & 57 & 5.4341 \\
(311000) 2003 WT157 & $77.7589 \pm 0.8922$ & $1$ & 17.4 & 19.074 & 40.8378 & 57 & 10.3563 \\
(312702) 2010 OD87 & $61.8052 \pm 0.7163$ & $2$ & 16.15 & 18.634 & 0.3475 & 120 & 26.9363 \\
(344511) 2002 RG85 & $38.9077 \pm 0.3209$ & $2-$ & 17.8 & 19.176 & 0.1222 & 179 & 14.0266 \\
(387562) 2001 SL15 & $70.0307 \pm 0.8144$ & $1+$ & 17.63 & 18.623 & 0.1767 & 111 & 3.5966 \\
(413291) 2003 UO164 & $17.5443 \pm 0.0915$ & $2-$ & 15.37 & 18.89 & 0.4100 & 85 & 2.2635 \\
(415204) 2012 HZ14 & $28.2020 \pm 0.7051$ & $1$ & 16.14 & 19.033 & 0.1880 & 170 & 2.4885 \\
(442649) 2012 TU124 & $58.3536 \pm 0.5590$ & $1$ & 18.05 & 18.228 & 0.2387 & 171 & 15.4212 \\
(443402) 2014 HU37 & $44.7353 \pm 0.7058$ & $1+$ & 16.79 & 18.624 & 0.6685 & 69 & 6.6857 \\
(547337) 2010 LF65 & $30.5449 \pm 0.6248$ & $2$ & 16.68 & 19.58 & 0.7176 & 57 & 6.5381 \\
(565258) 2017 DG1 & $2.3137 \pm 0.0026$ & $2$ & 17.04 & 18.56 & 0.4057 & 64 & 34.3100 \\
(573142) 2008 YK11 & $25.5075 \pm 0.1804$ & $2$ & 17.57 & 18.989 & 0.7353 & 74 & 2.9745 \\
(614054) 2008 SC29 & $56.8403 \pm 0.6984$ & $2-$ & 18.37 & 19.109 & 0.2579 & 162 & 7.5052 \\
(739003) 2017 DG118 & $102.0560 \pm 0.8148$ & $1+$ & 15.89 & 18.619 & 1.0100 & 96 & 2.7268 \\
(829373) 2006 OB40 & $48.3685 \pm 0.5001$ & $2-$ & 17.46 & 19.049 & 0.2747 & 226 & 7.3110 \\
(855712) 2011 GK57 & $59.7218 \pm 0.7456$ & $1$ & 18.59 & 18.974 & 0.2569 & 105 & 10.8725 \\

\end{longtable}

\clearpage

\section{\HL{Lightcurve figures}}\label{appendix:figure}

\begin{figure*}[h]
    \centering
    \includegraphics[width=0.45\linewidth]{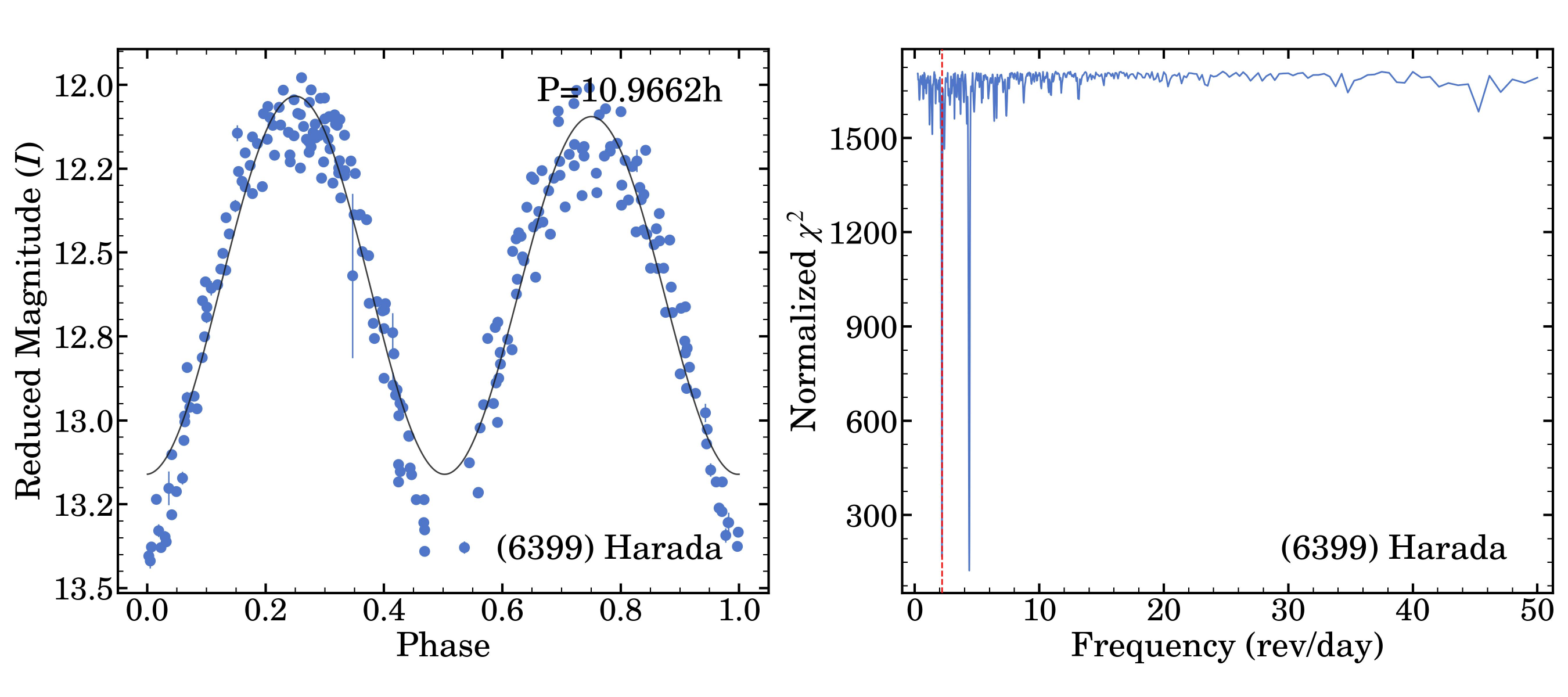}
    \includegraphics[width=0.45\linewidth]{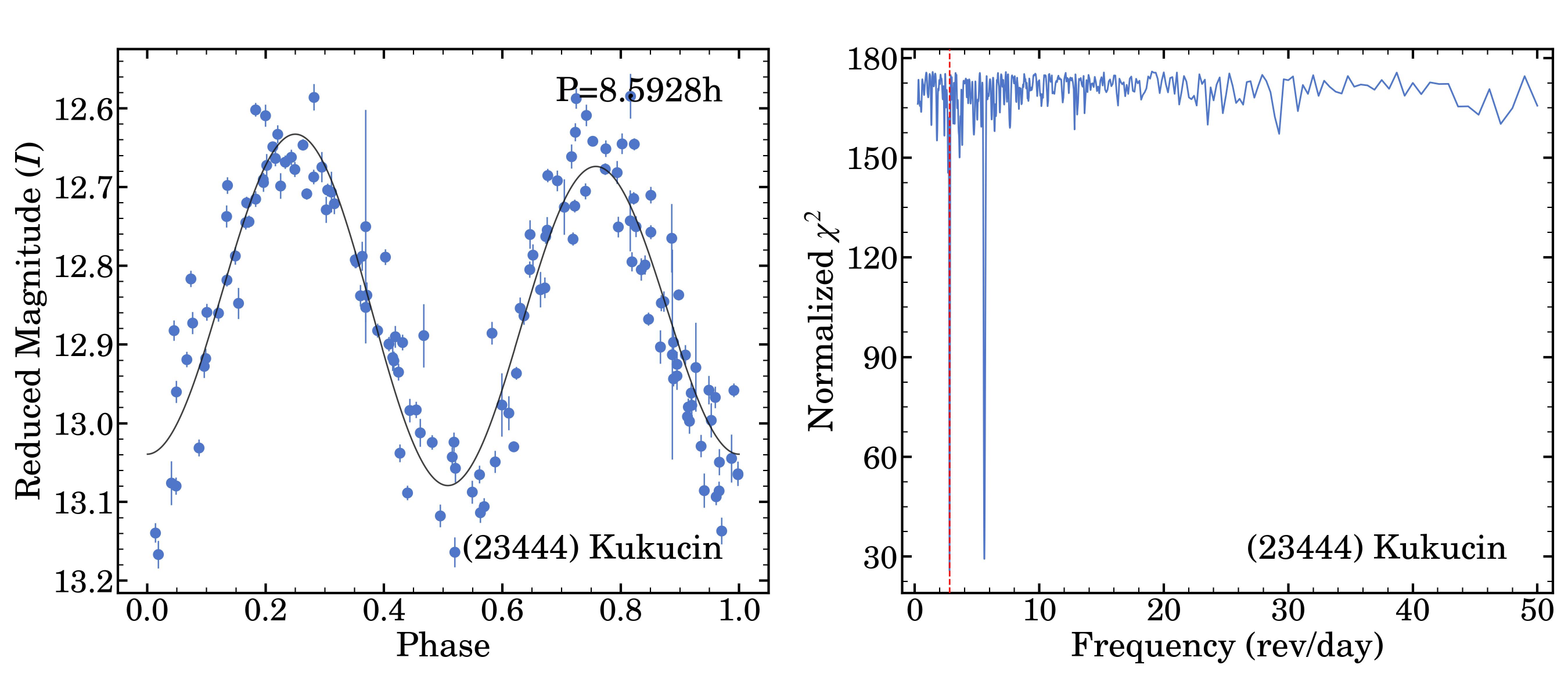}
    \includegraphics[width=0.45\linewidth]{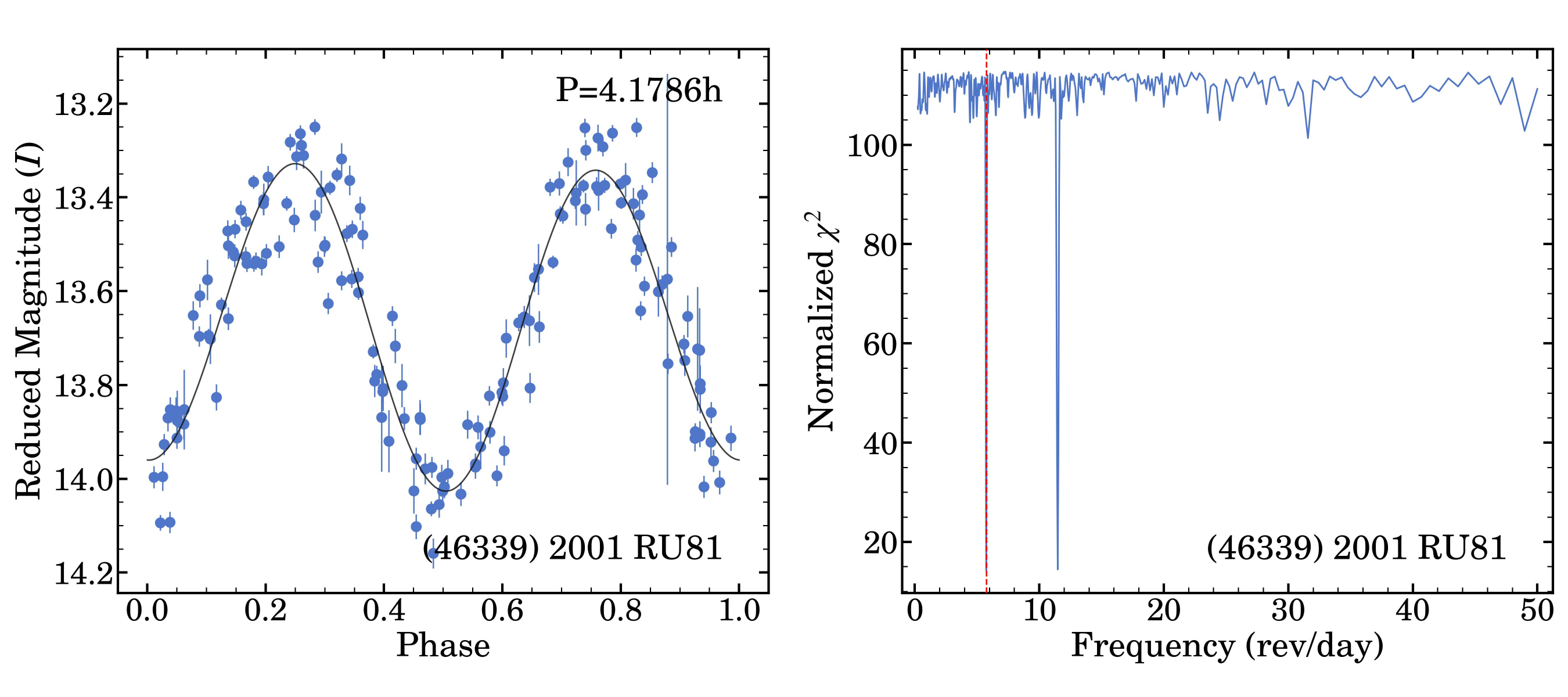}
    \includegraphics[width=0.45\linewidth]{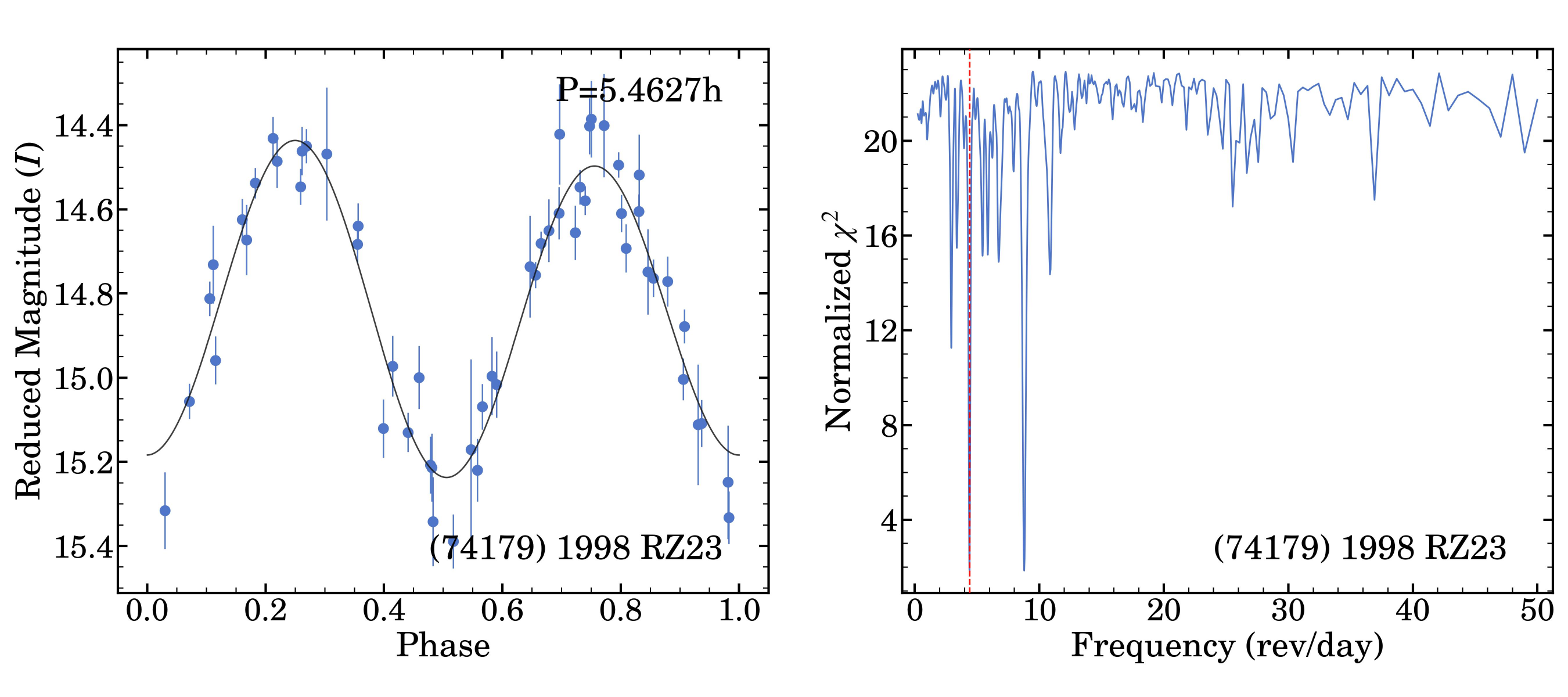}
    \includegraphics[width=0.45\linewidth]{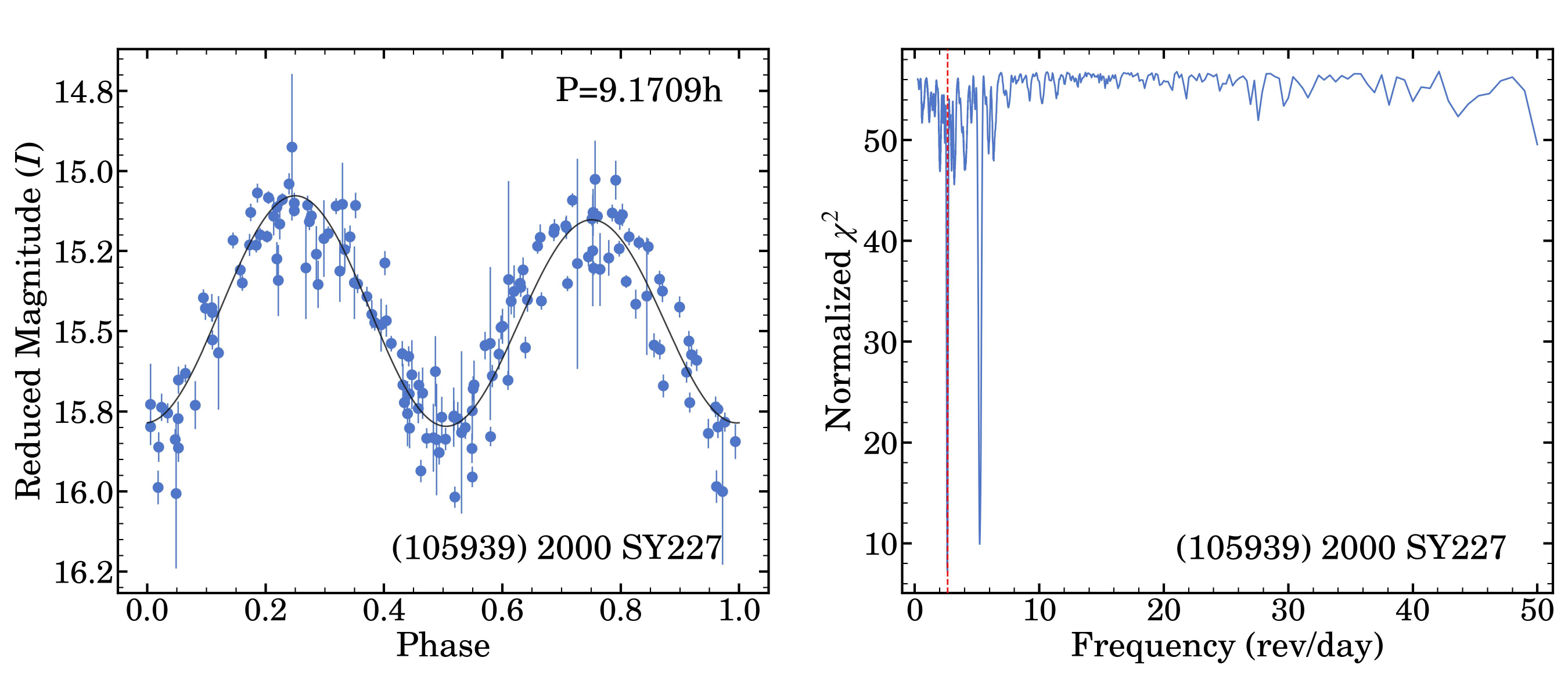}
    \includegraphics[width=0.45\linewidth]{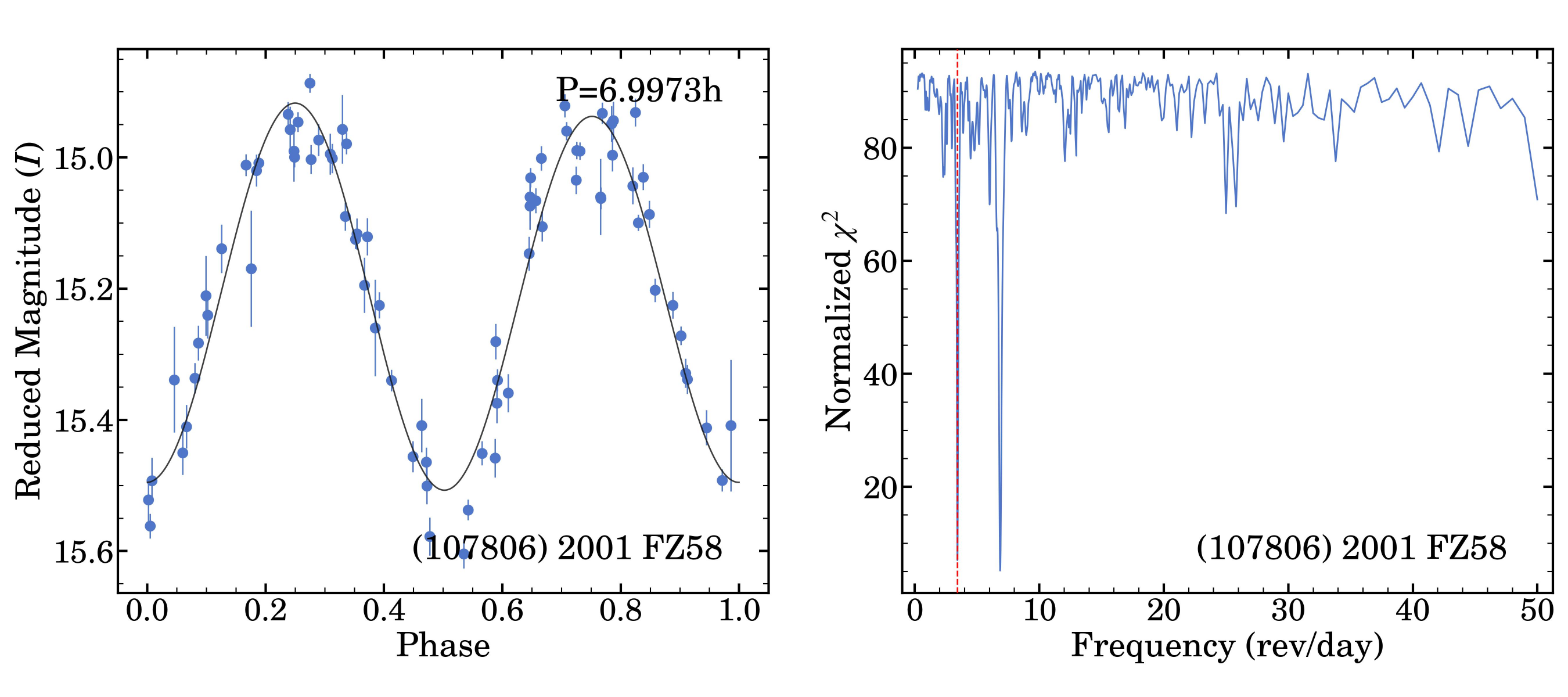}
    \caption{Folded lightcurves (left) and normalized $\chi^2$ as a function of frequency (right) for asteroids with rotation period quality code U = 3.}
    \label{fig:appendix3}
\end{figure*}

This appendix presents the folded light curves and the corresponding normalized $\chi^2$ distributions as a function of frequency for asteroids with different rotation-period quality codes. Figures~\ref{fig:appendix3},Figures~\ref{fig:appendix3-} and Figures~\ref{fig:appendix2+} show the results for asteroids with U = 3, U = 3−, and U = 2+, respectively.

\clearpage

\begin{figure*}
    \centering
    \includegraphics[width=0.45\linewidth]{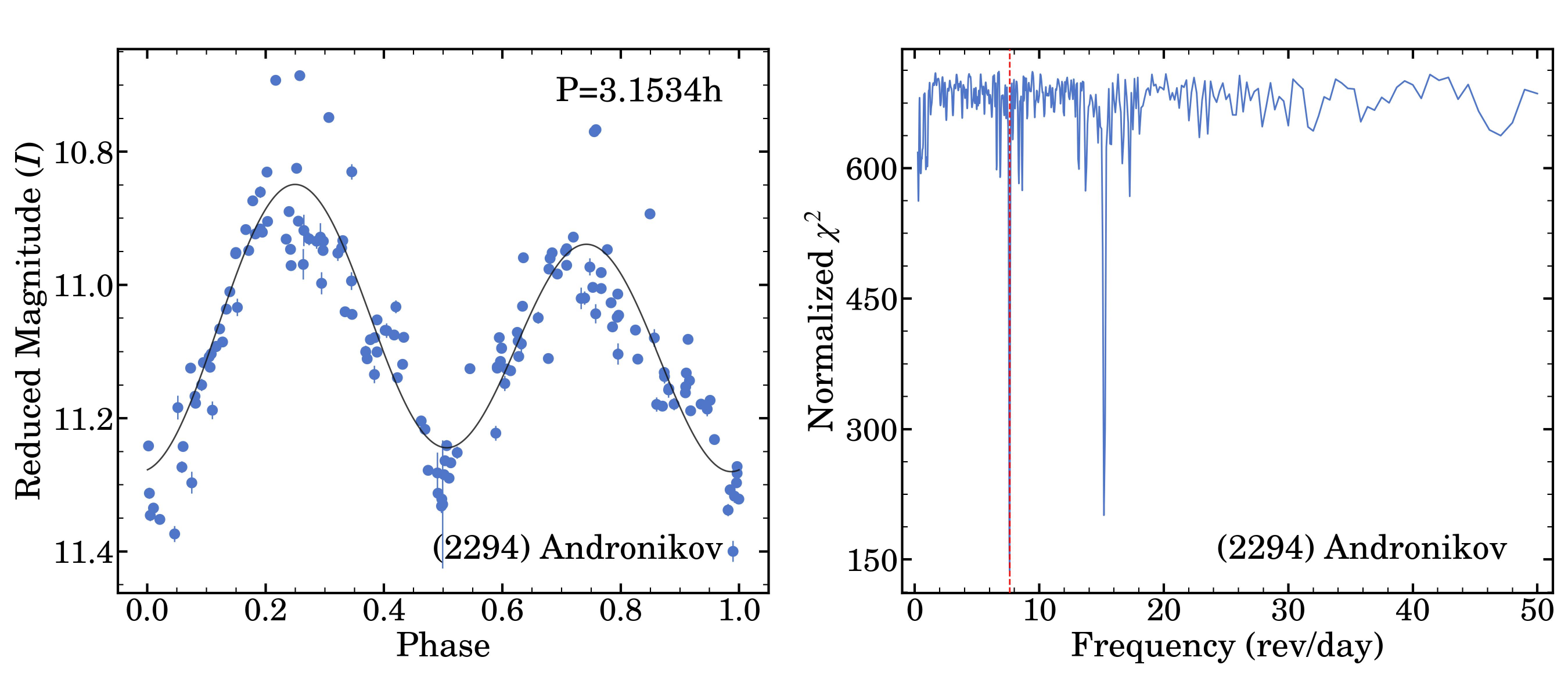}
    \includegraphics[width=0.45\linewidth]{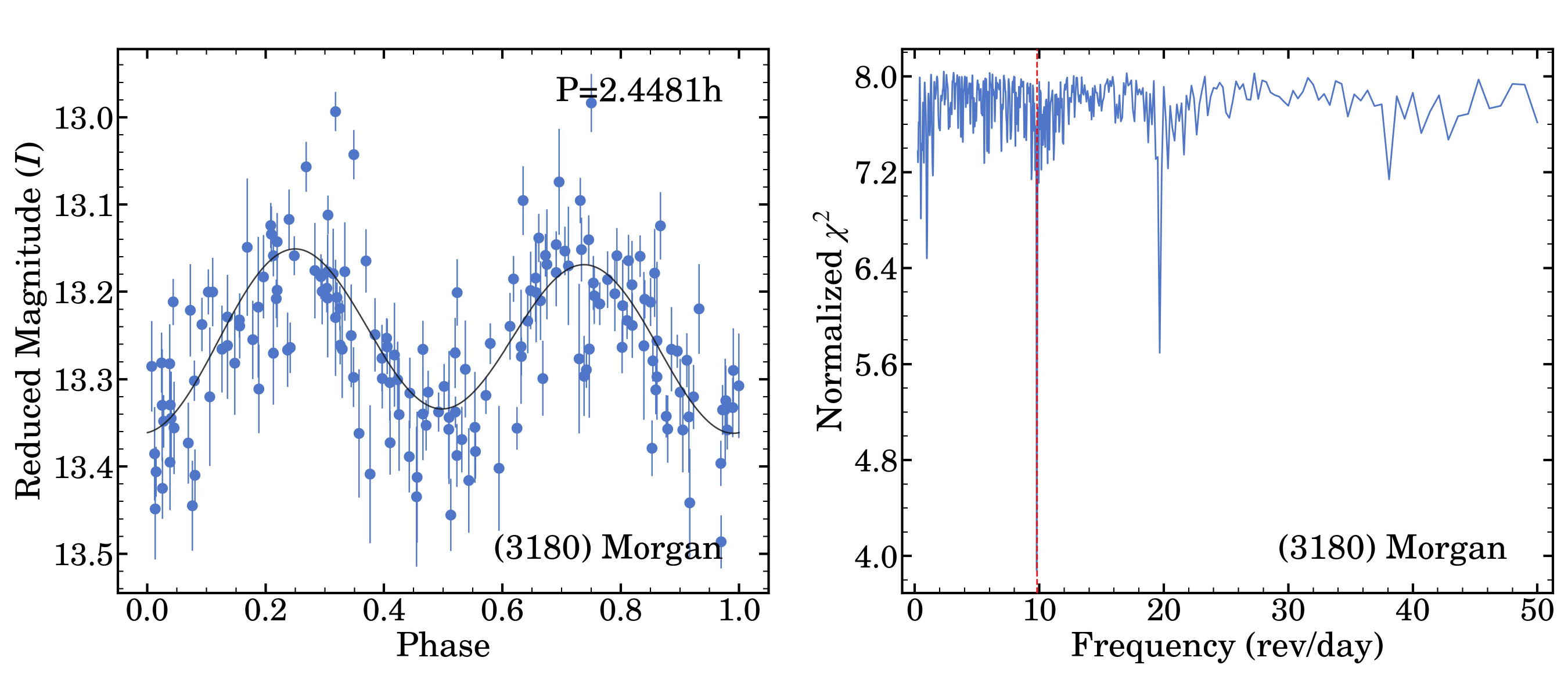}
    \includegraphics[width=0.45\linewidth]{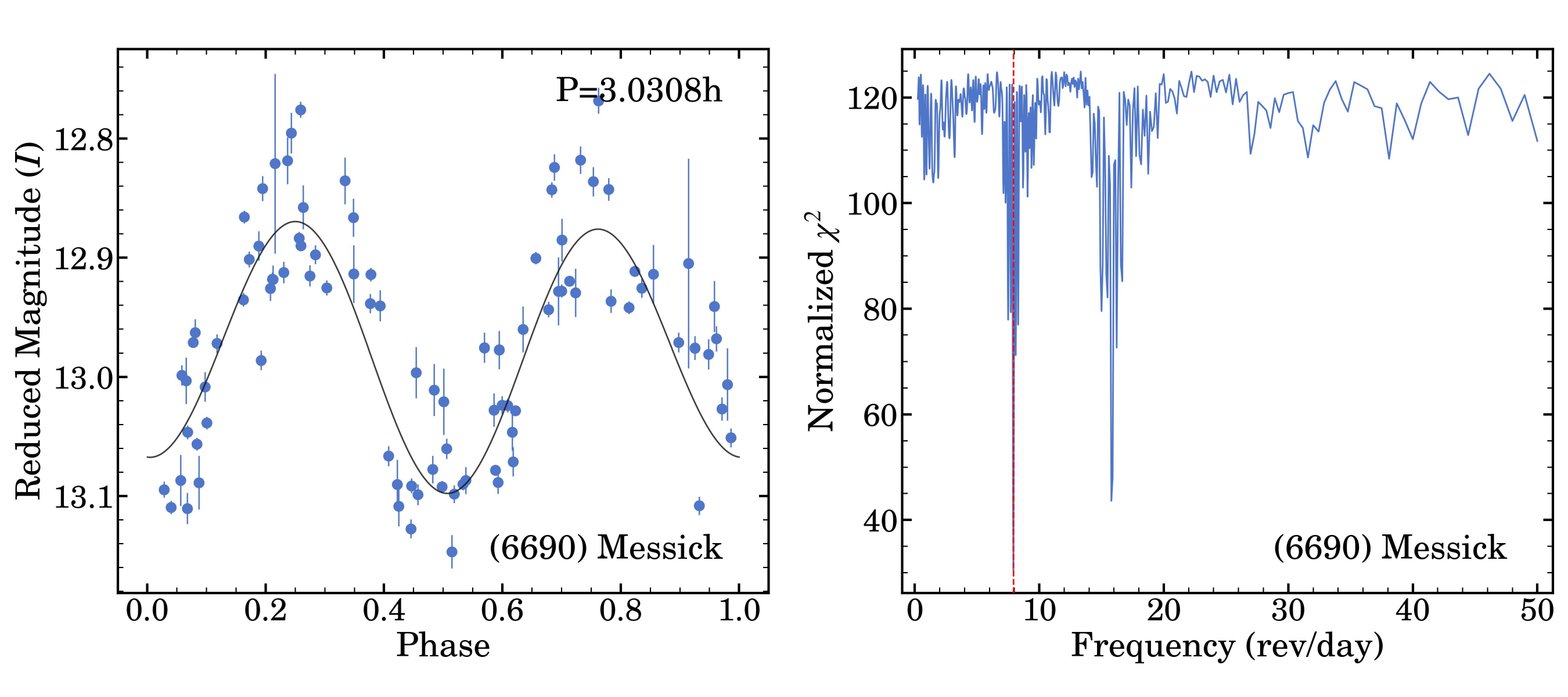}
    \includegraphics[width=0.45\linewidth]{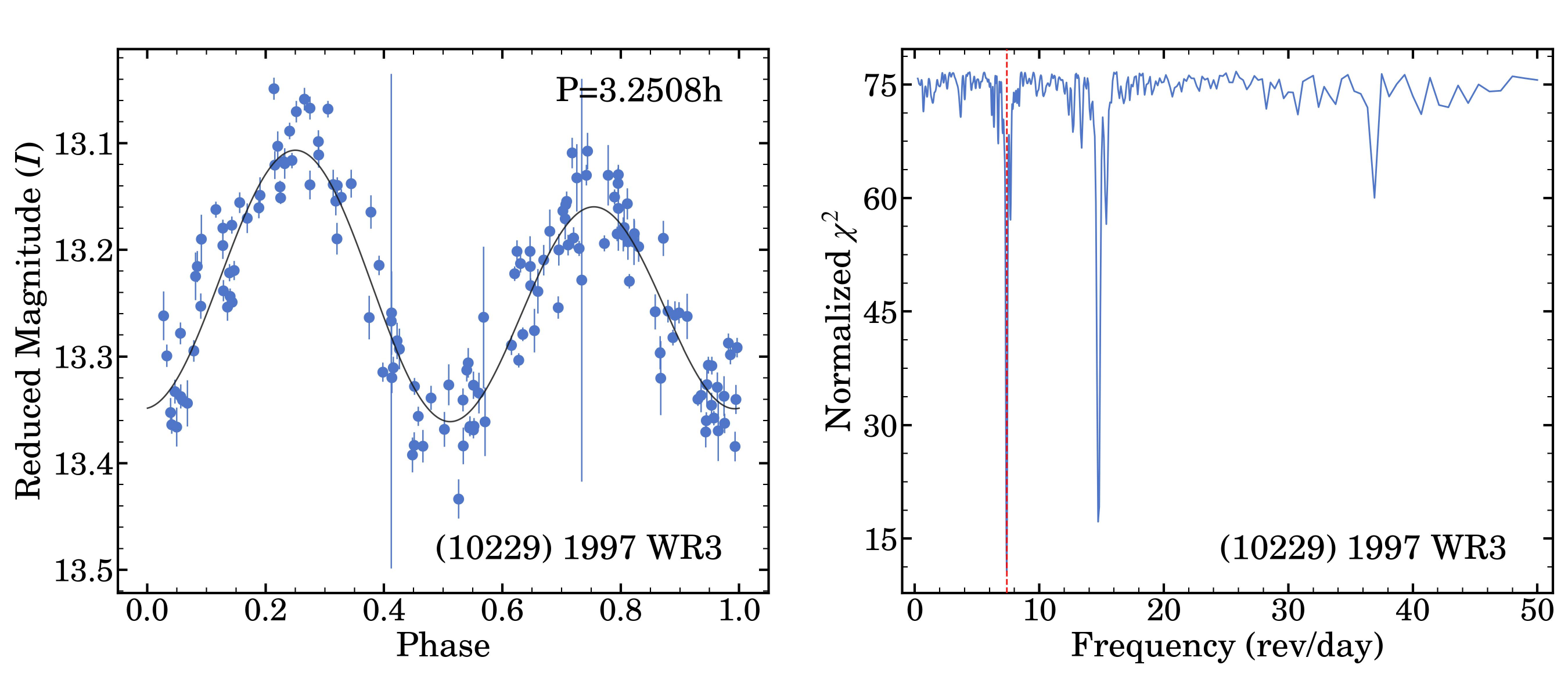}
    \includegraphics[width=0.45\linewidth]{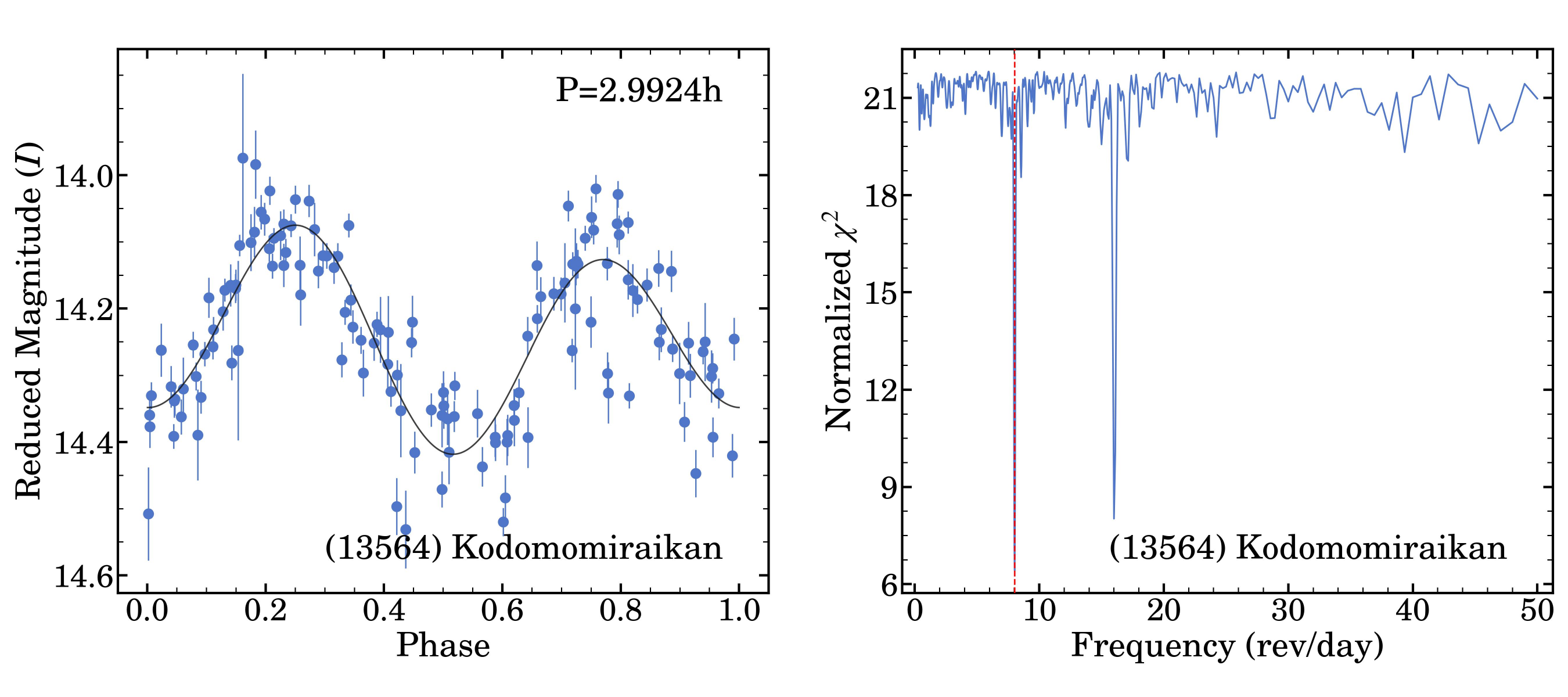}
    \includegraphics[width=0.45\linewidth]{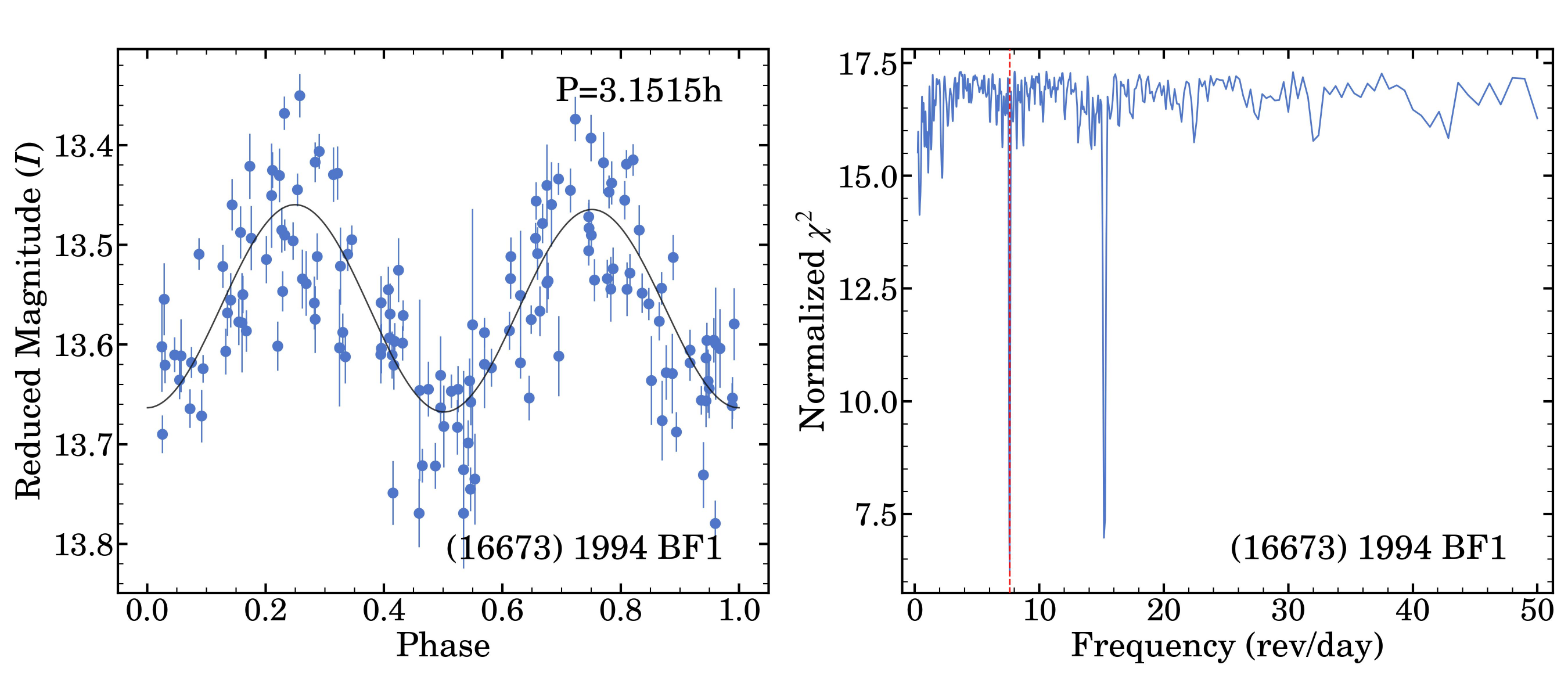}
    \includegraphics[width=0.45\linewidth]{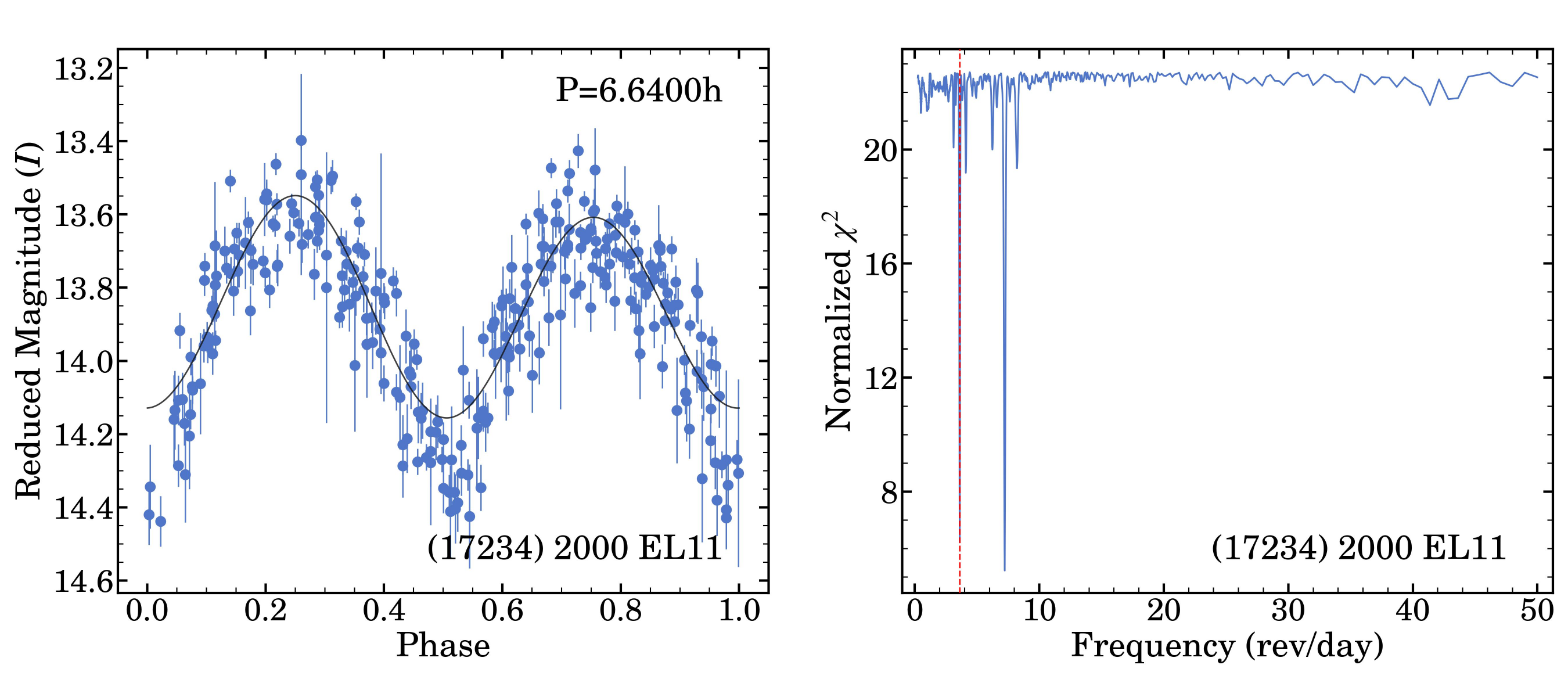}
    \includegraphics[width=0.45\linewidth]{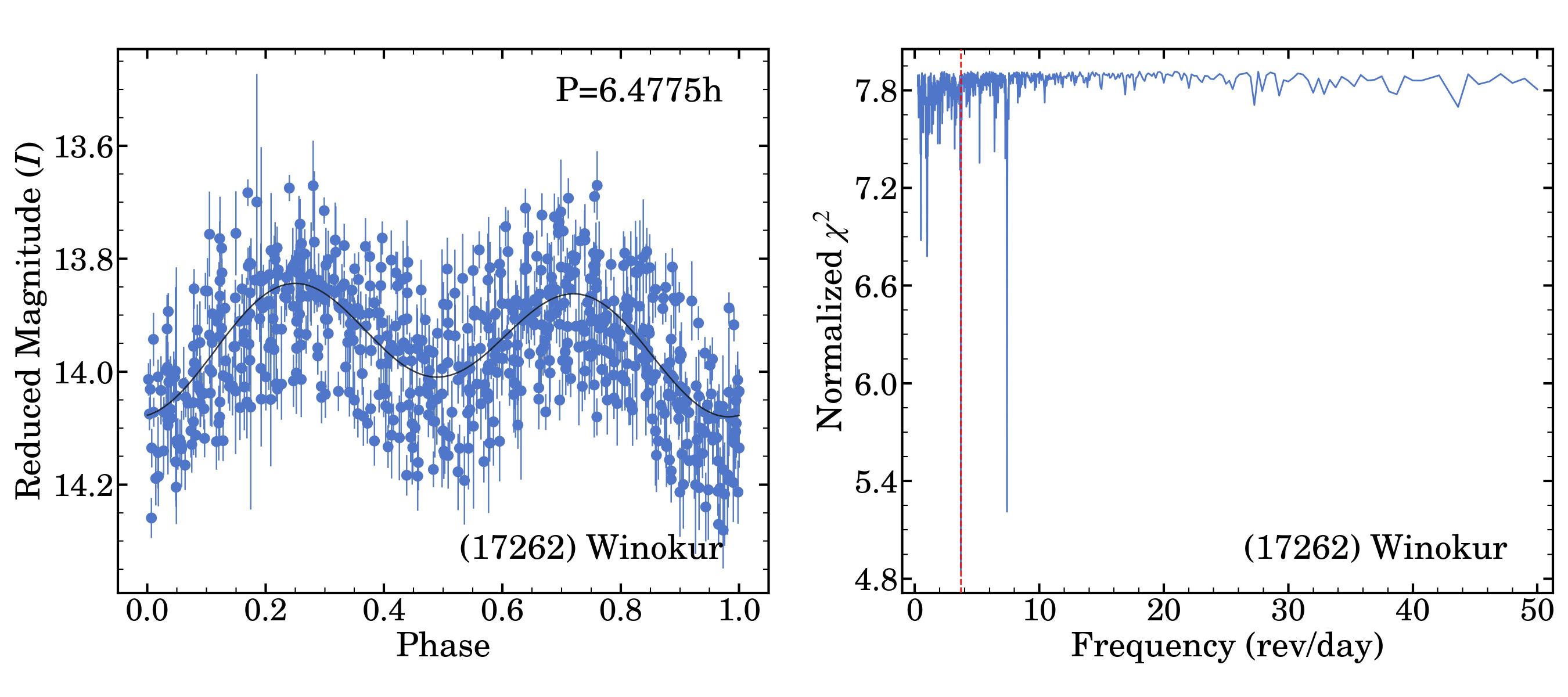}
    \includegraphics[width=0.45\linewidth]{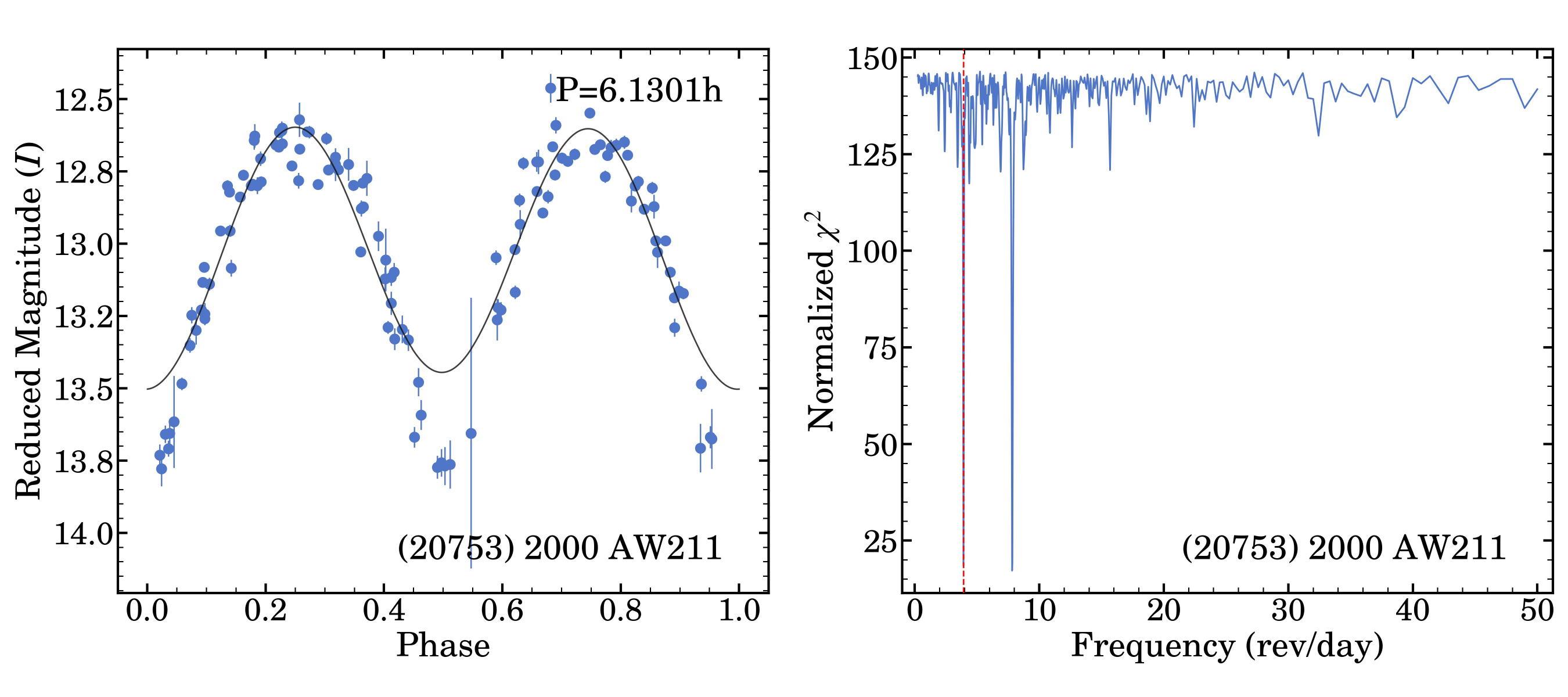}
    \includegraphics[width=0.45\linewidth]{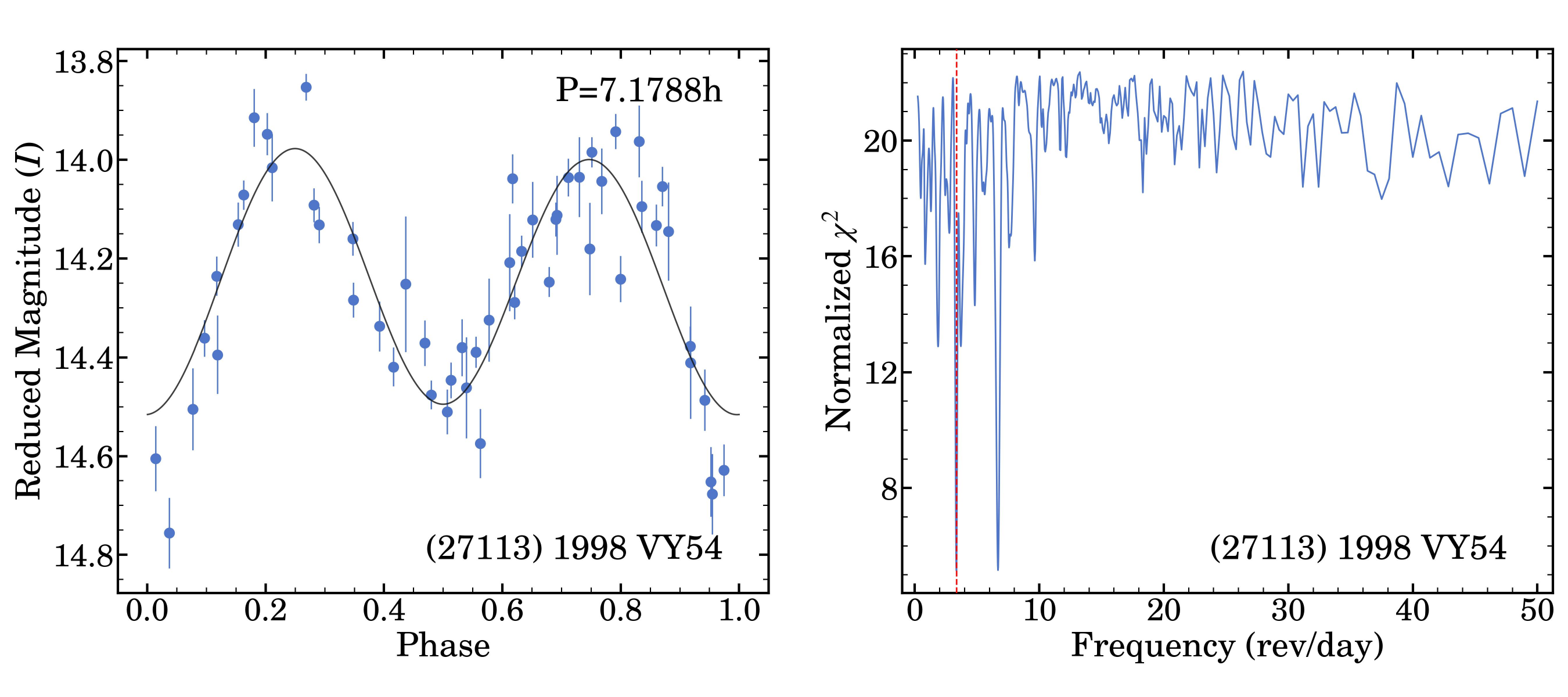}
    \includegraphics[width=0.45\linewidth]{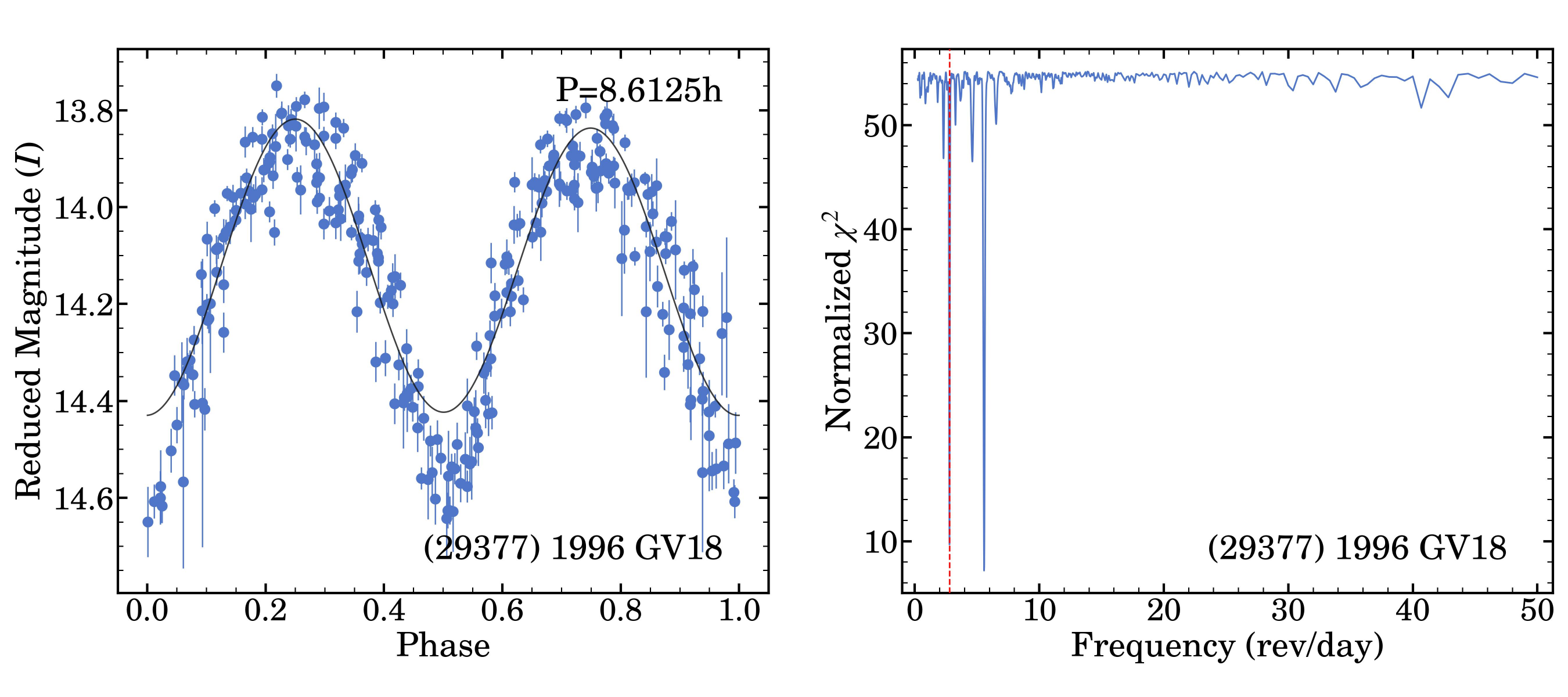}
    \includegraphics[width=0.45\linewidth]{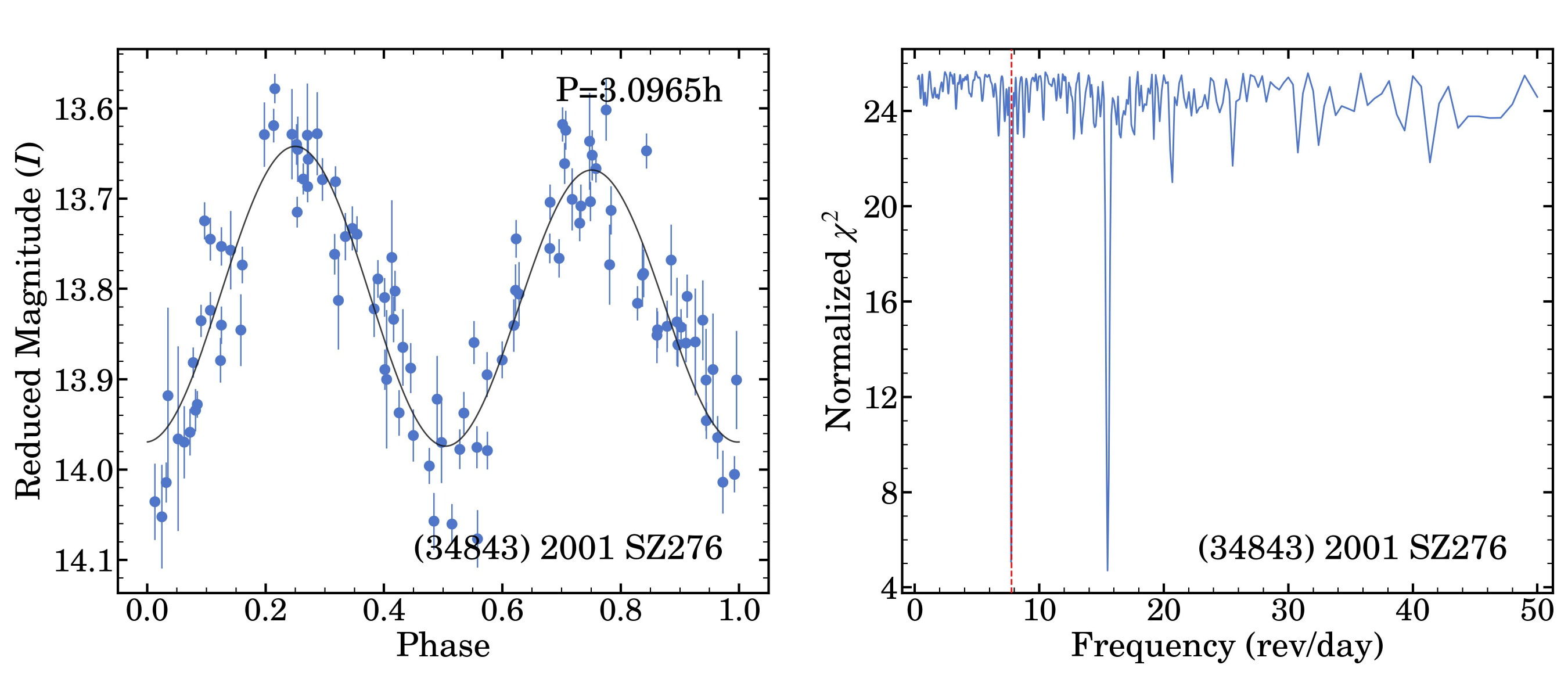}
    \caption{Folded lightcurves (left) and normalized $\chi^2$ as a function of frequency (right) for asteroids with rotation period quality code U = 3-.}
    \label{fig:appendix3-}
\end{figure*}

\begin{figure*}
    \addtocounter{figure}{-1}
    \centering
    \includegraphics[width=0.45\linewidth]{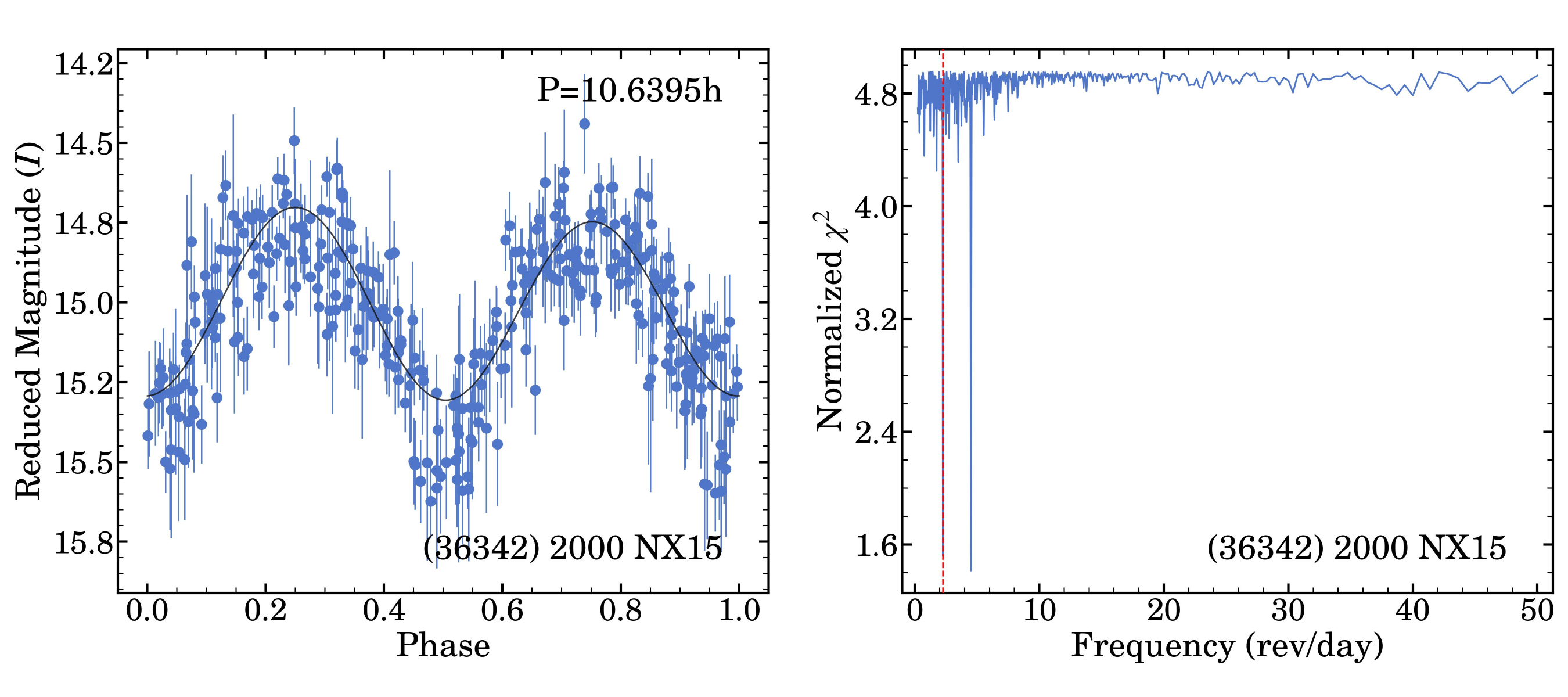}
    \includegraphics[width=0.45\linewidth]{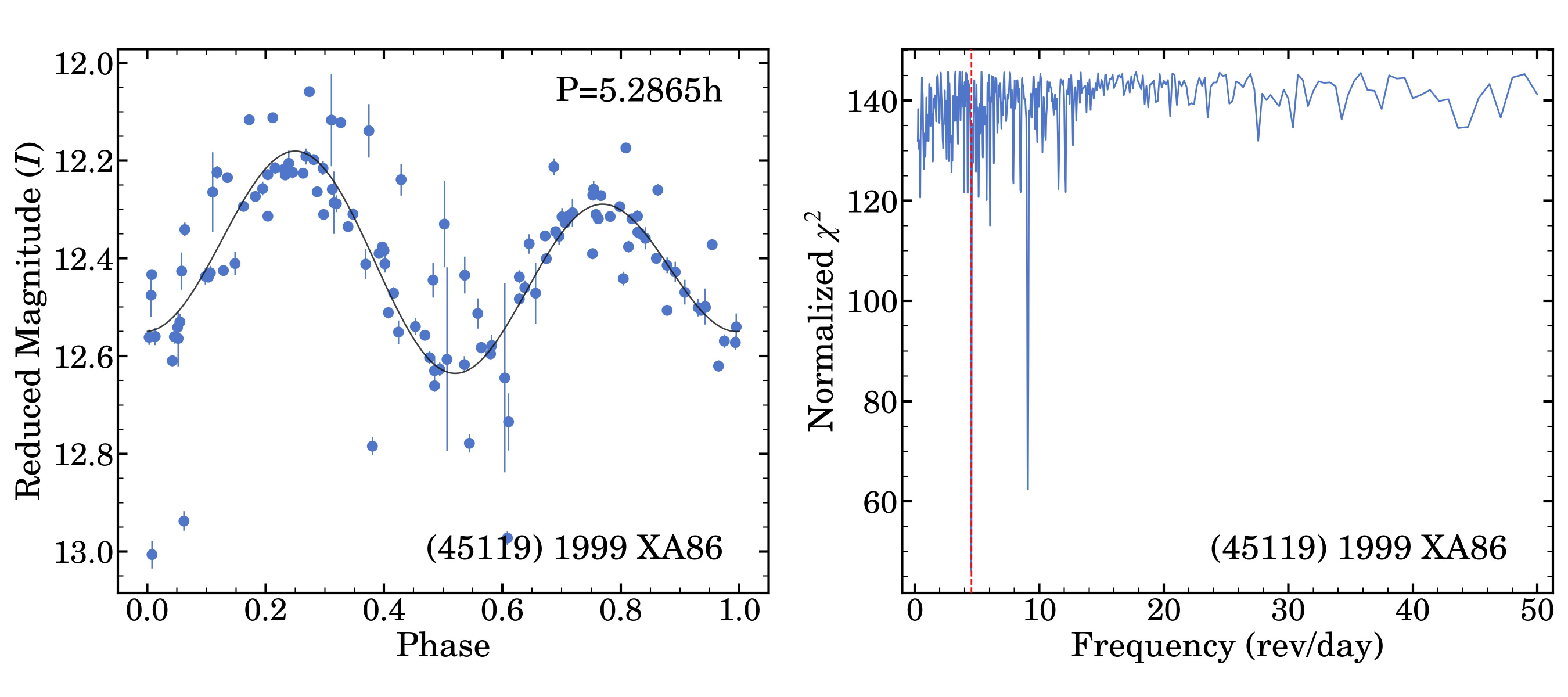}
    \includegraphics[width=0.45\linewidth]{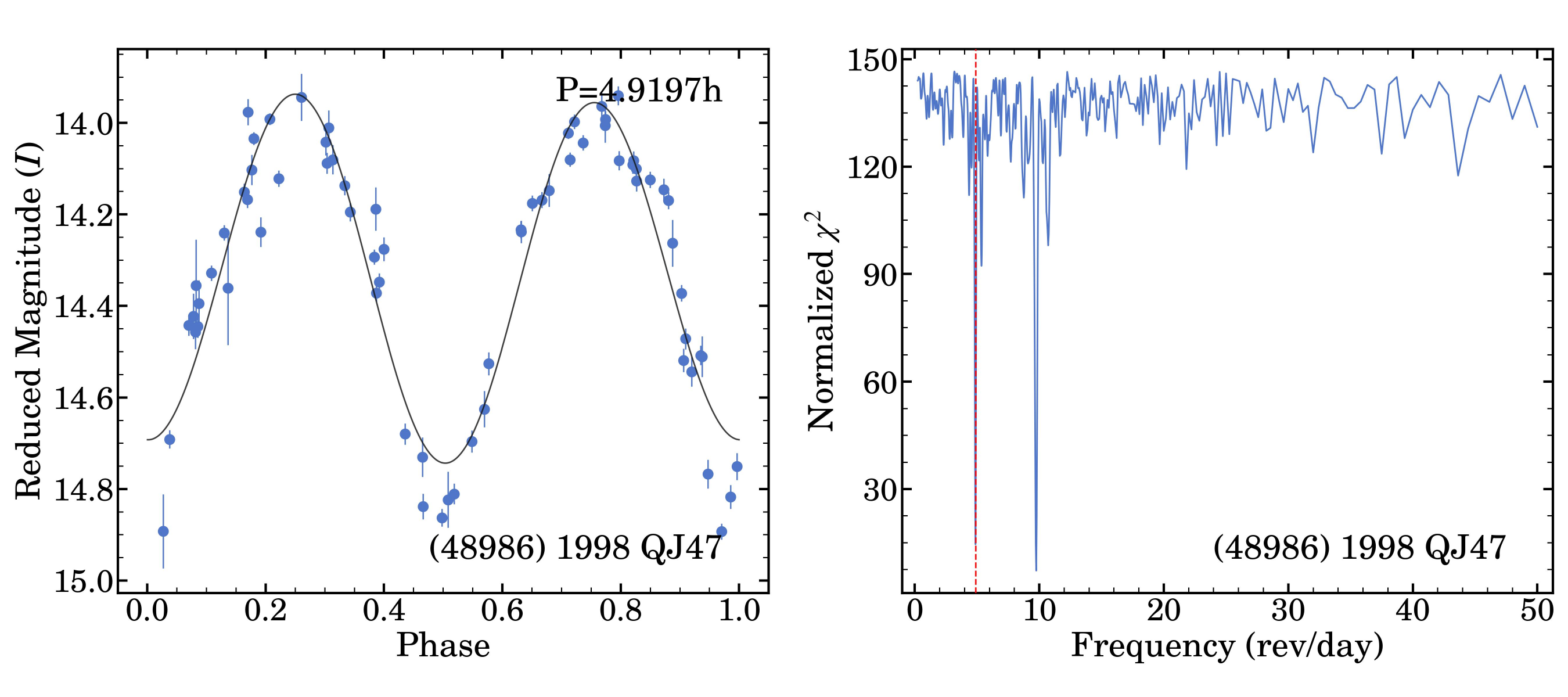}
    \includegraphics[width=0.45\linewidth]{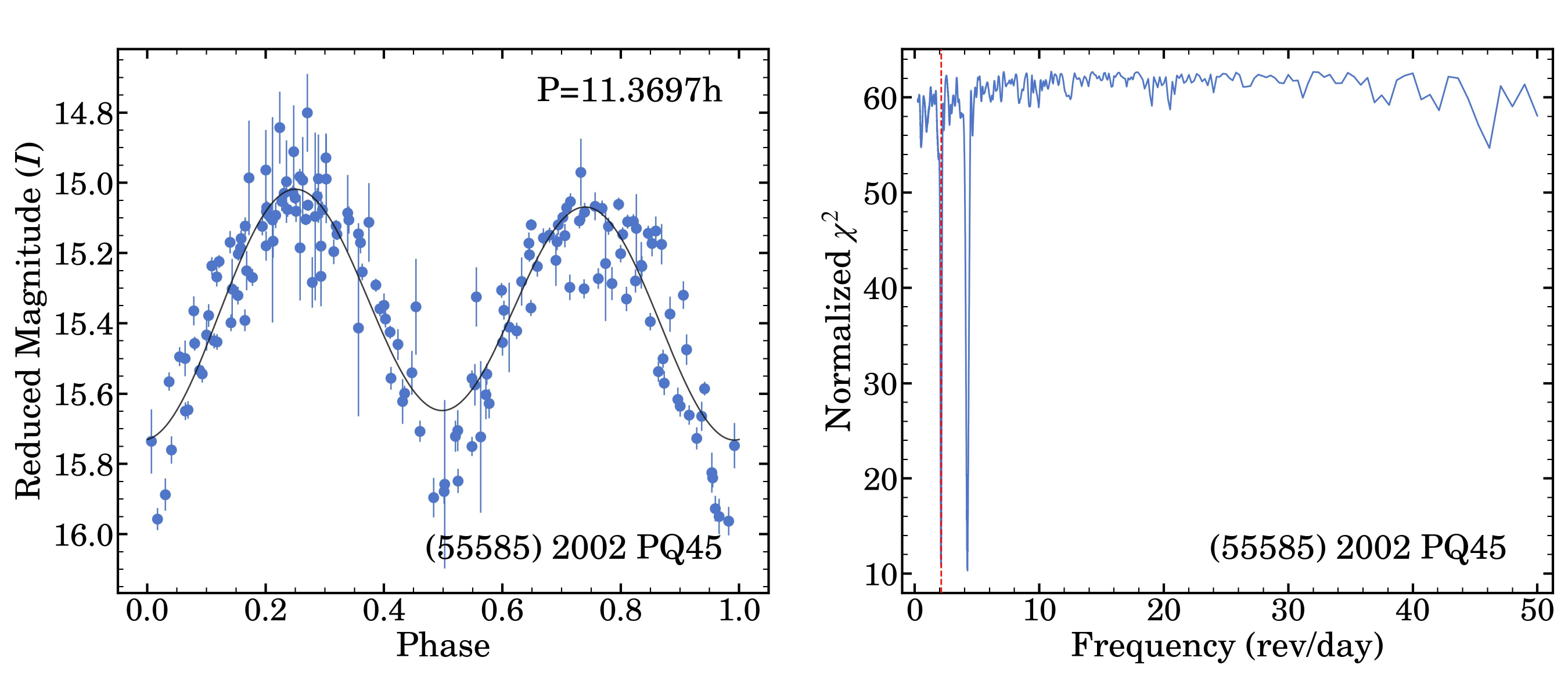}
    \includegraphics[width=0.45\linewidth]{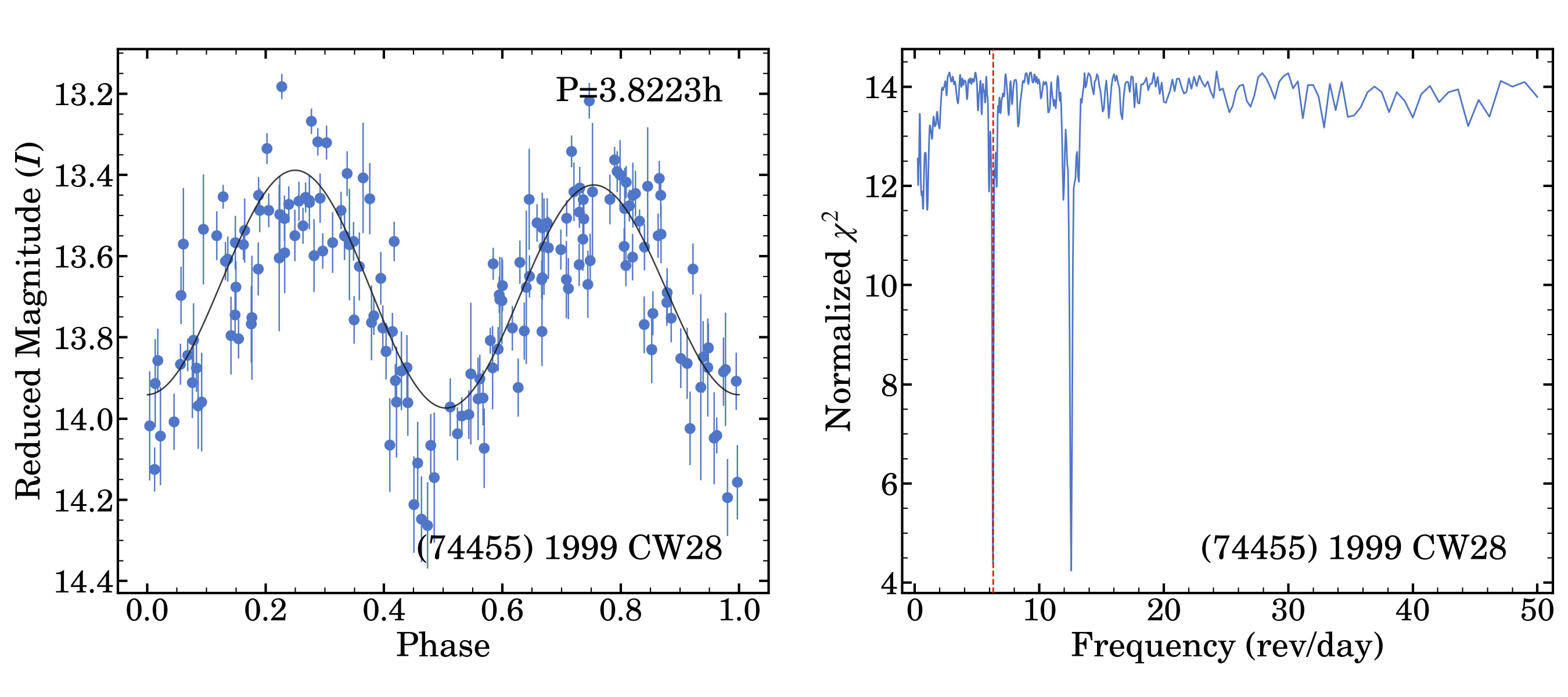}
    \includegraphics[width=0.45\linewidth]{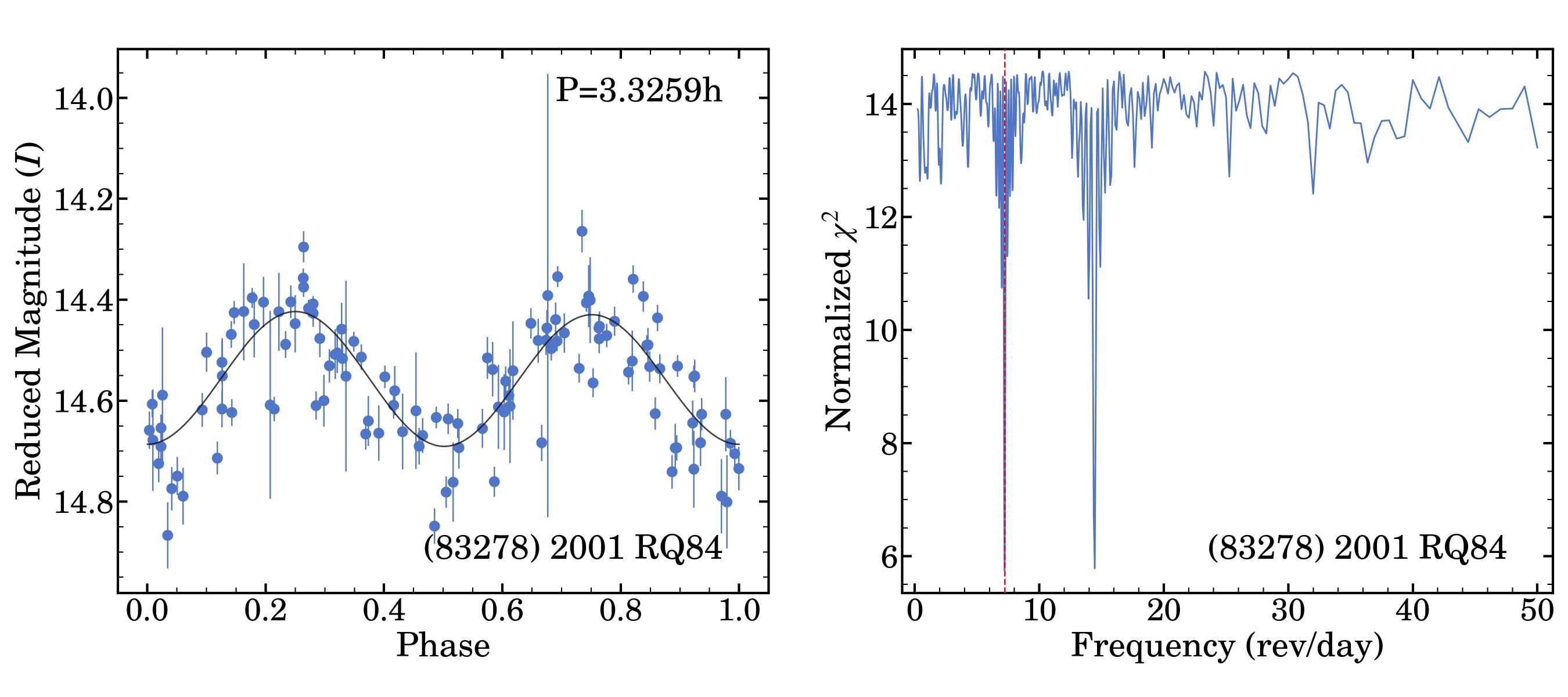}
    \includegraphics[width=0.45\linewidth]{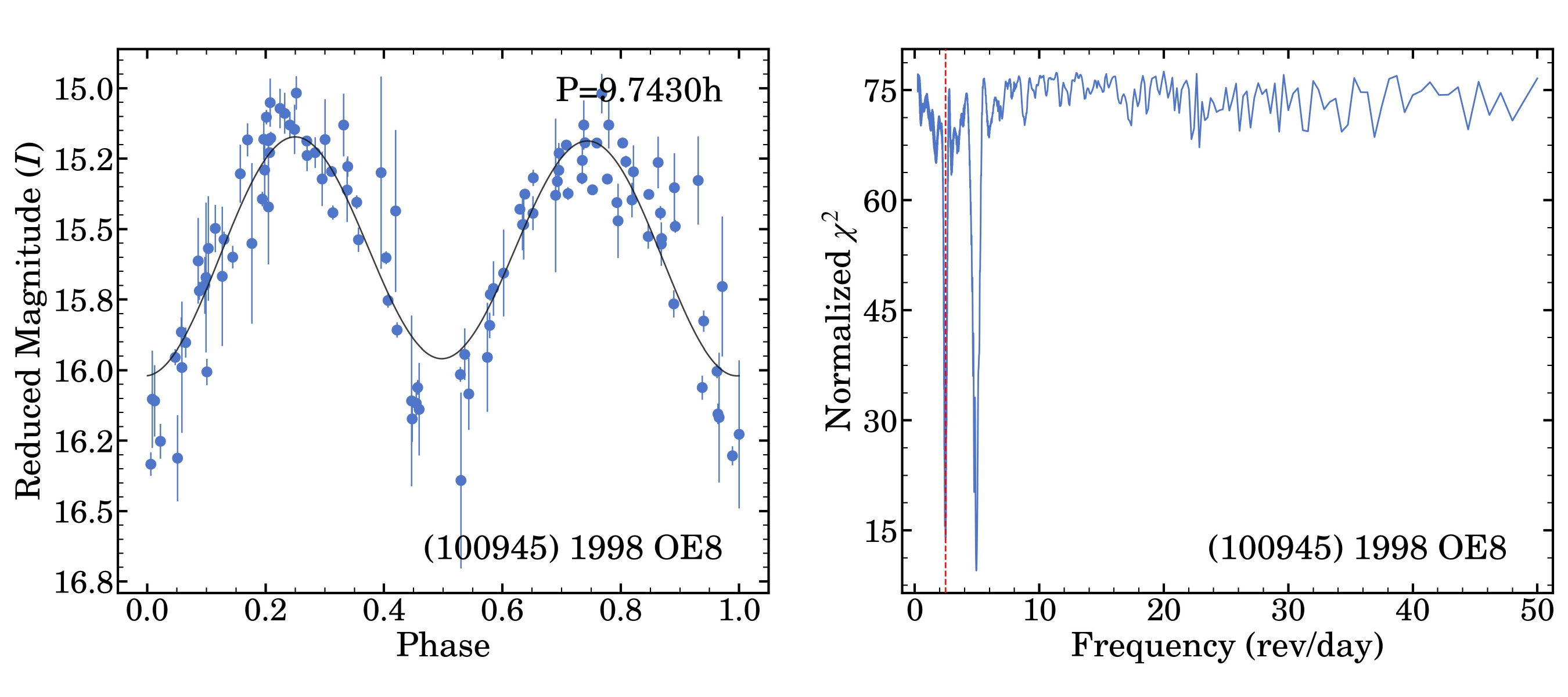}
    \includegraphics[width=0.45\linewidth]{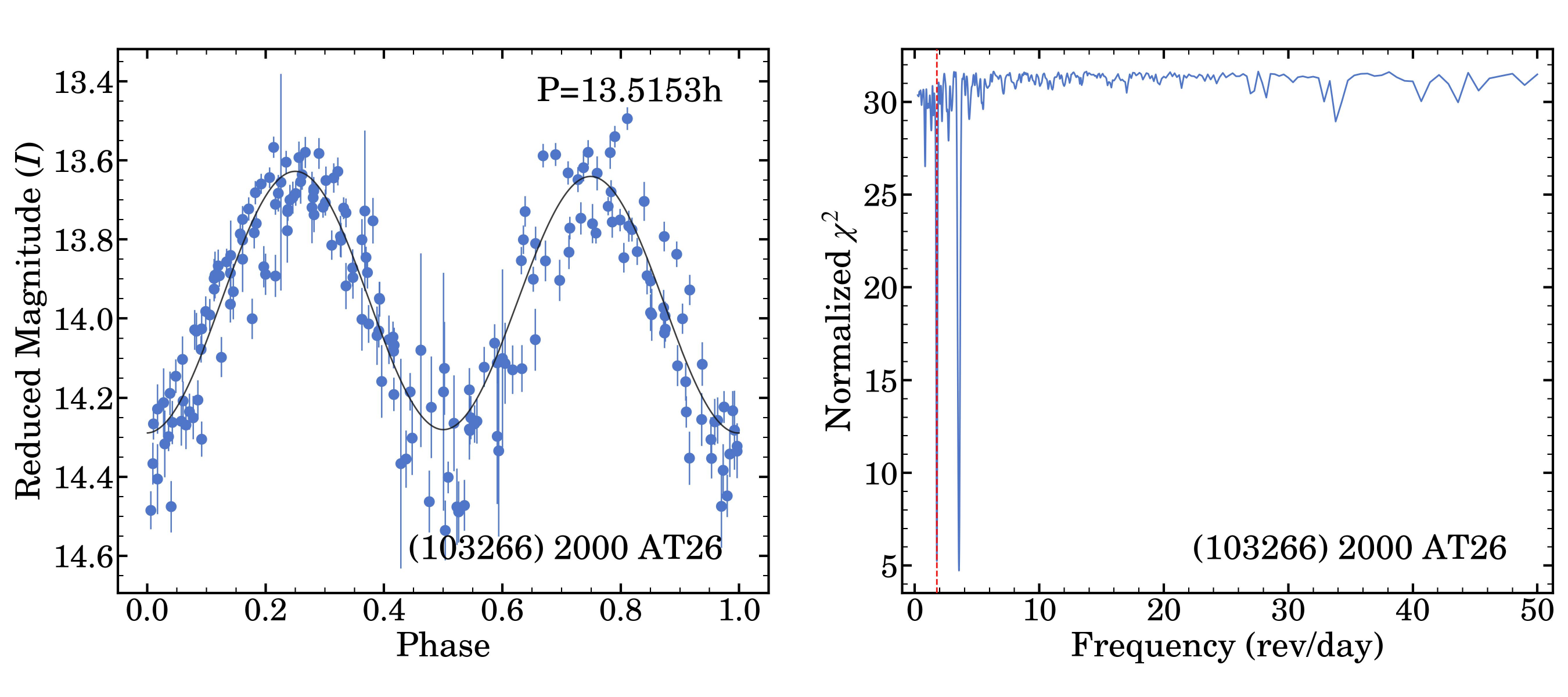}
    \includegraphics[width=0.45\linewidth]{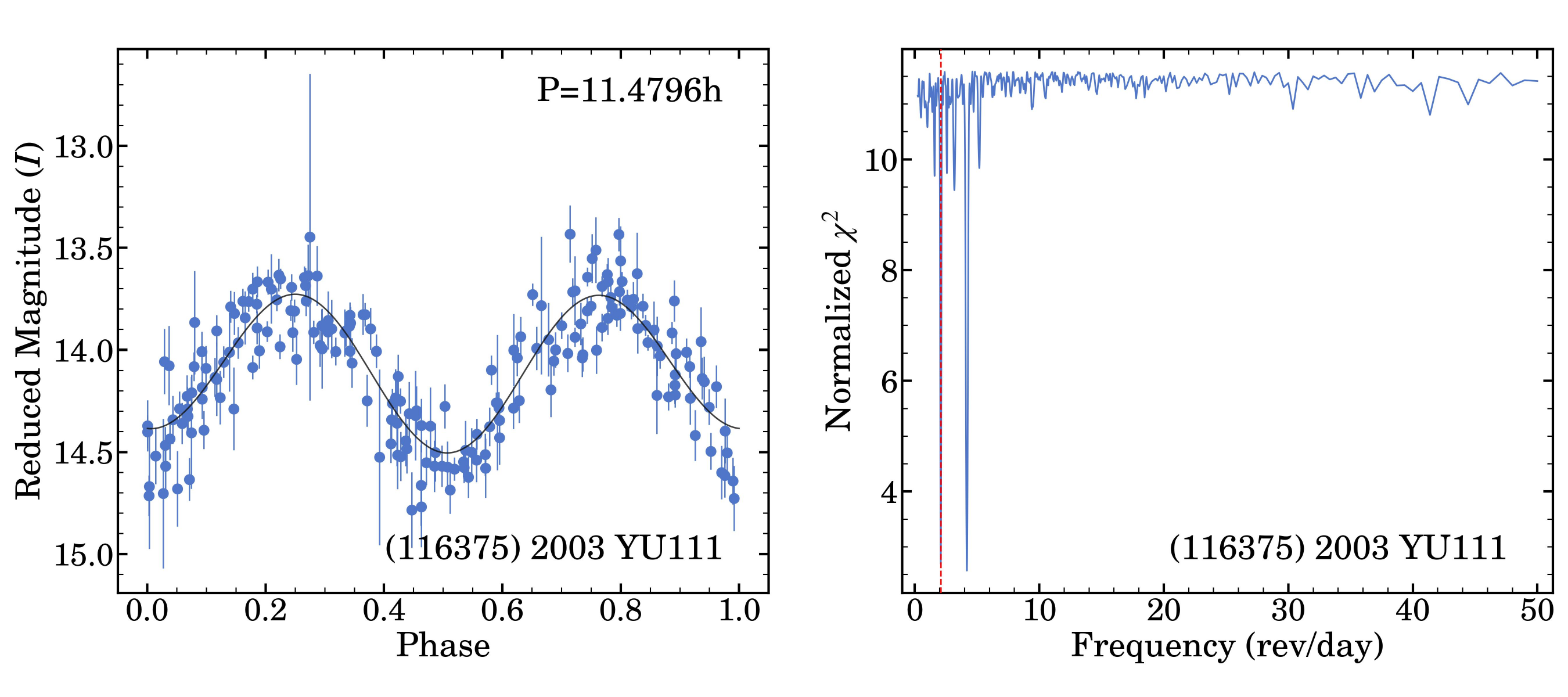}
    \includegraphics[width=0.45\linewidth]{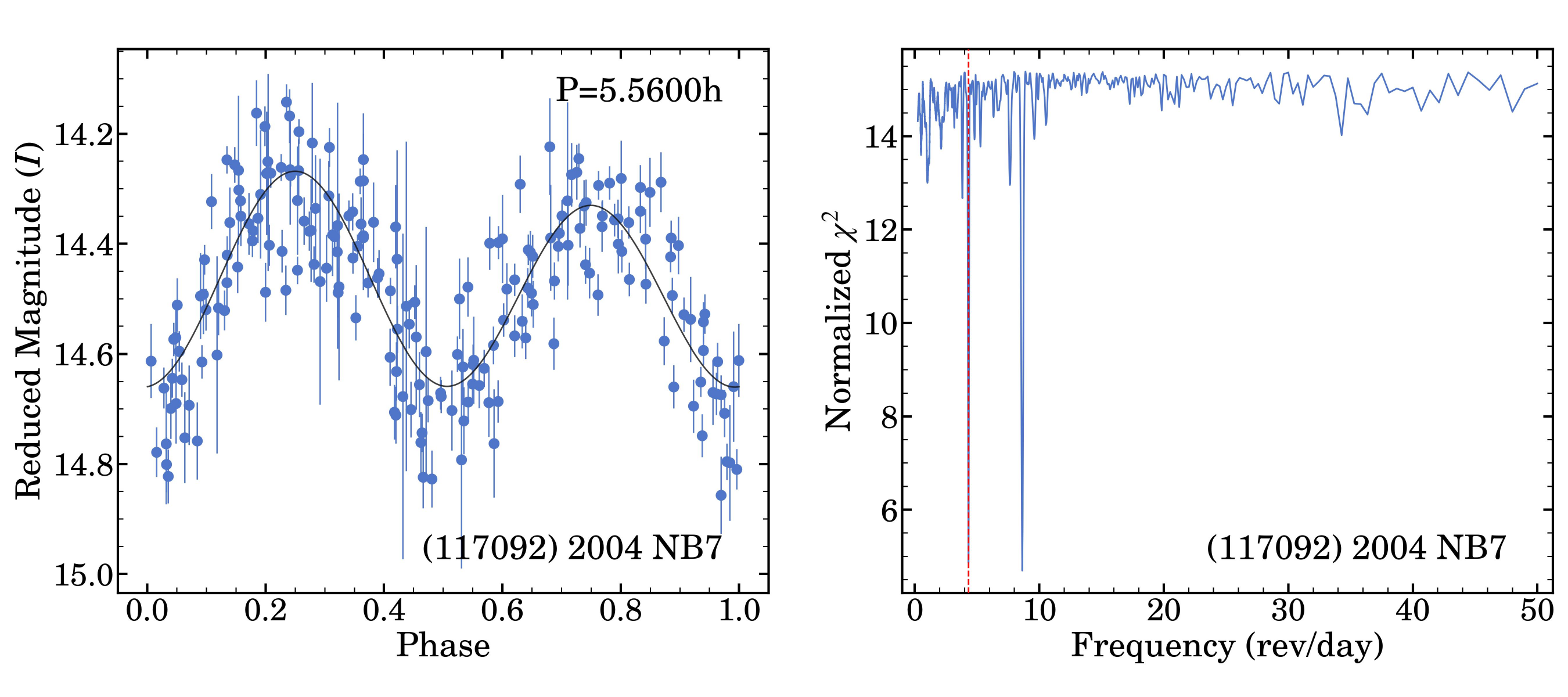}
    \includegraphics[width=0.45\linewidth]{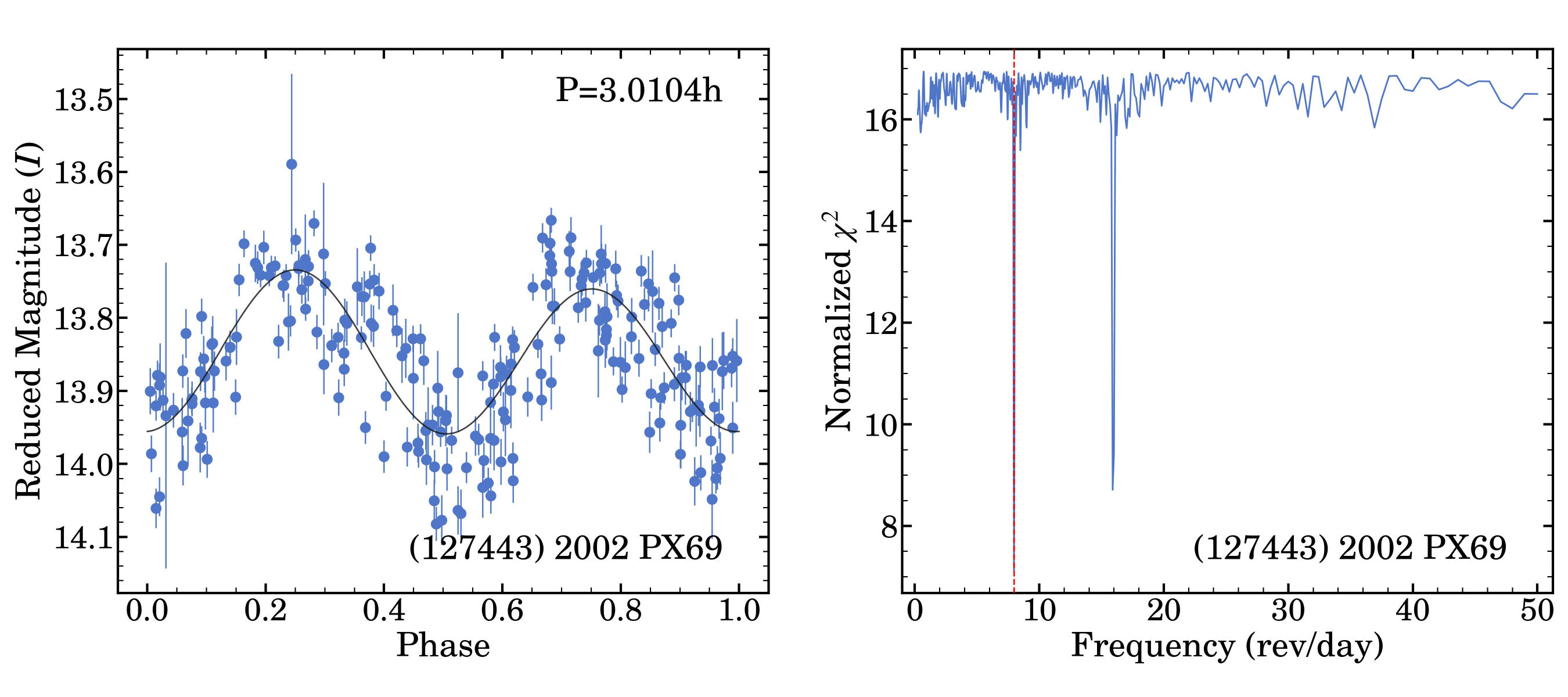}
    \includegraphics[width=0.45\linewidth]{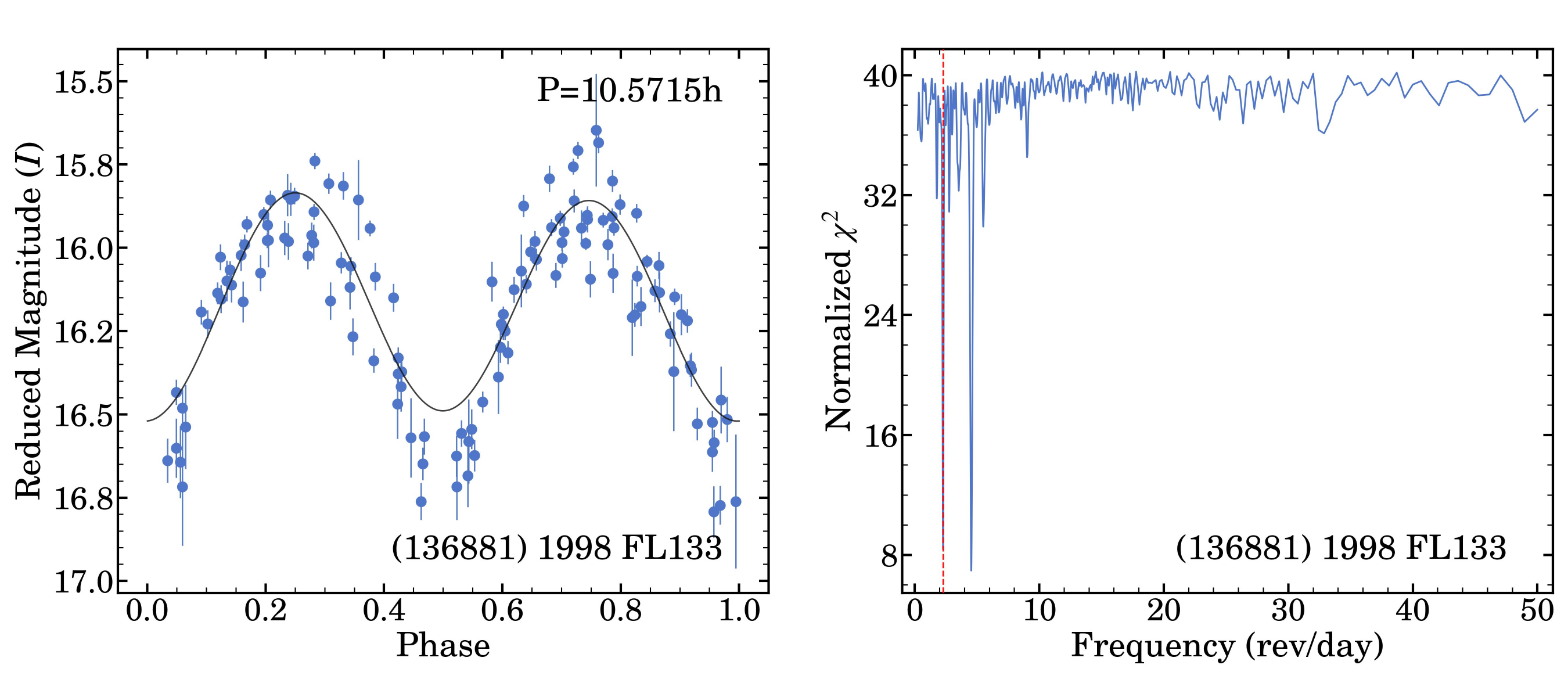}
    \caption{Continued (U = 3-).}
\end{figure*}

\begin{figure*}
    \addtocounter{figure}{-1}
    \centering
    \includegraphics[width=0.45\linewidth]{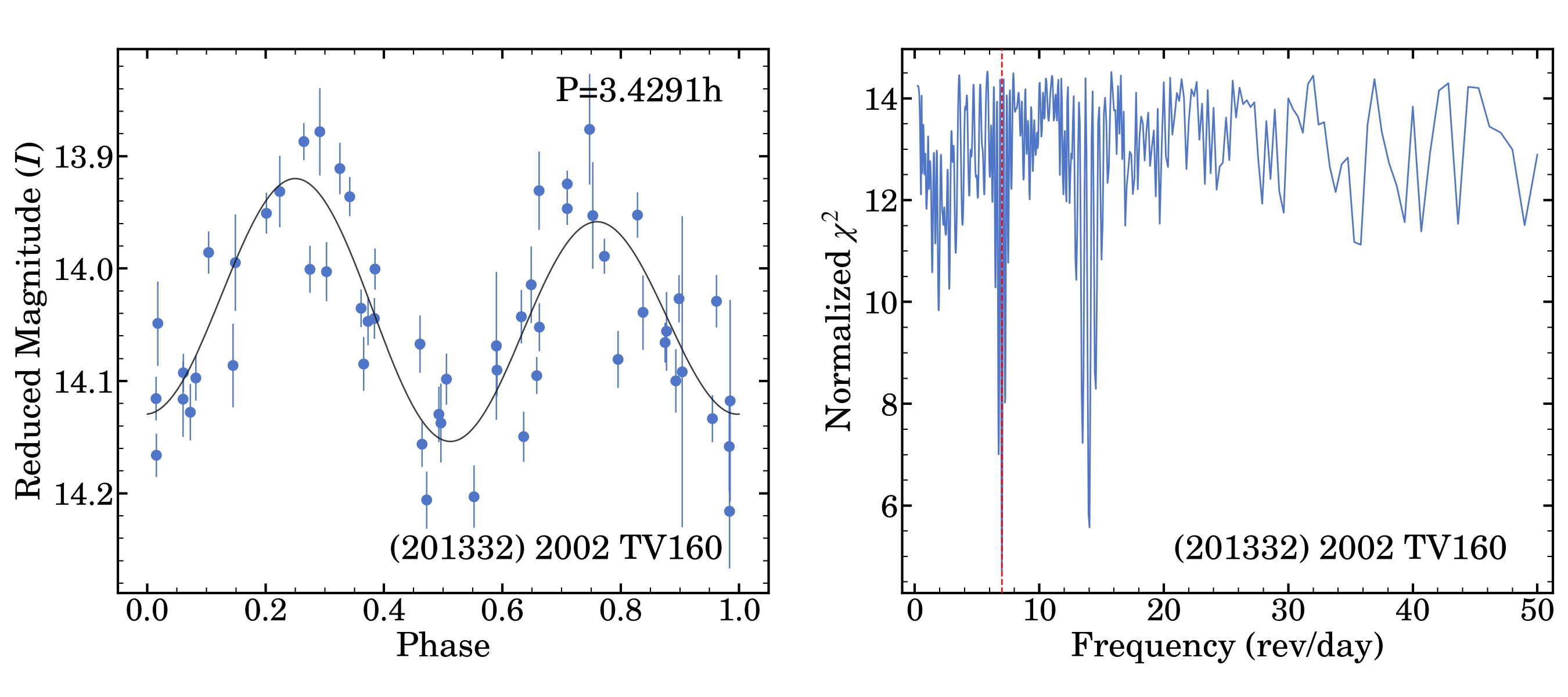}
    \includegraphics[width=0.45\linewidth]{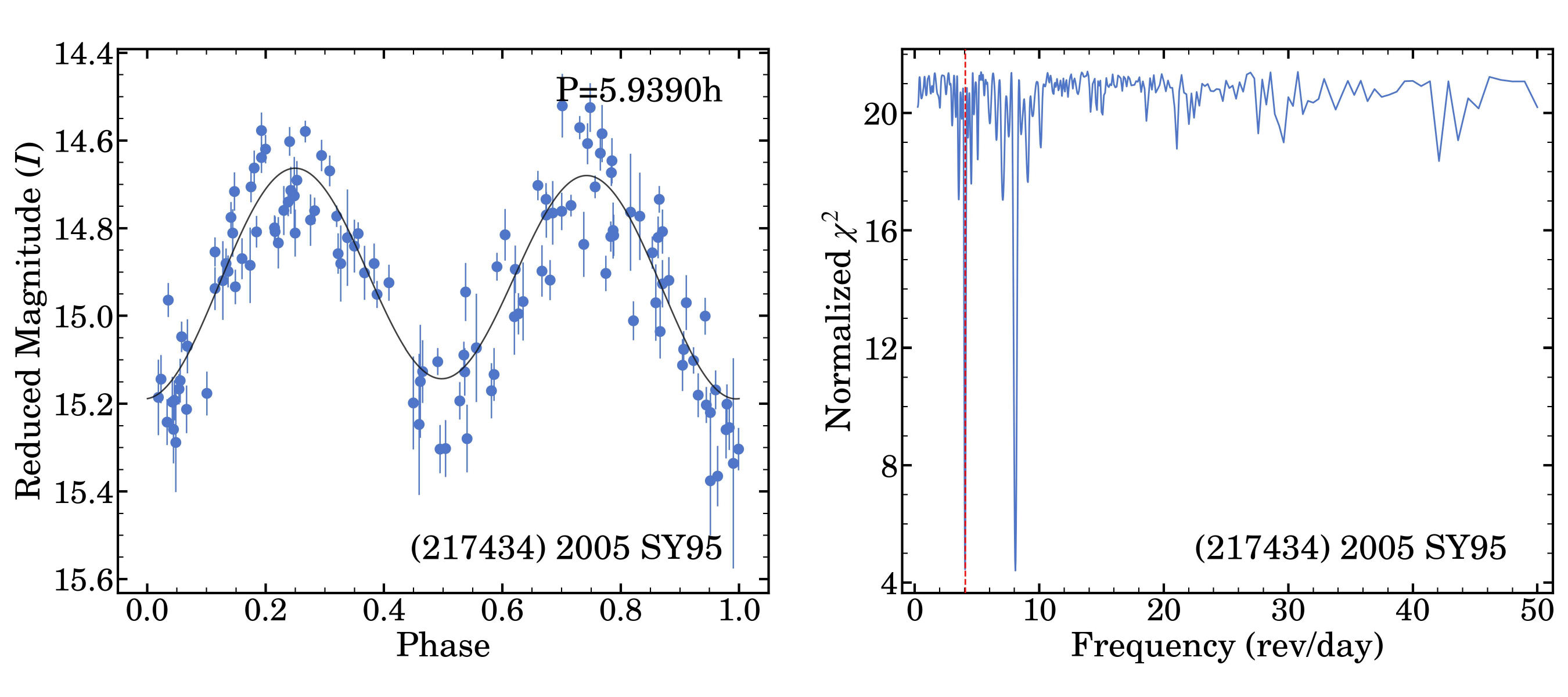}
    \includegraphics[width=0.45\linewidth]{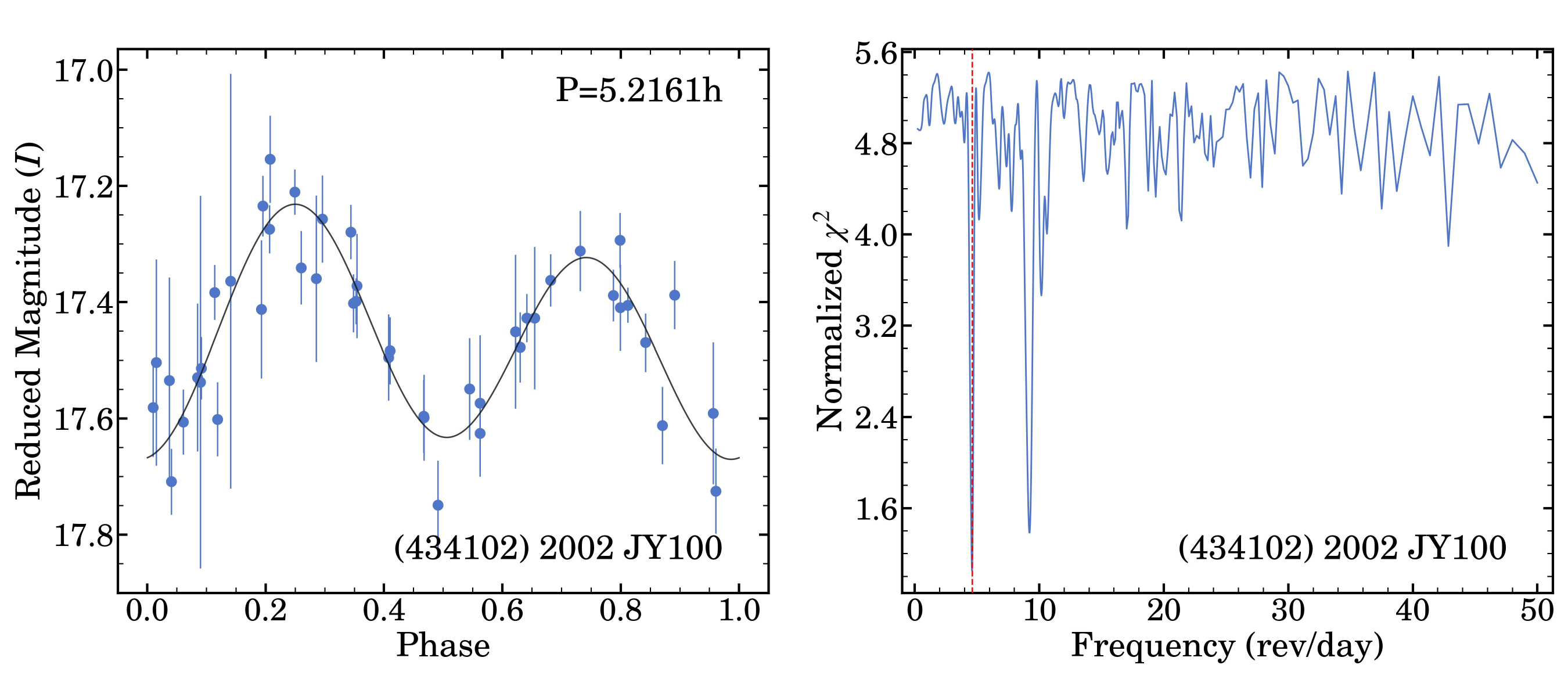}
    \includegraphics[width=0.45\linewidth]{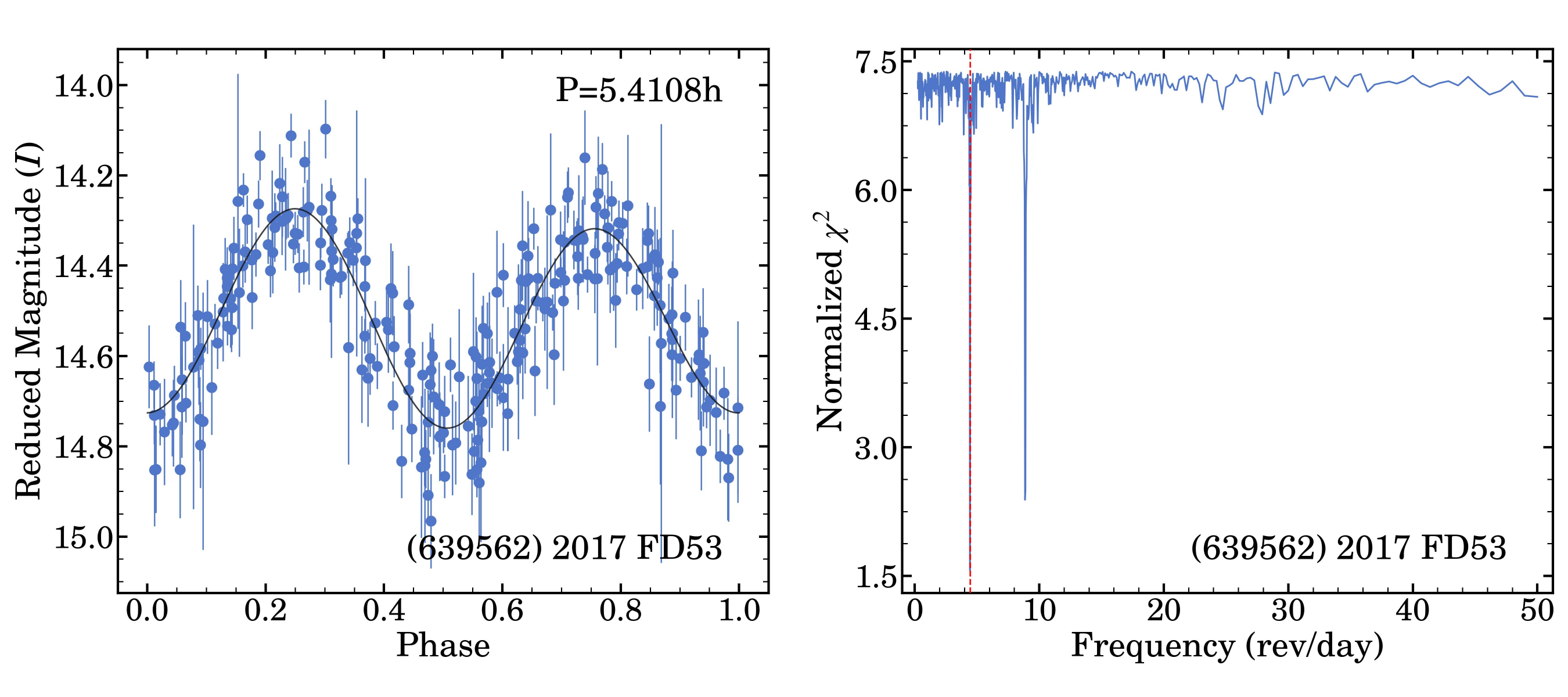}
    \includegraphics[width=0.45\linewidth]{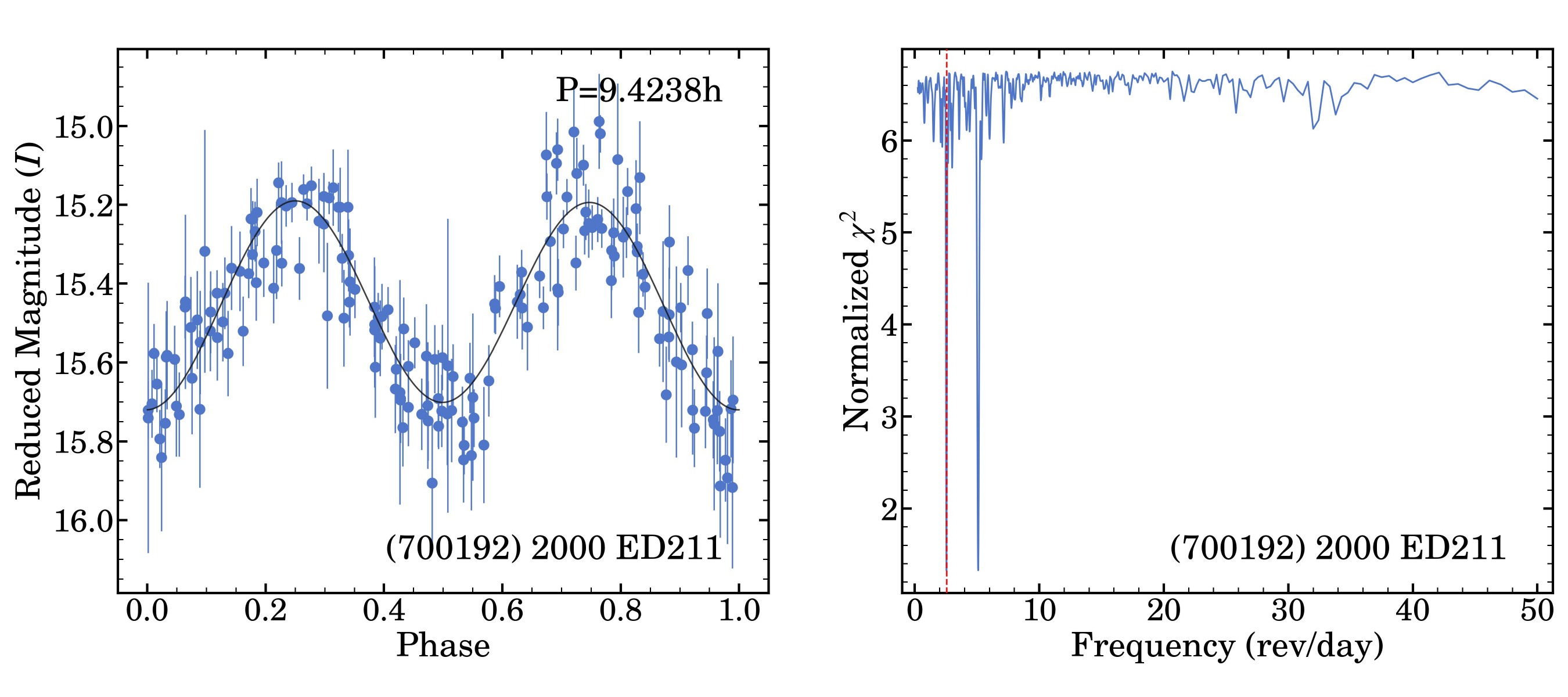}
    \caption{Continued (U = 3-).}
\end{figure*}

\clearpage

\begin{figure*}
    \centering
    \includegraphics[width=0.45\linewidth]{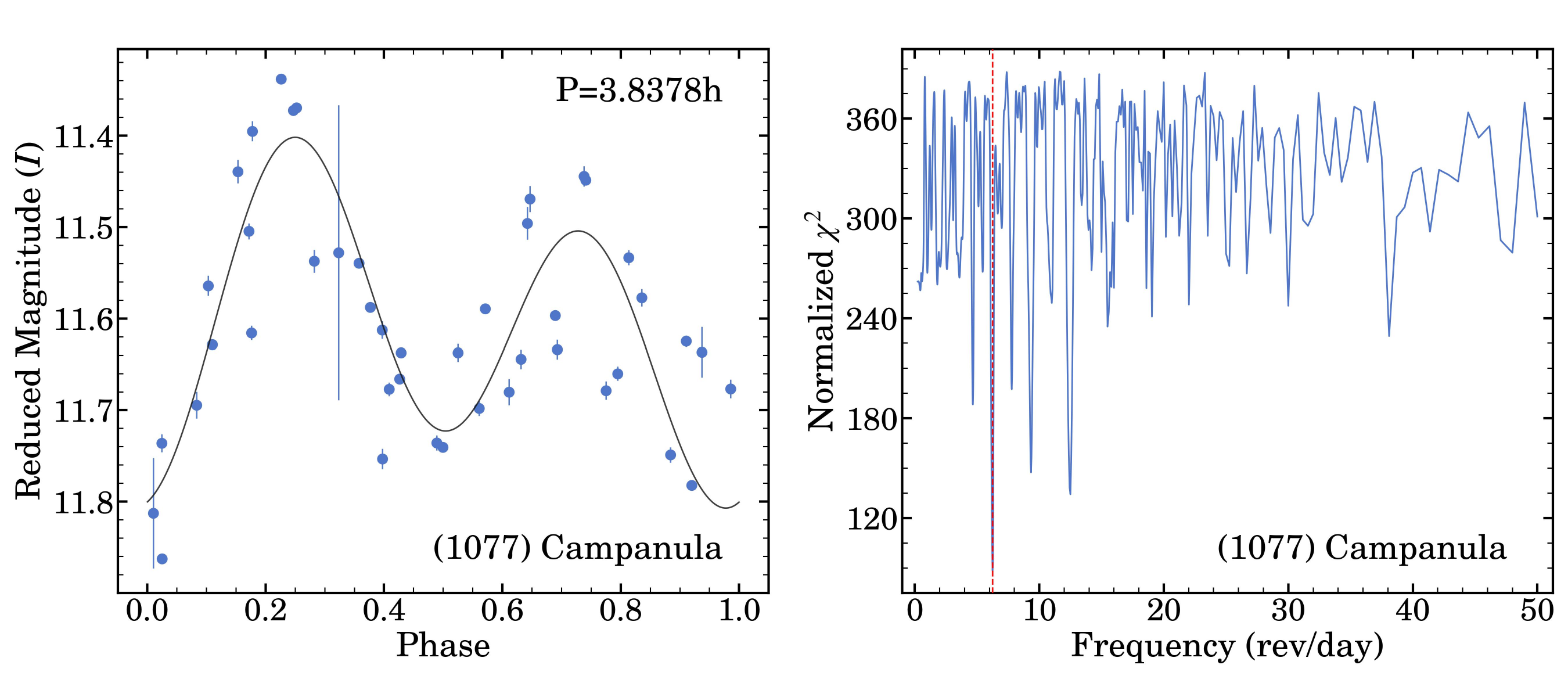}
    \includegraphics[width=0.45\linewidth]{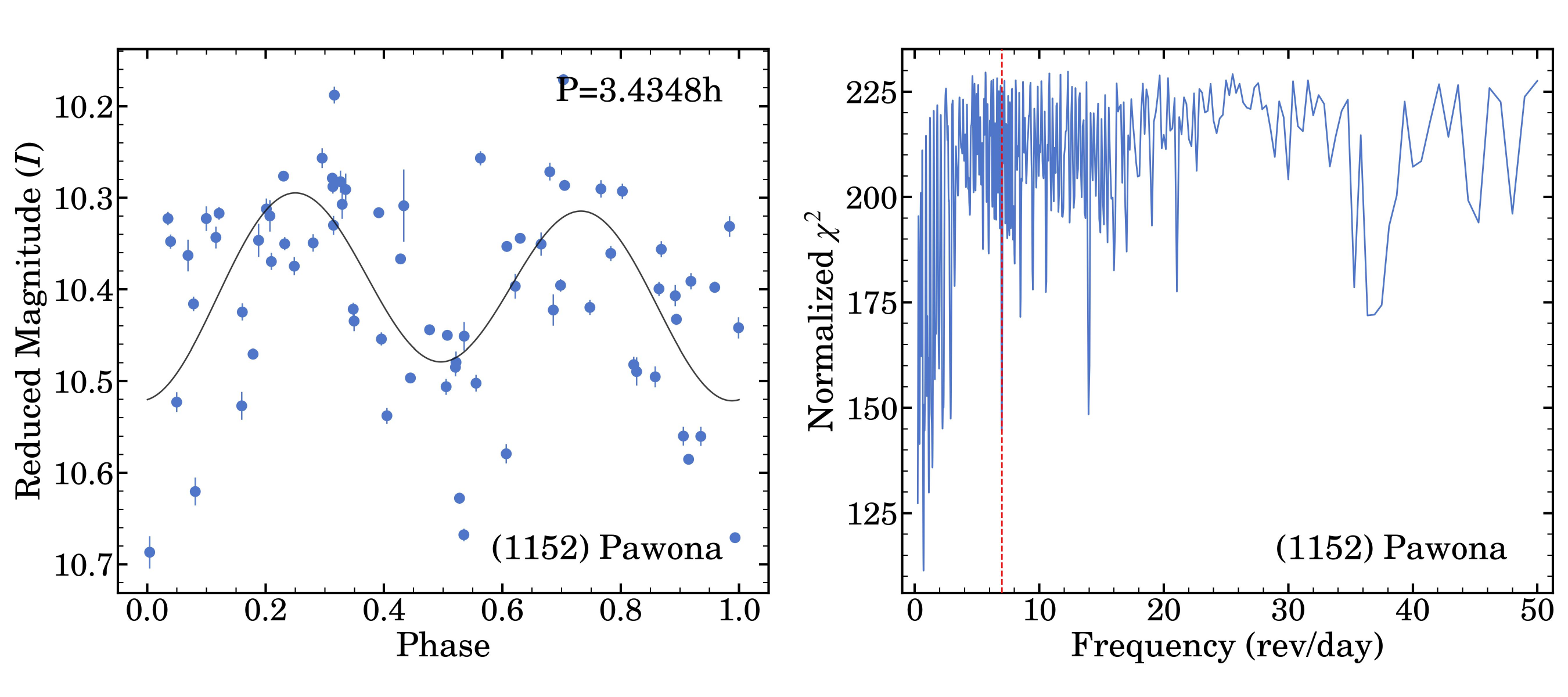}
    \includegraphics[width=0.45\linewidth]{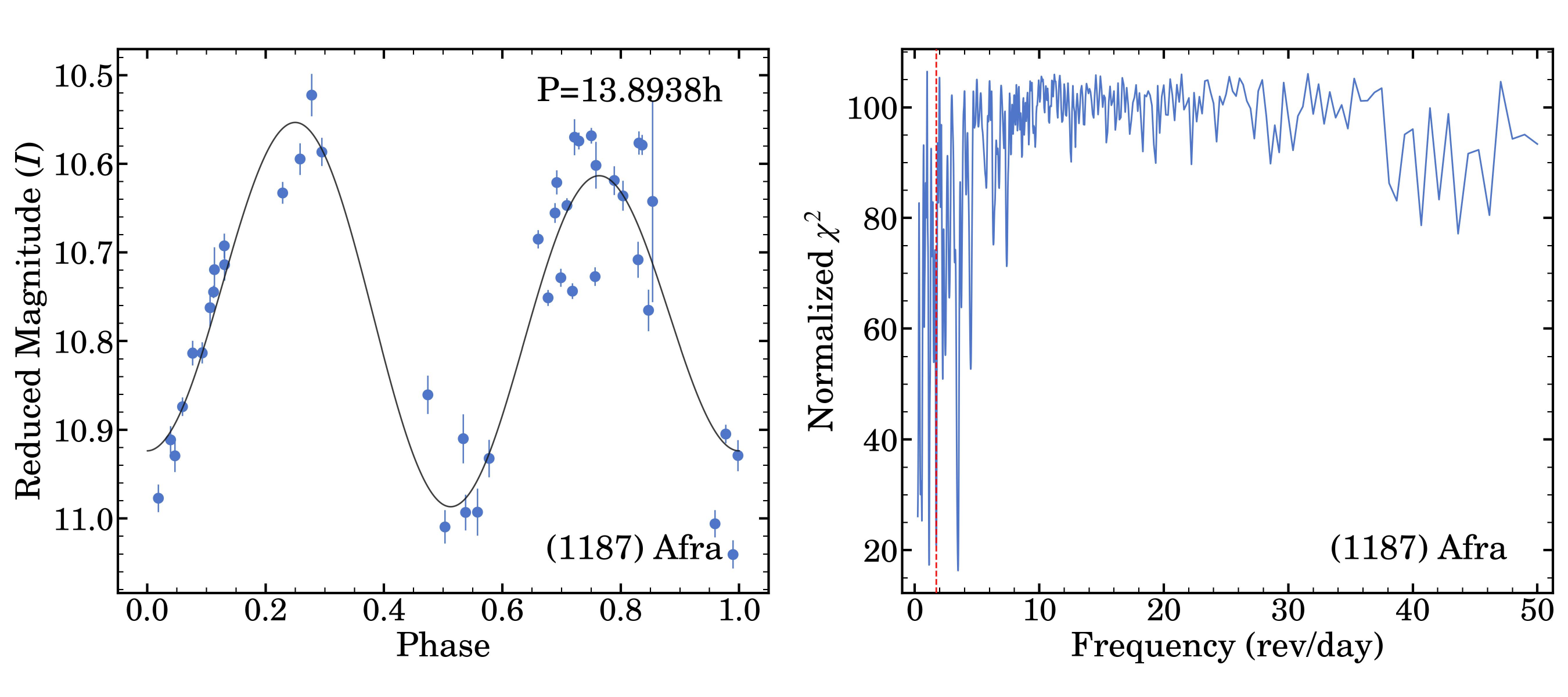}
    \includegraphics[width=0.45\linewidth]{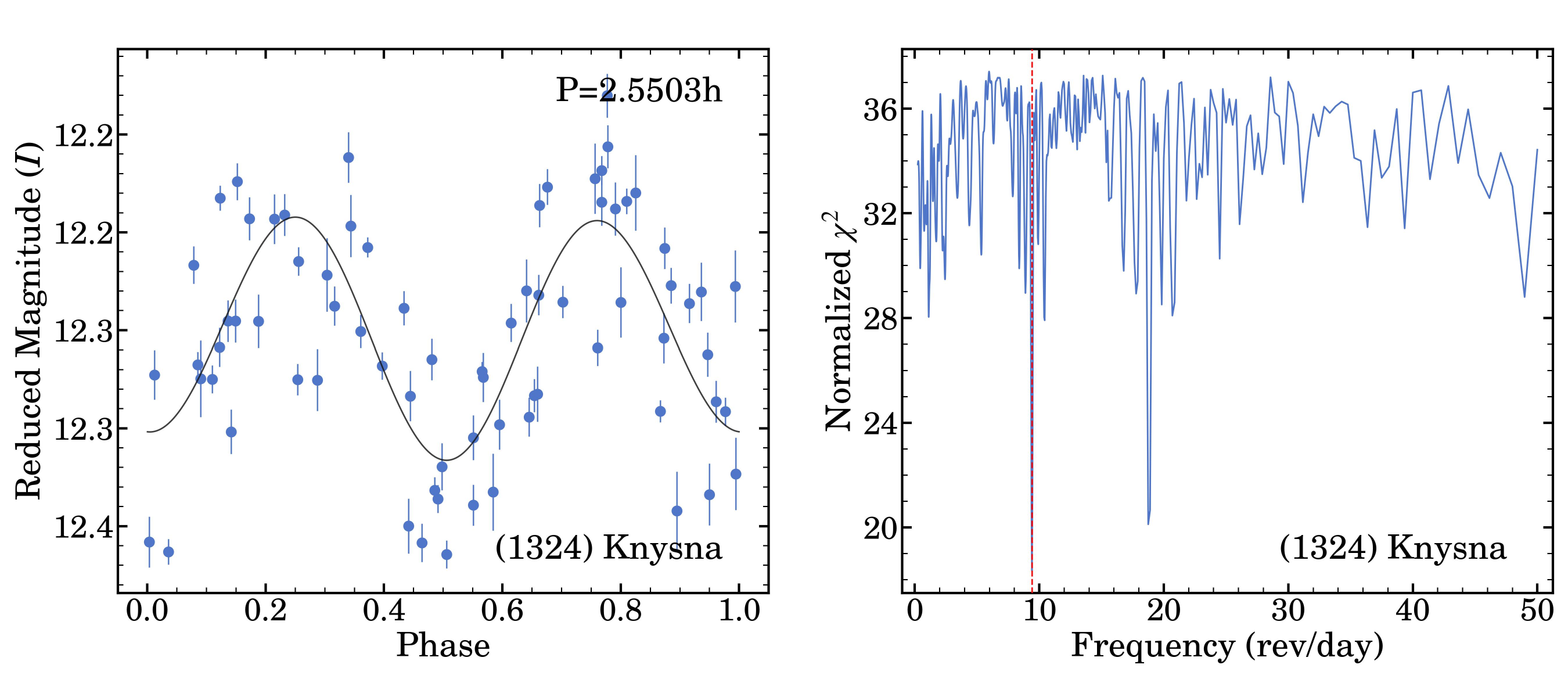}
    \includegraphics[width=0.45\linewidth]{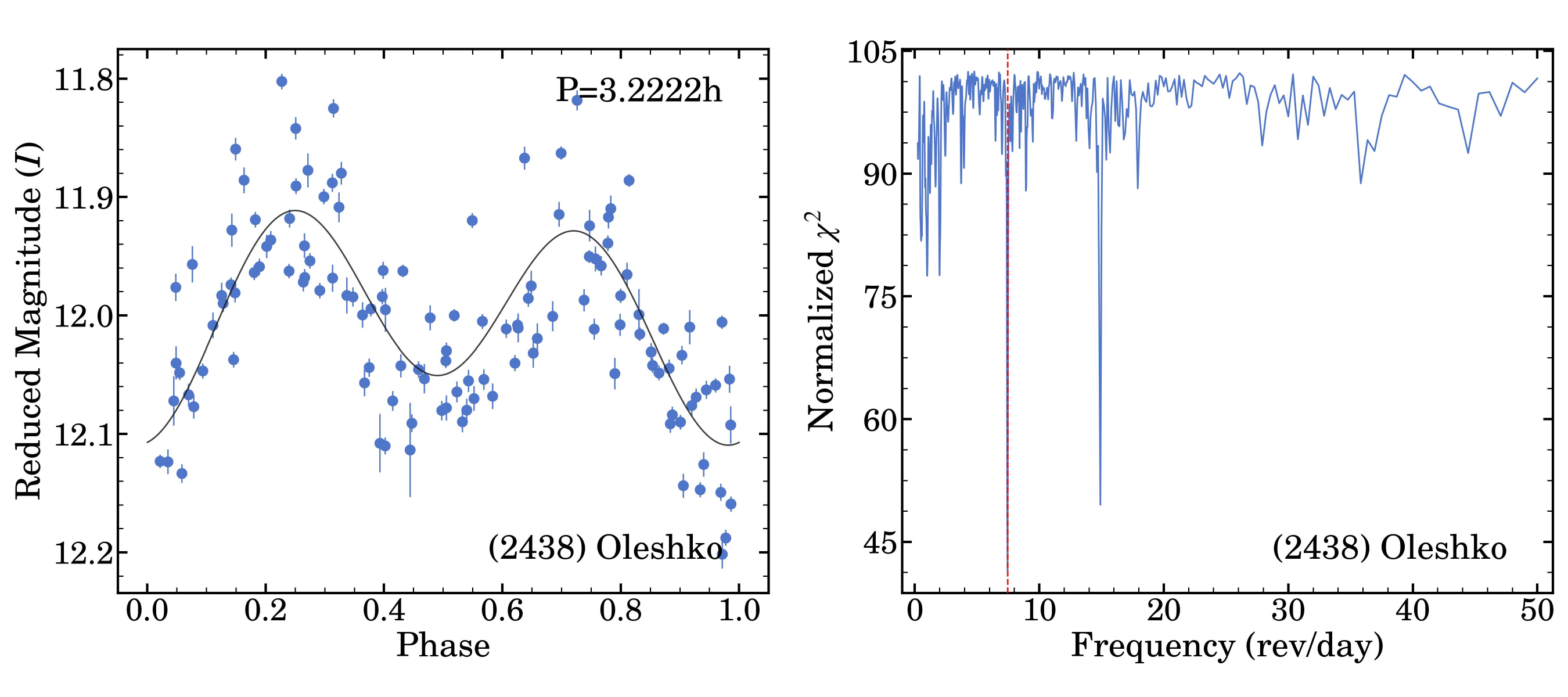}
    \includegraphics[width=0.45\linewidth]{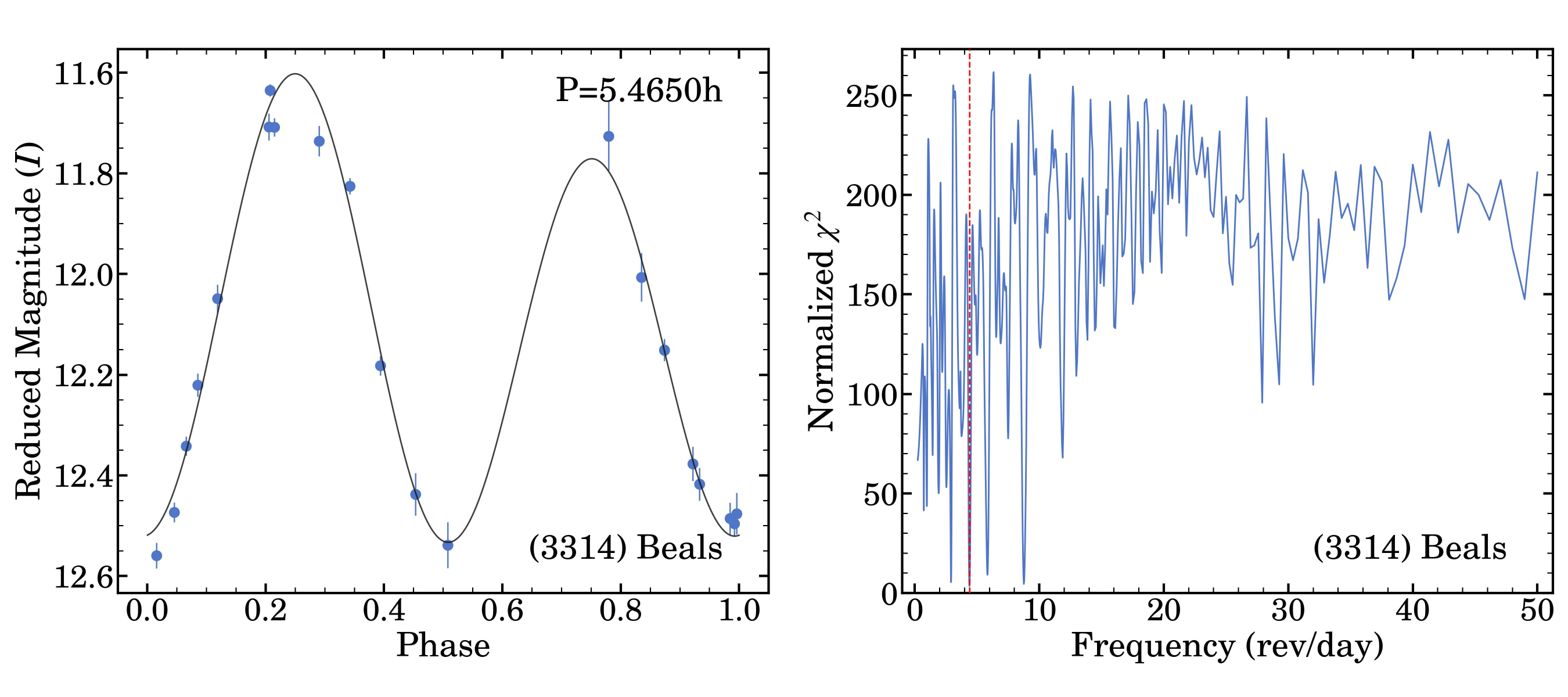}
    \includegraphics[width=0.45\linewidth]{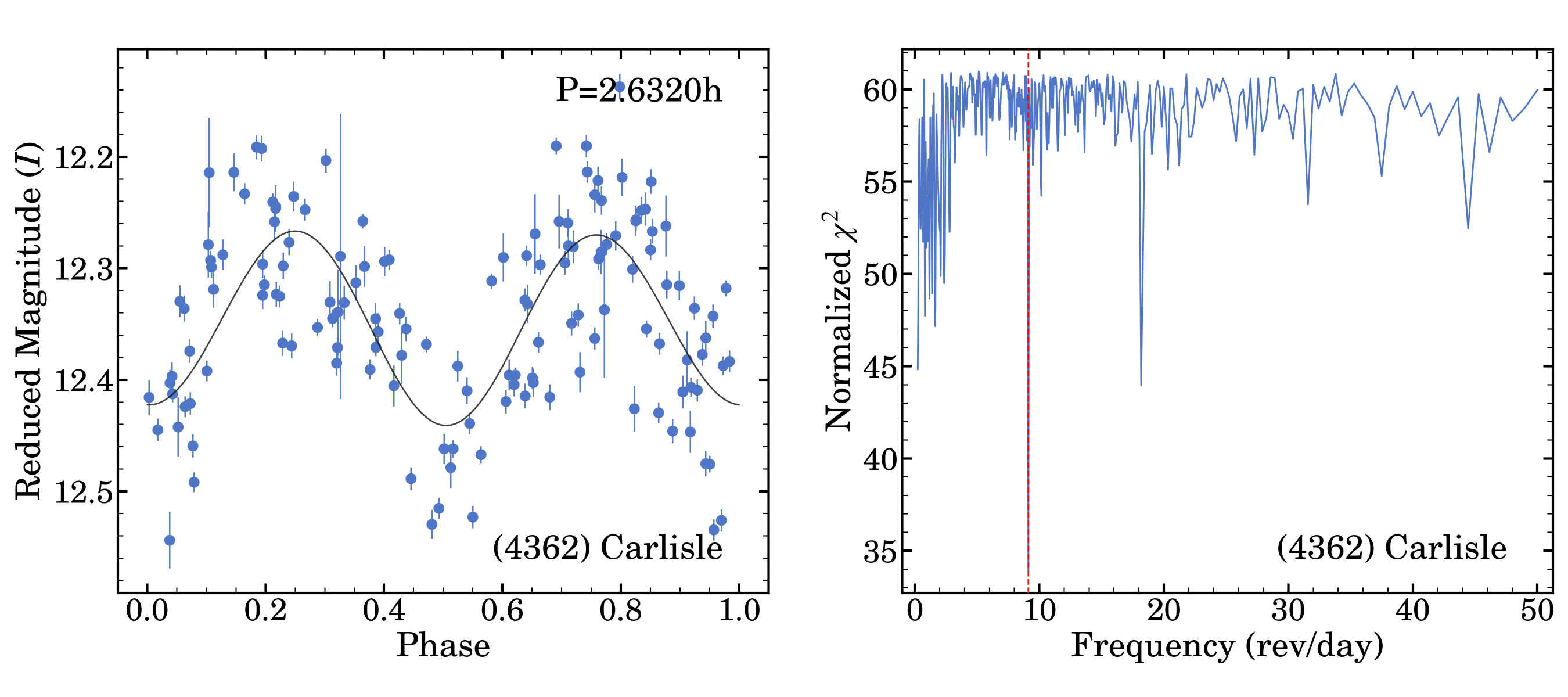}
    \includegraphics[width=0.45\linewidth]{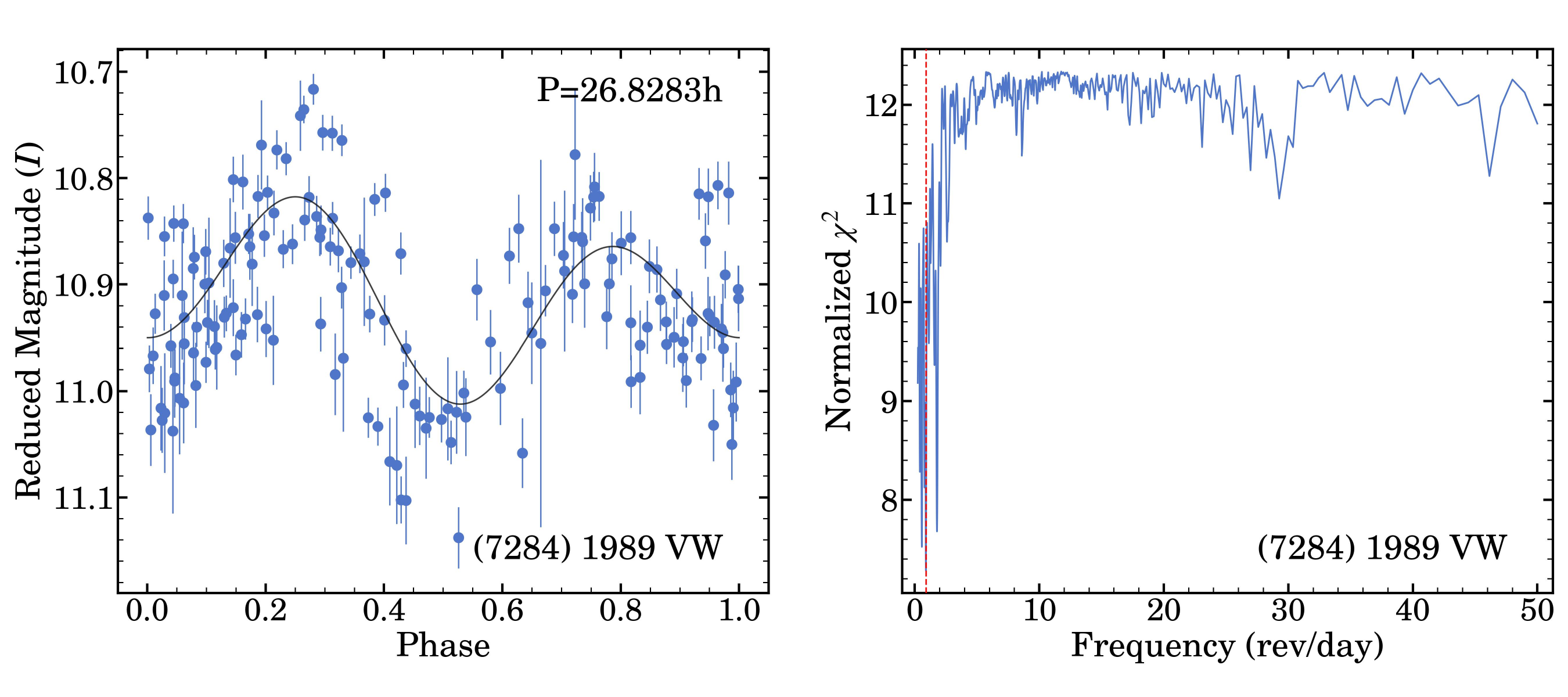}
    \includegraphics[width=0.45\linewidth]{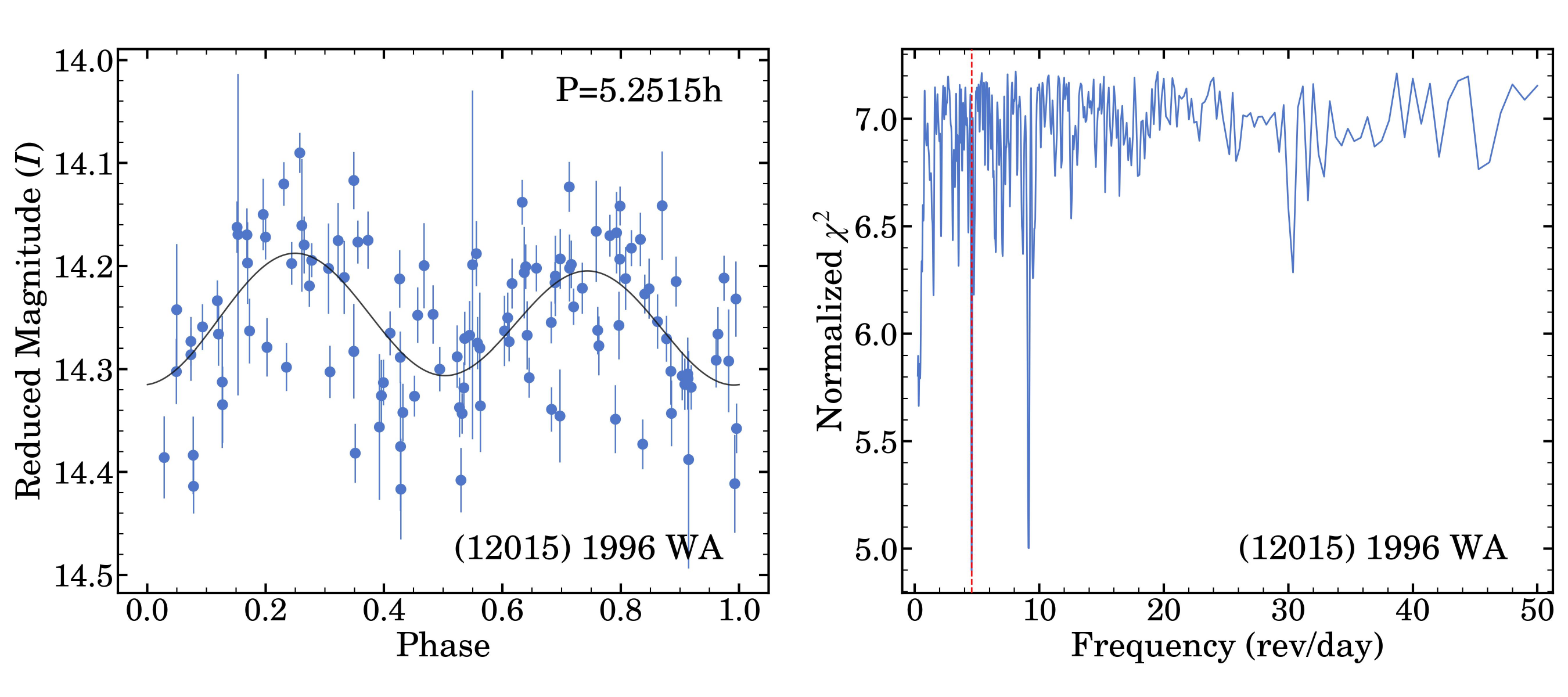}
    \includegraphics[width=0.45\linewidth]{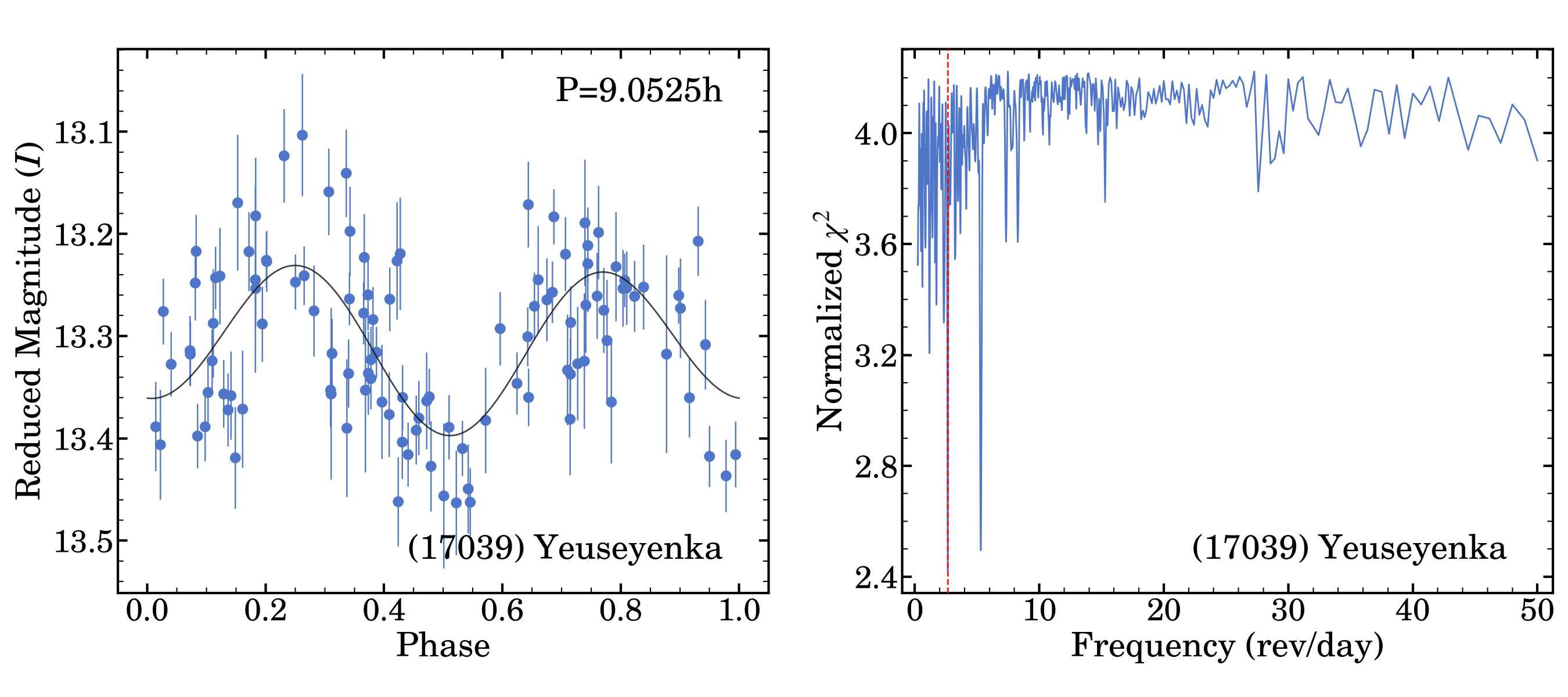}
    \includegraphics[width=0.45\linewidth]{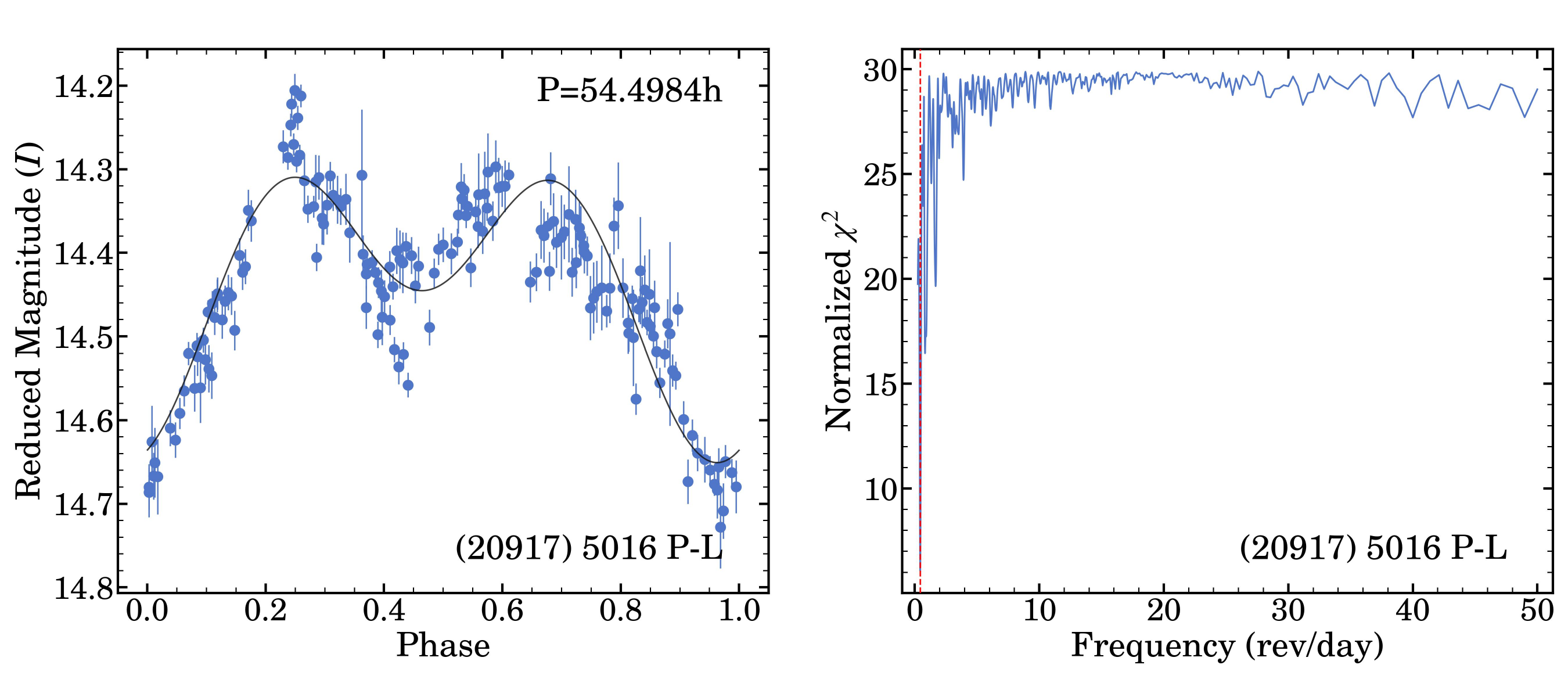}
    \includegraphics[width=0.45\linewidth]{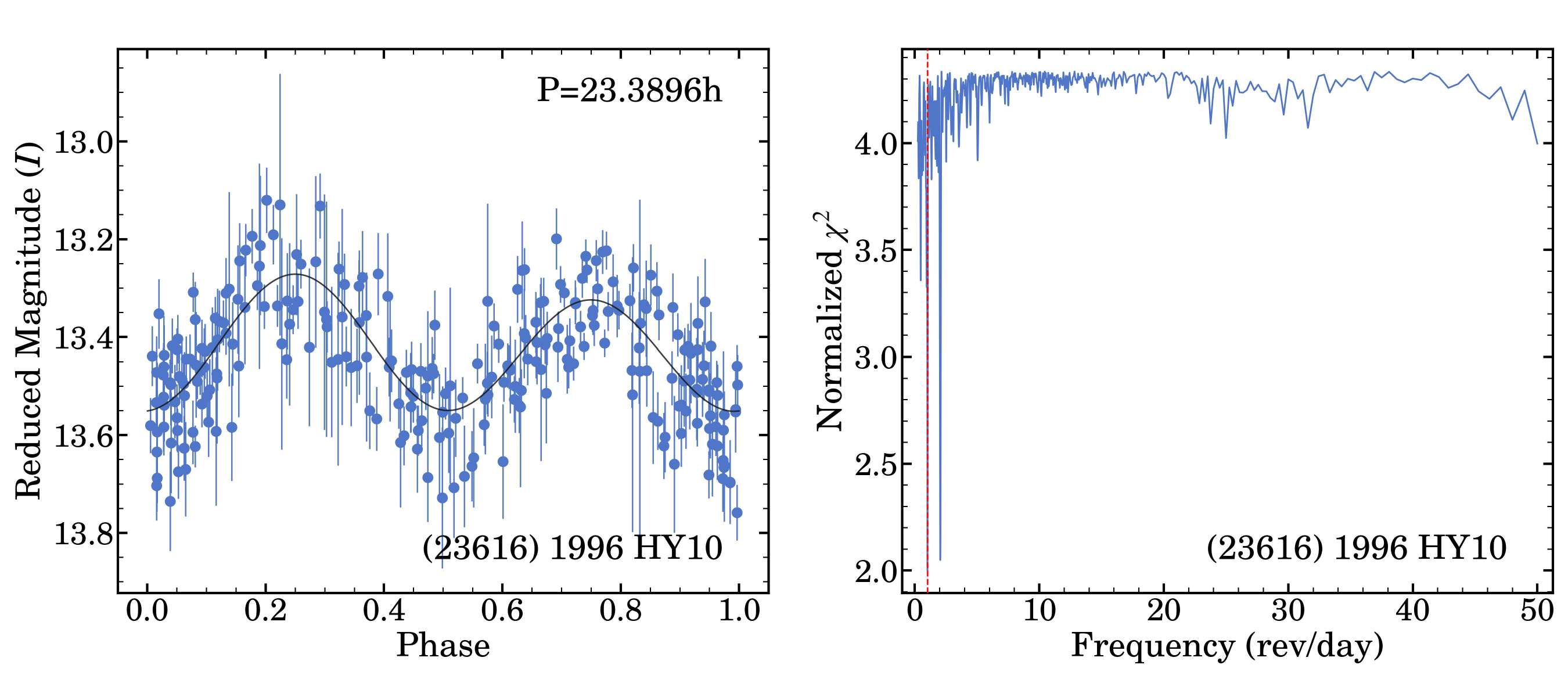}
    \caption{Folded lightcurves (left) and normalized $\chi^2$ as a function of frequency (right) for asteroids with rotation period quality code U = 2+.}
    \label{fig:appendix2+}
\end{figure*}

\begin{figure*}
    \addtocounter{figure}{-1}
    \centering
    \includegraphics[width=0.45\linewidth]{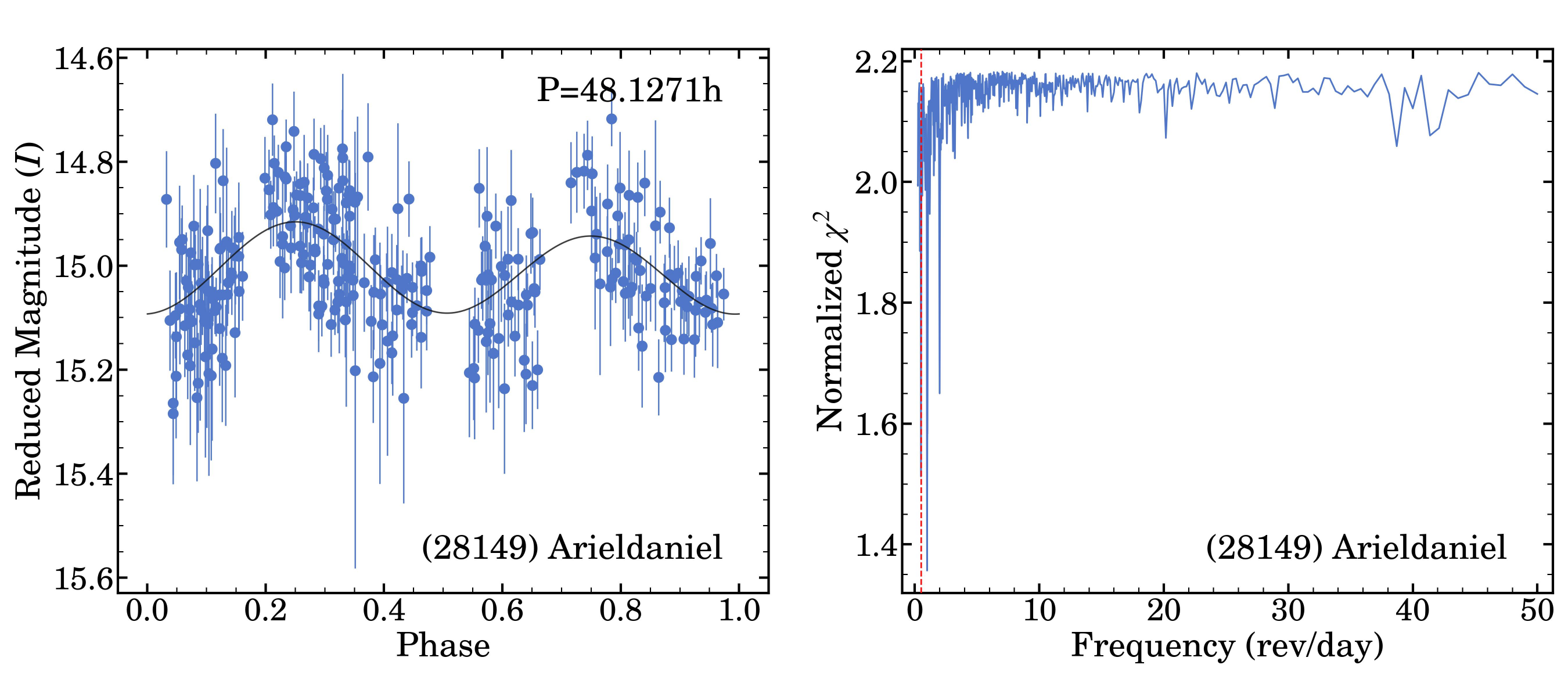}
    \includegraphics[width=0.45\linewidth]{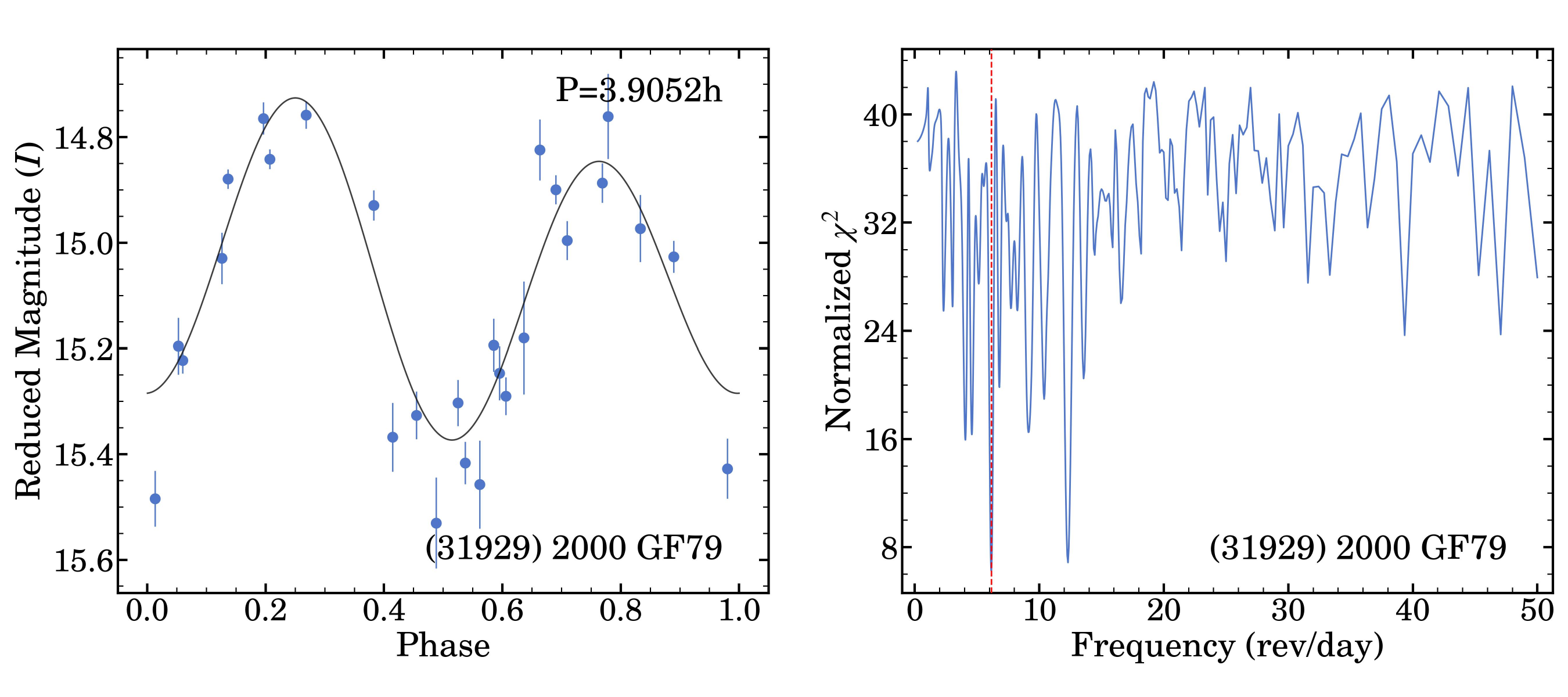}
    \includegraphics[width=0.45\linewidth]{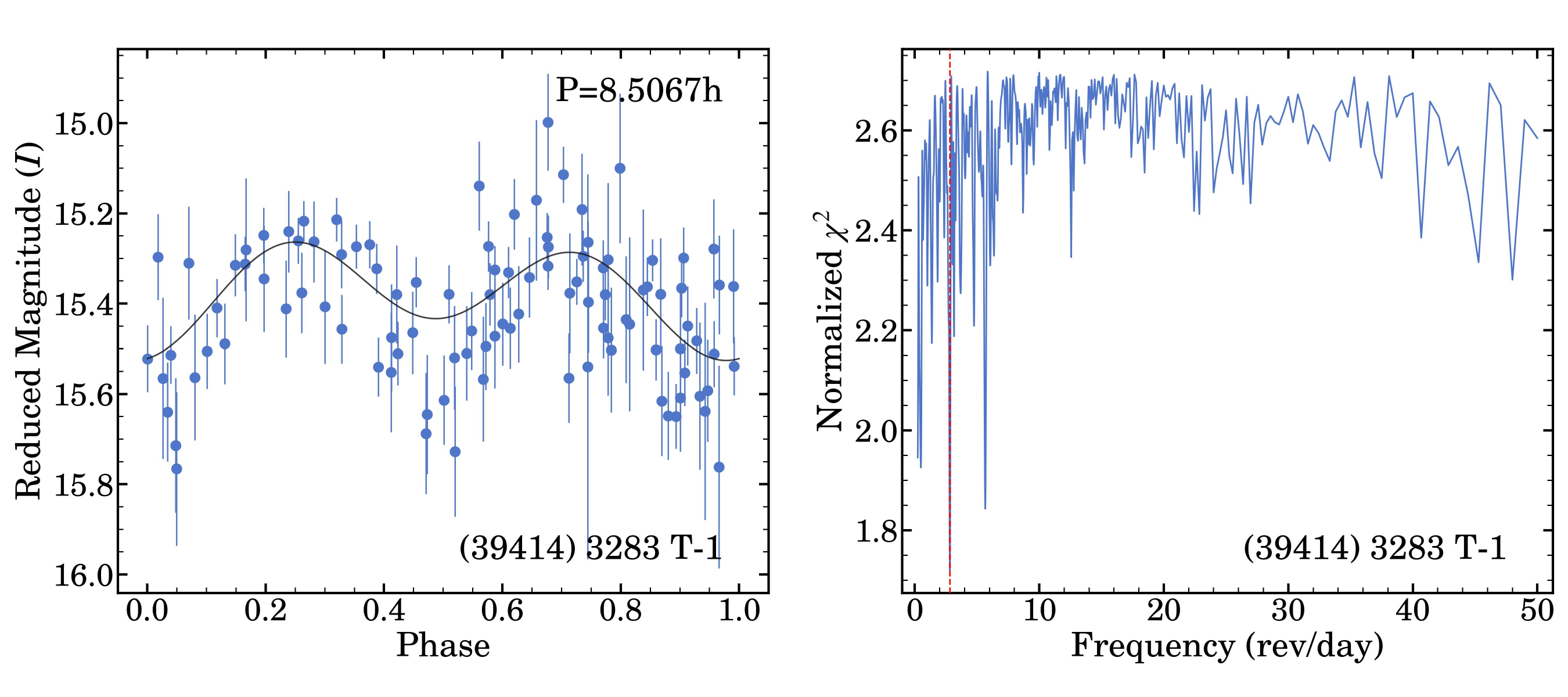}
    \includegraphics[width=0.45\linewidth]{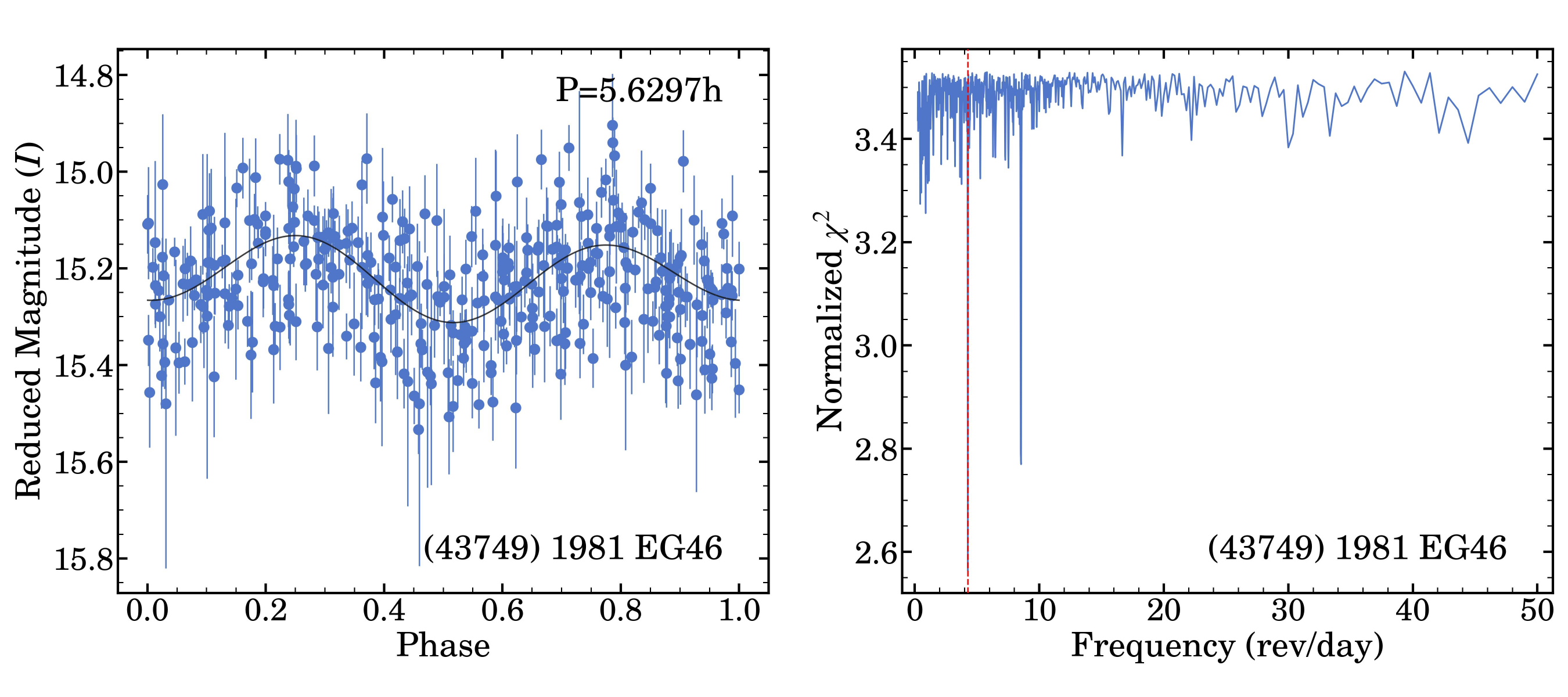}
    \includegraphics[width=0.45\linewidth]{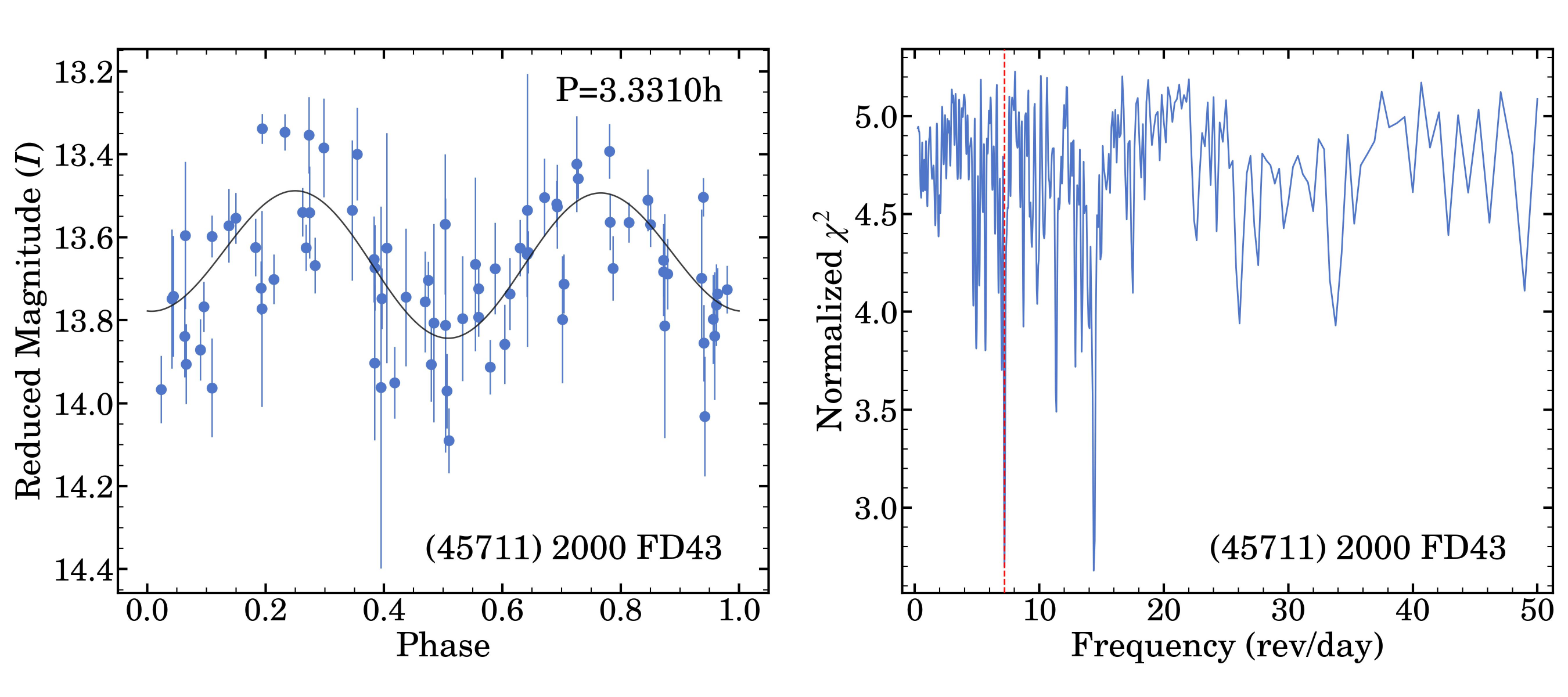}
    \includegraphics[width=0.45\linewidth]{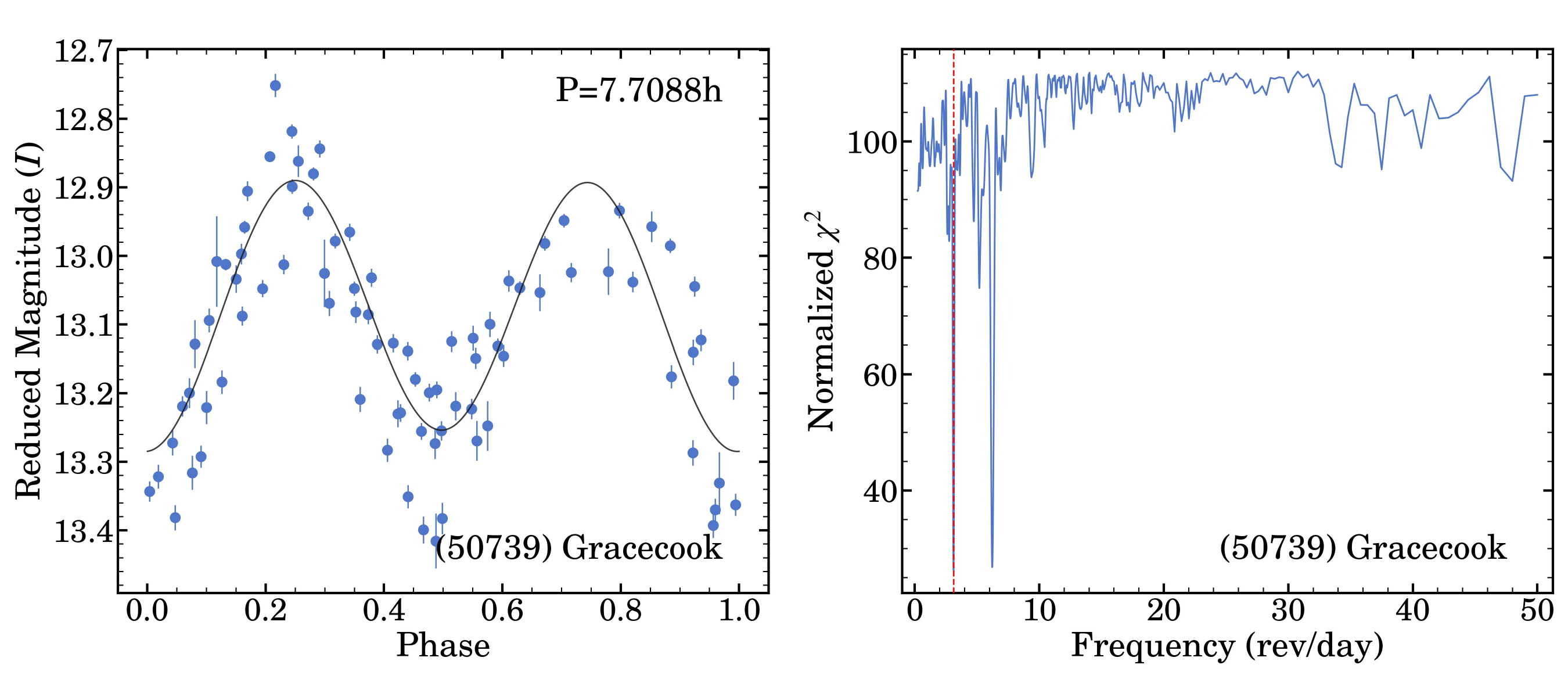}
    \includegraphics[width=0.45\linewidth]{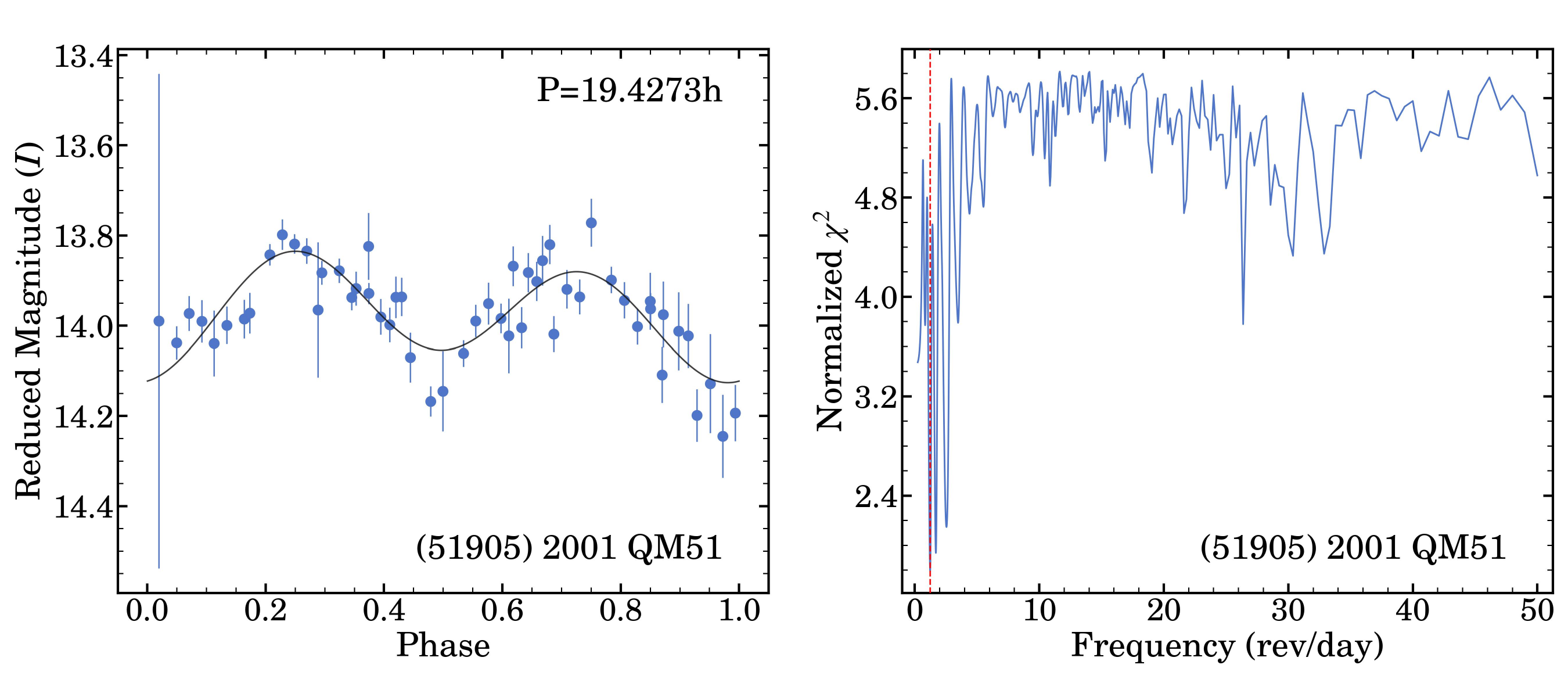}
    \includegraphics[width=0.45\linewidth]{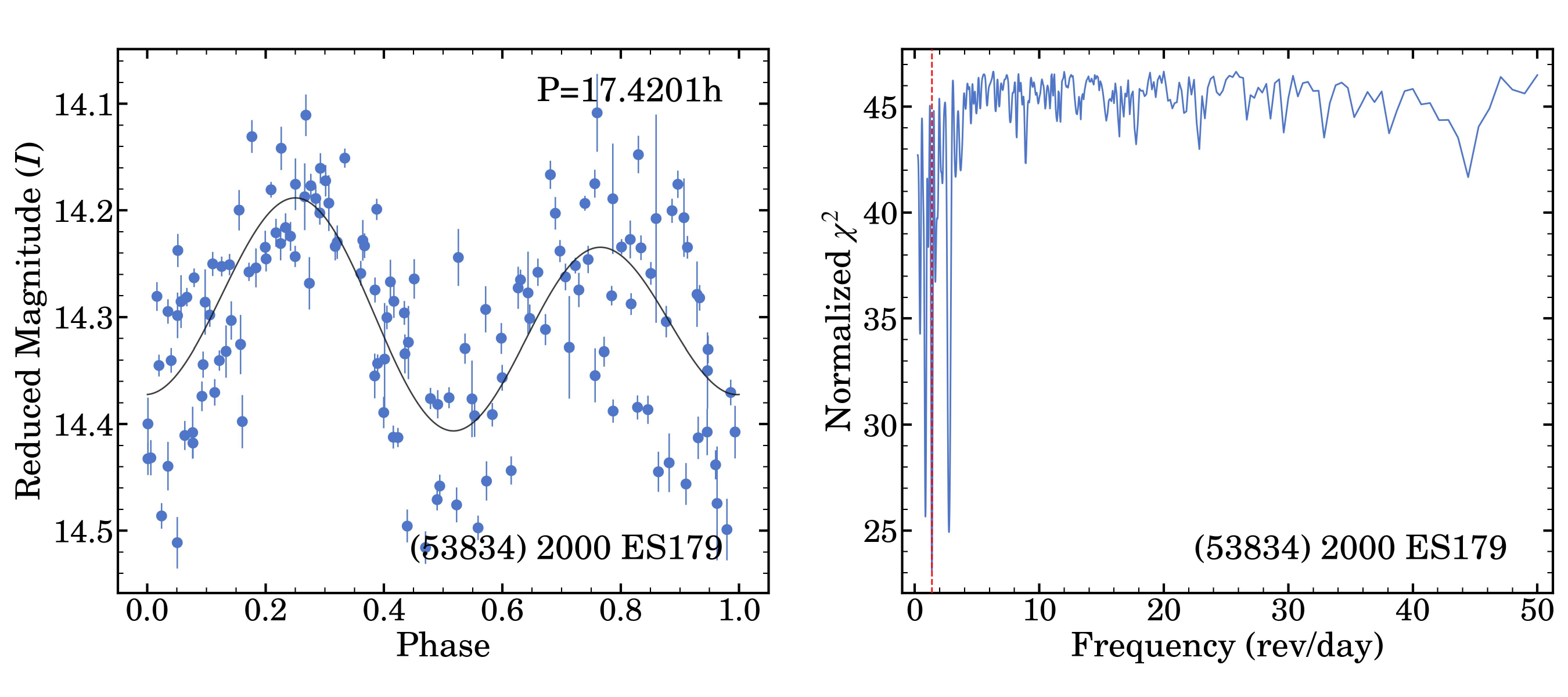}
    \includegraphics[width=0.45\linewidth]{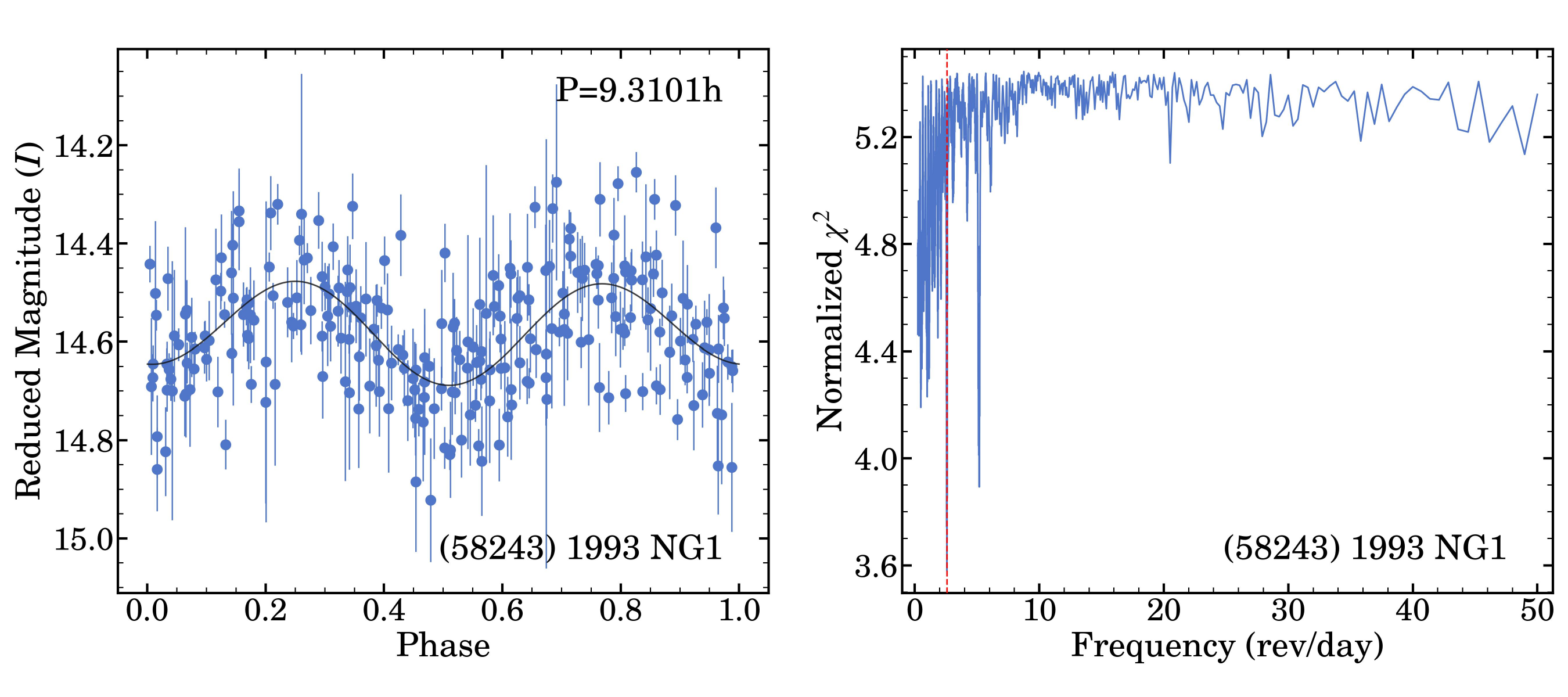}
    \includegraphics[width=0.45\linewidth]{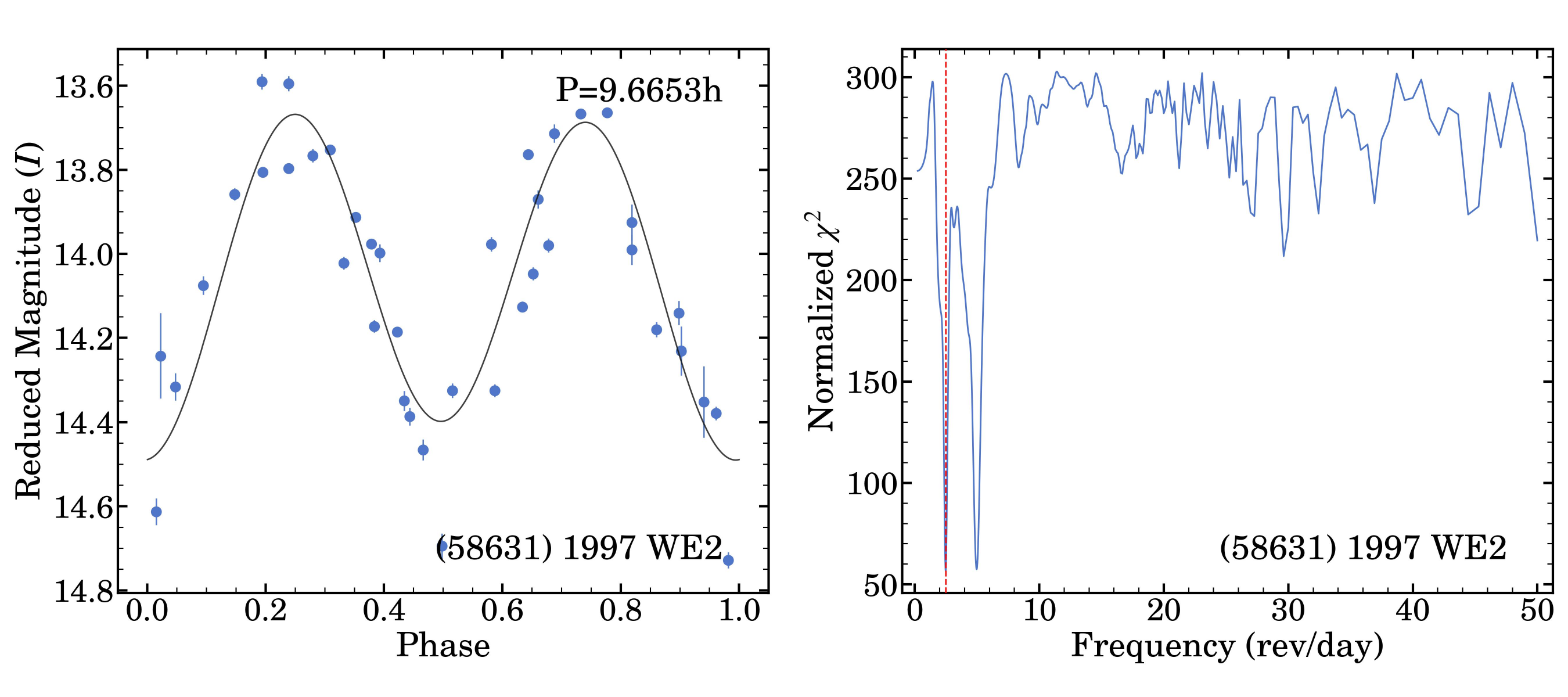}
    \includegraphics[width=0.45\linewidth]{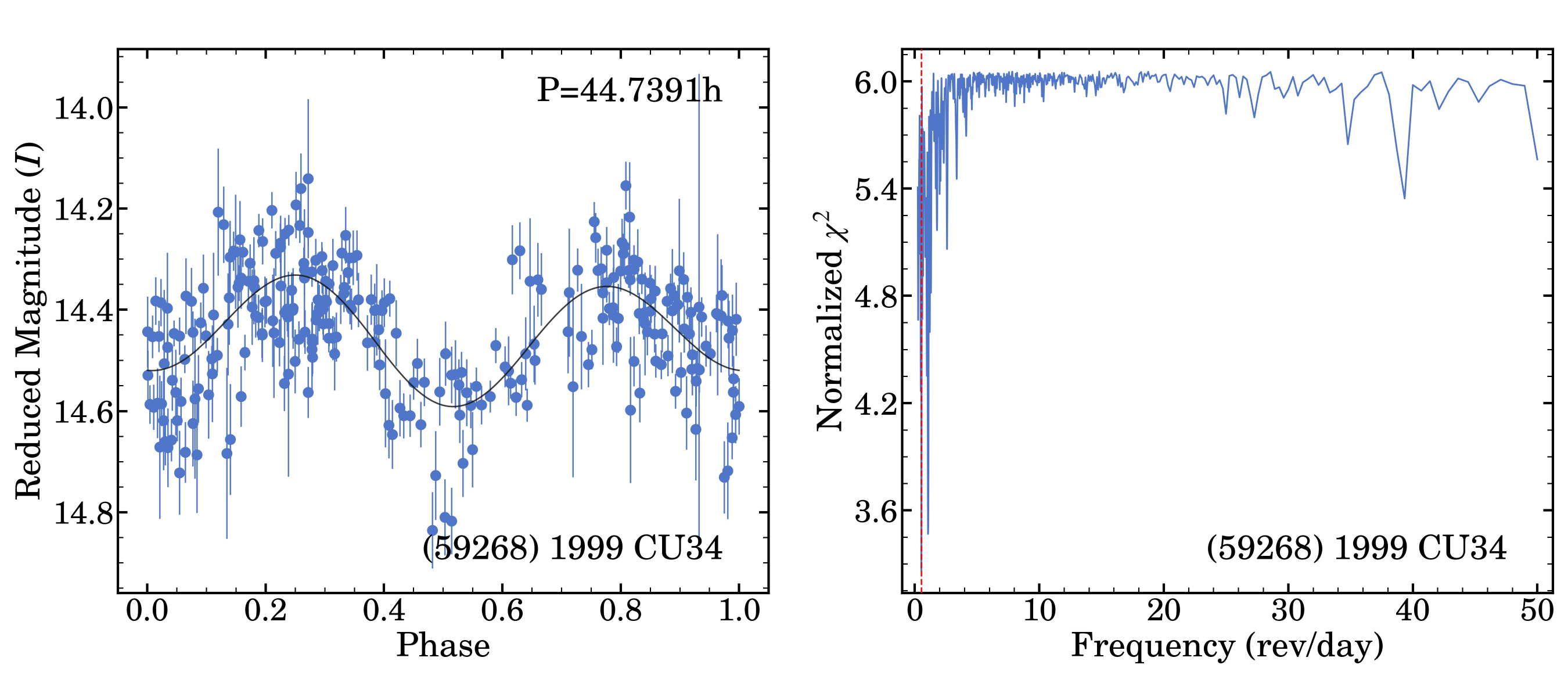}
    \includegraphics[width=0.45\linewidth]{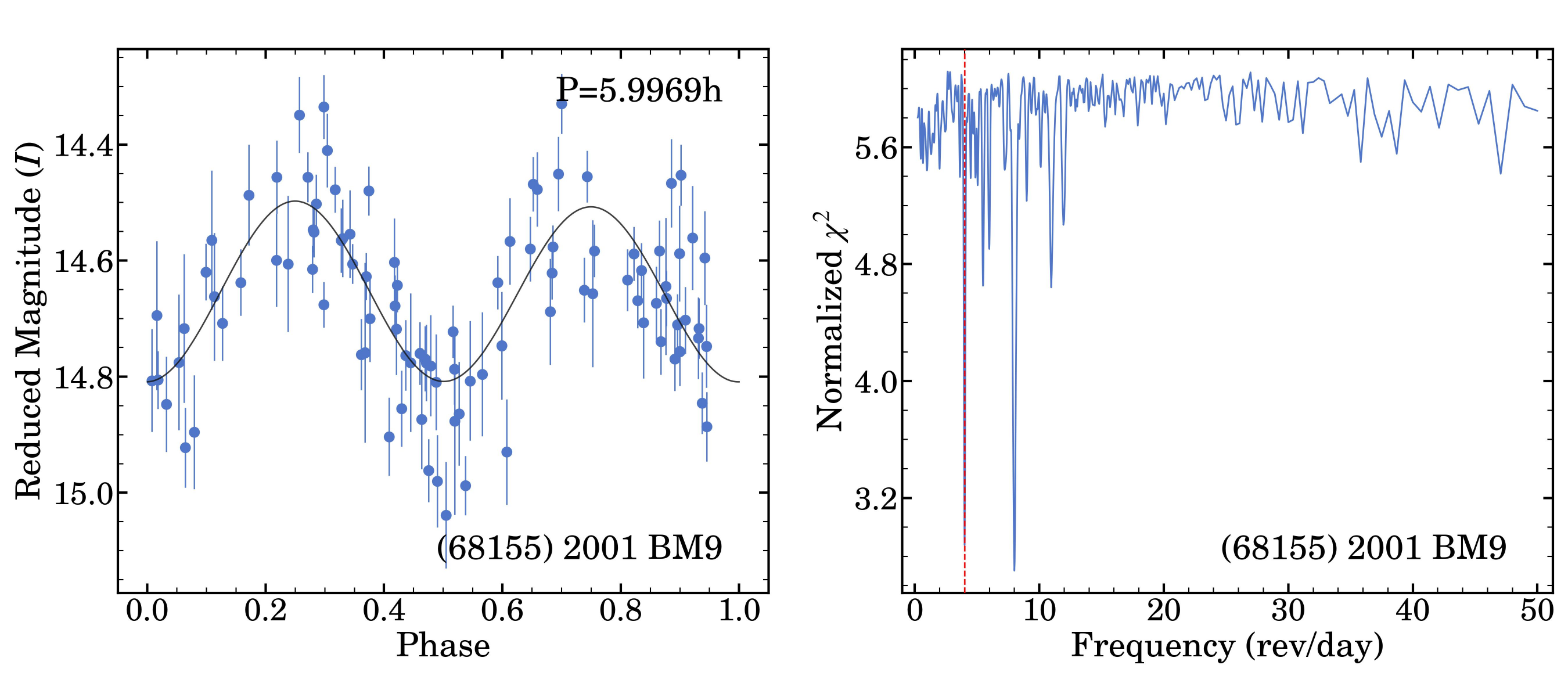}
    \caption{Continued (U = 2+).}
\end{figure*}

\begin{figure*}
    \addtocounter{figure}{-1}
    \centering
    \includegraphics[width=0.45\linewidth]{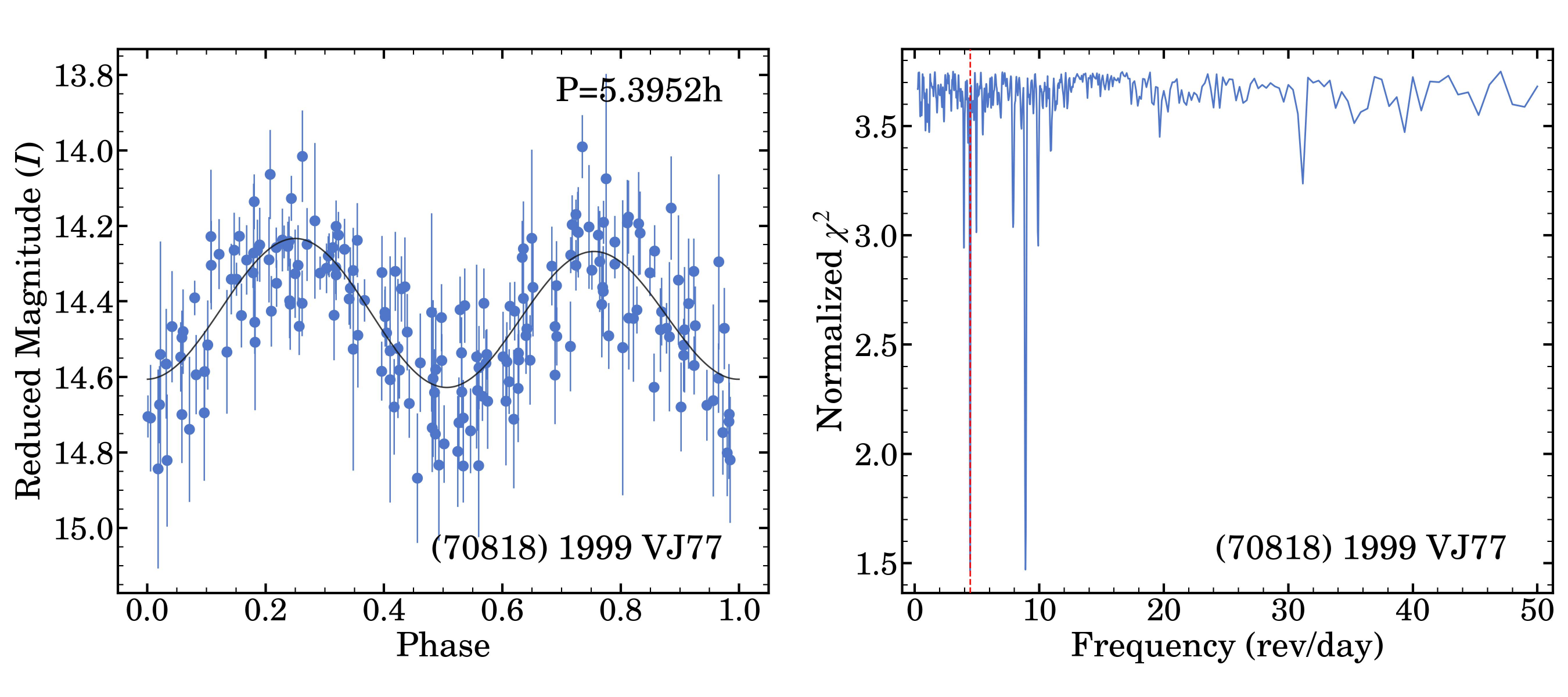}
    \includegraphics[width=0.45\linewidth]{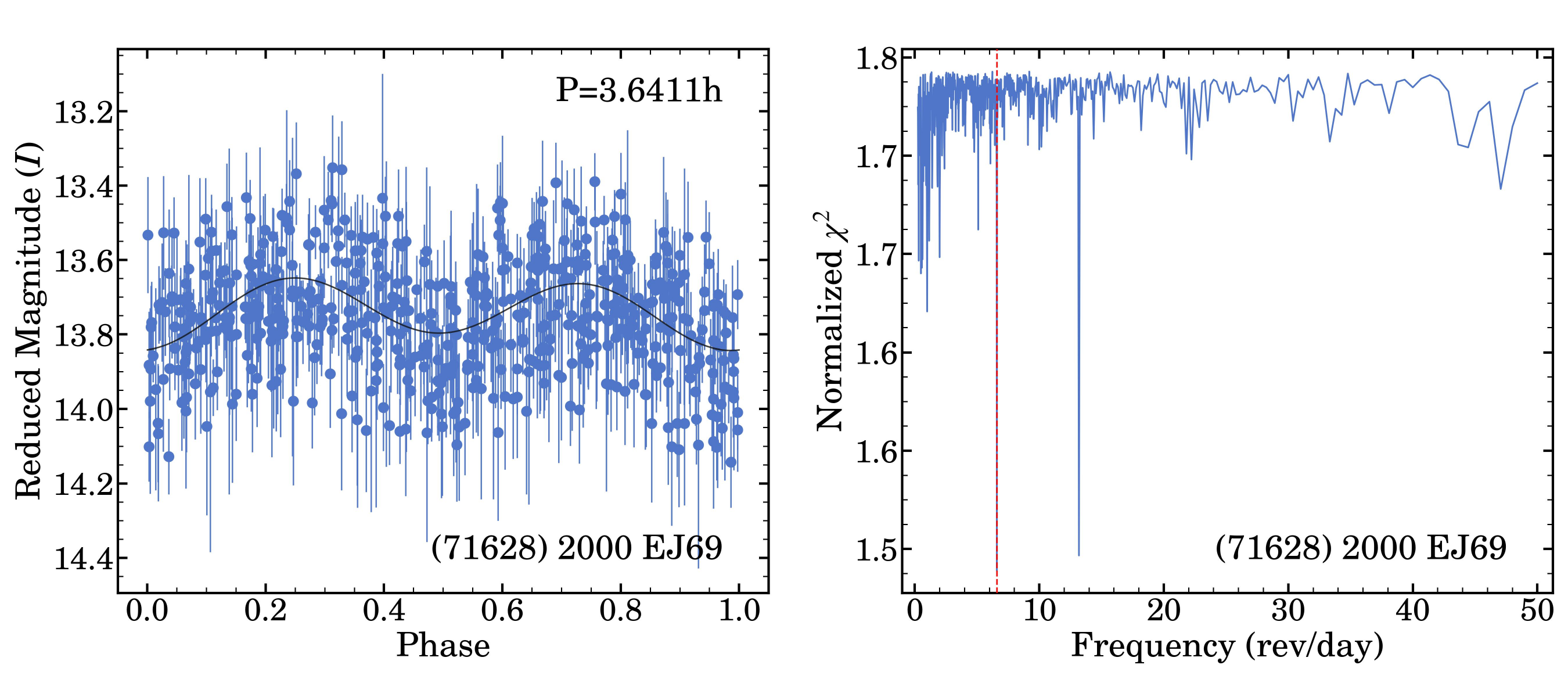}
    \includegraphics[width=0.45\linewidth]{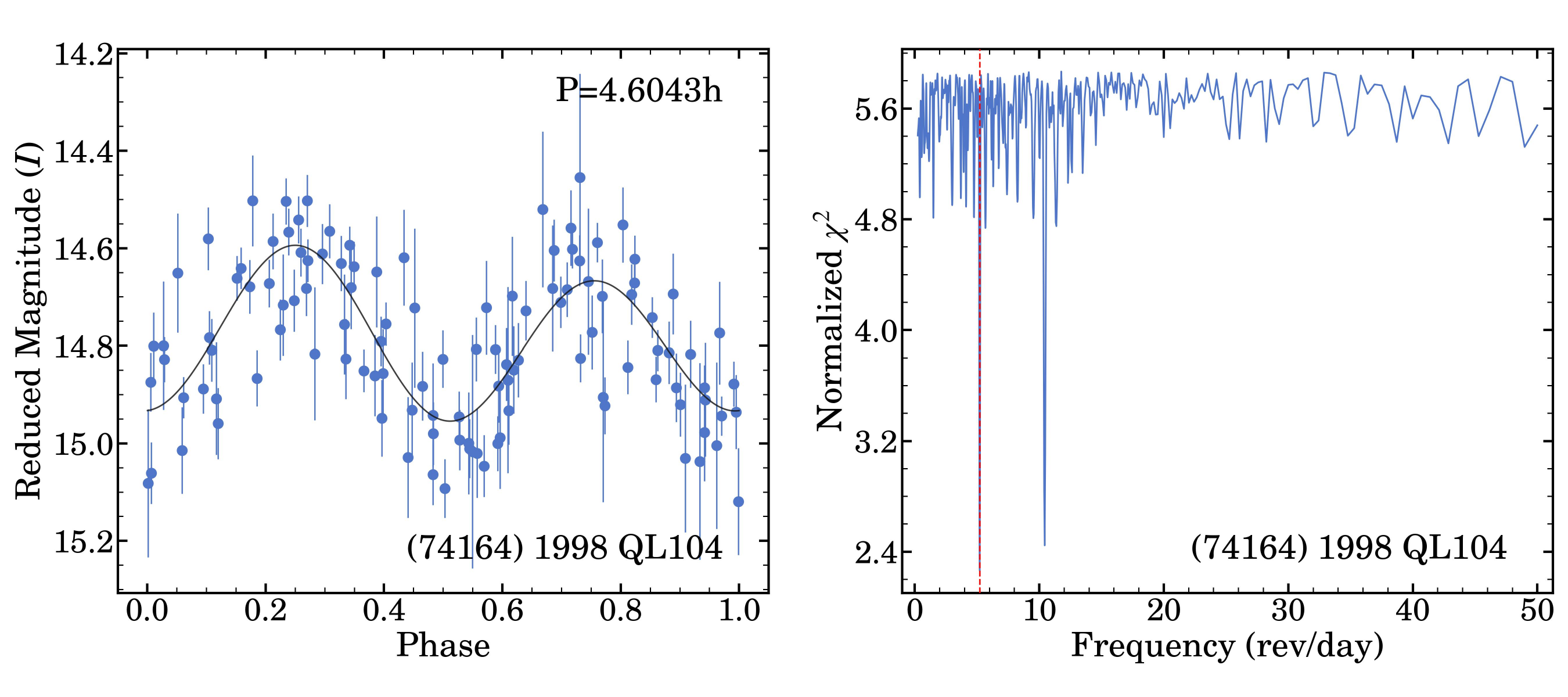}
    \includegraphics[width=0.45\linewidth]{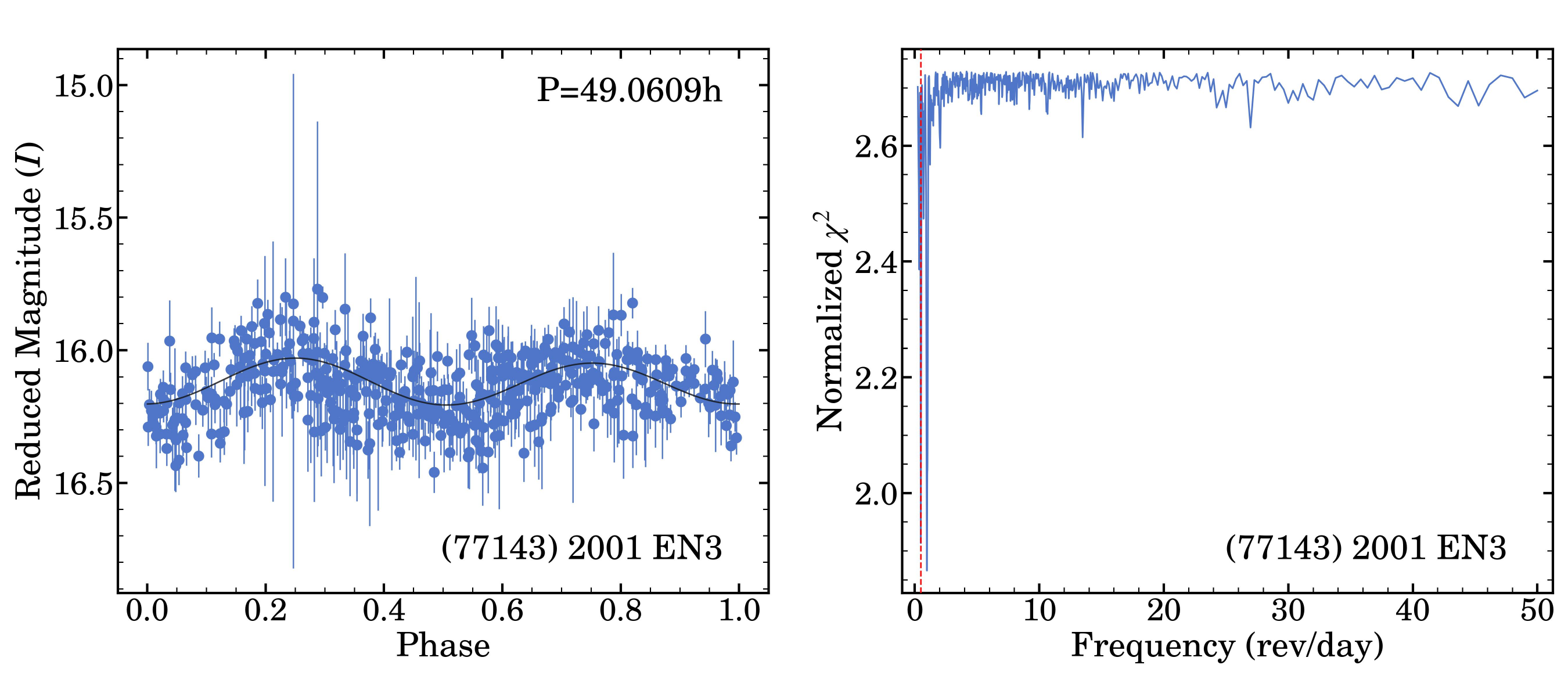}
    \includegraphics[width=0.45\linewidth]{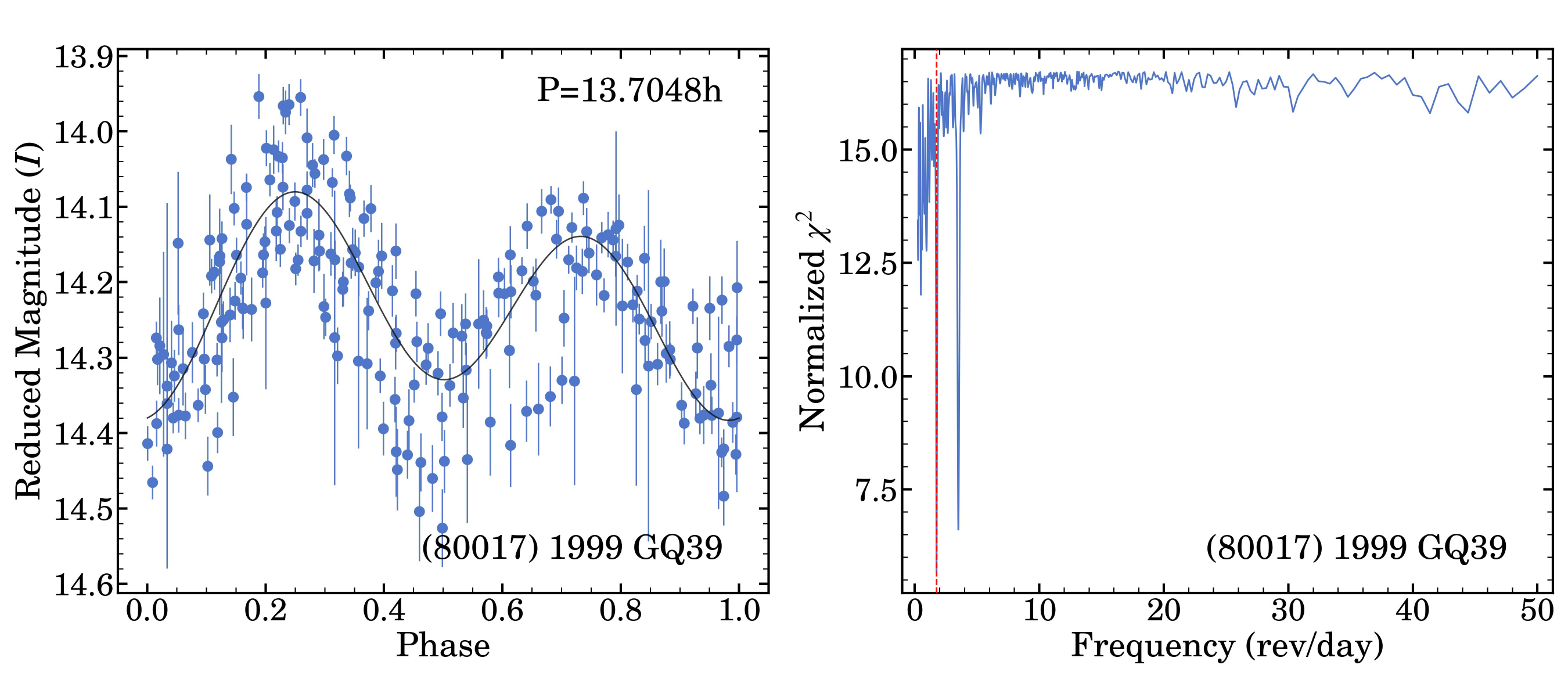}
    \includegraphics[width=0.45\linewidth]{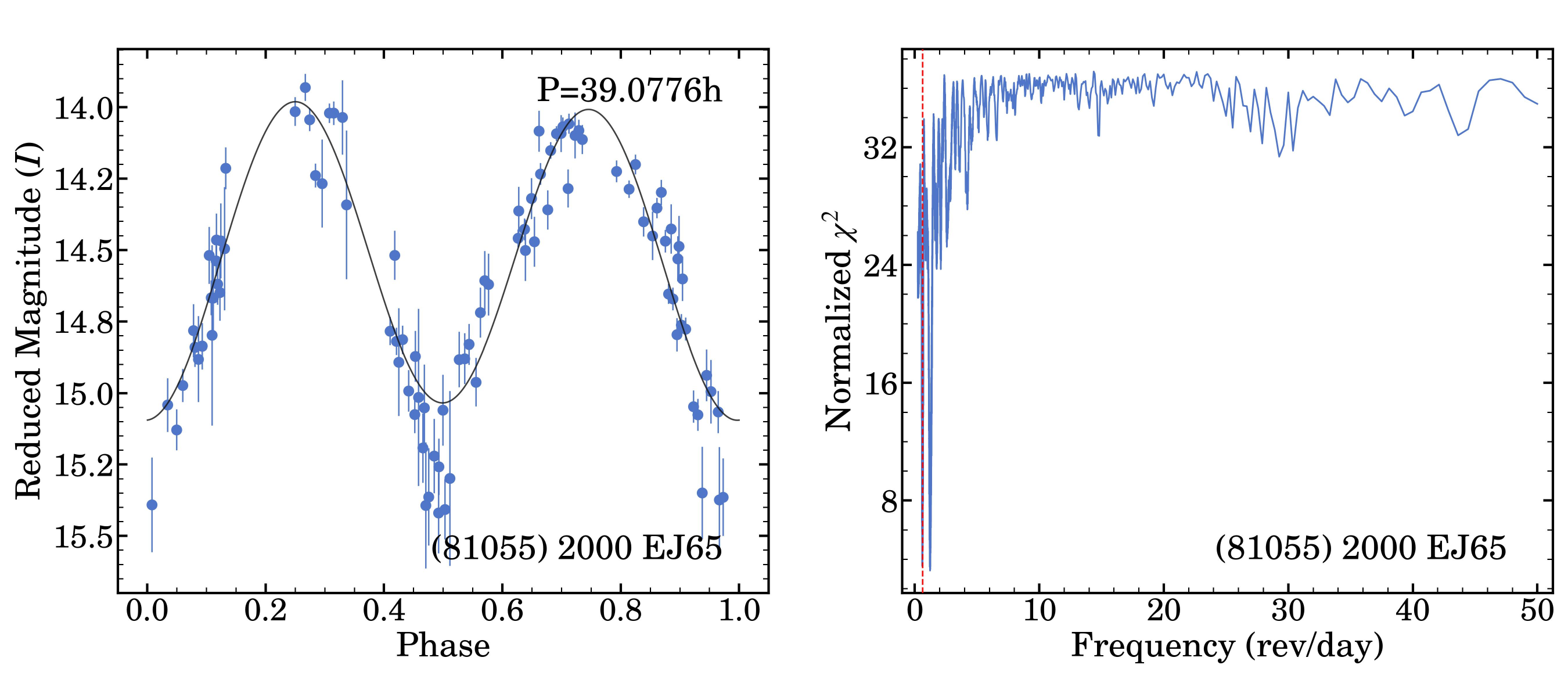}
    \includegraphics[width=0.45\linewidth]{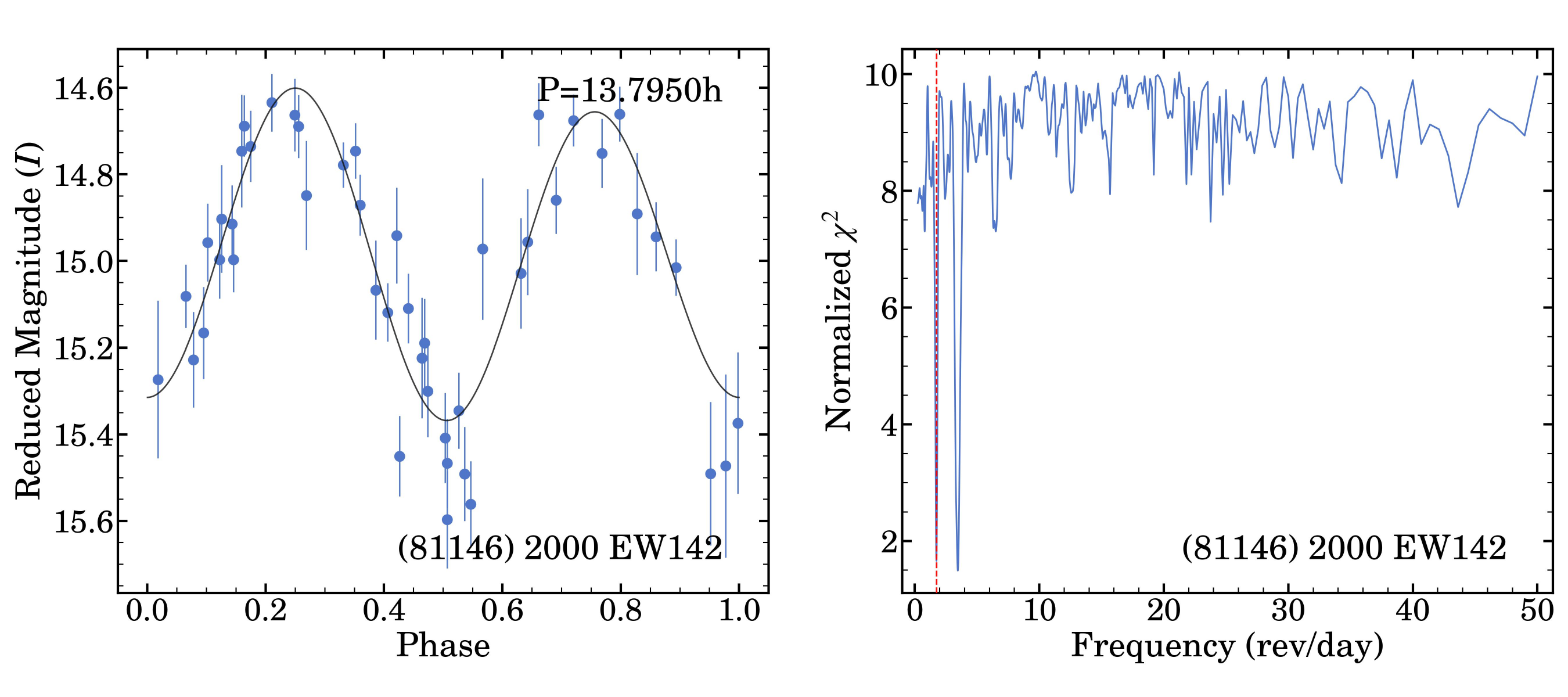}
    \includegraphics[width=0.45\linewidth]{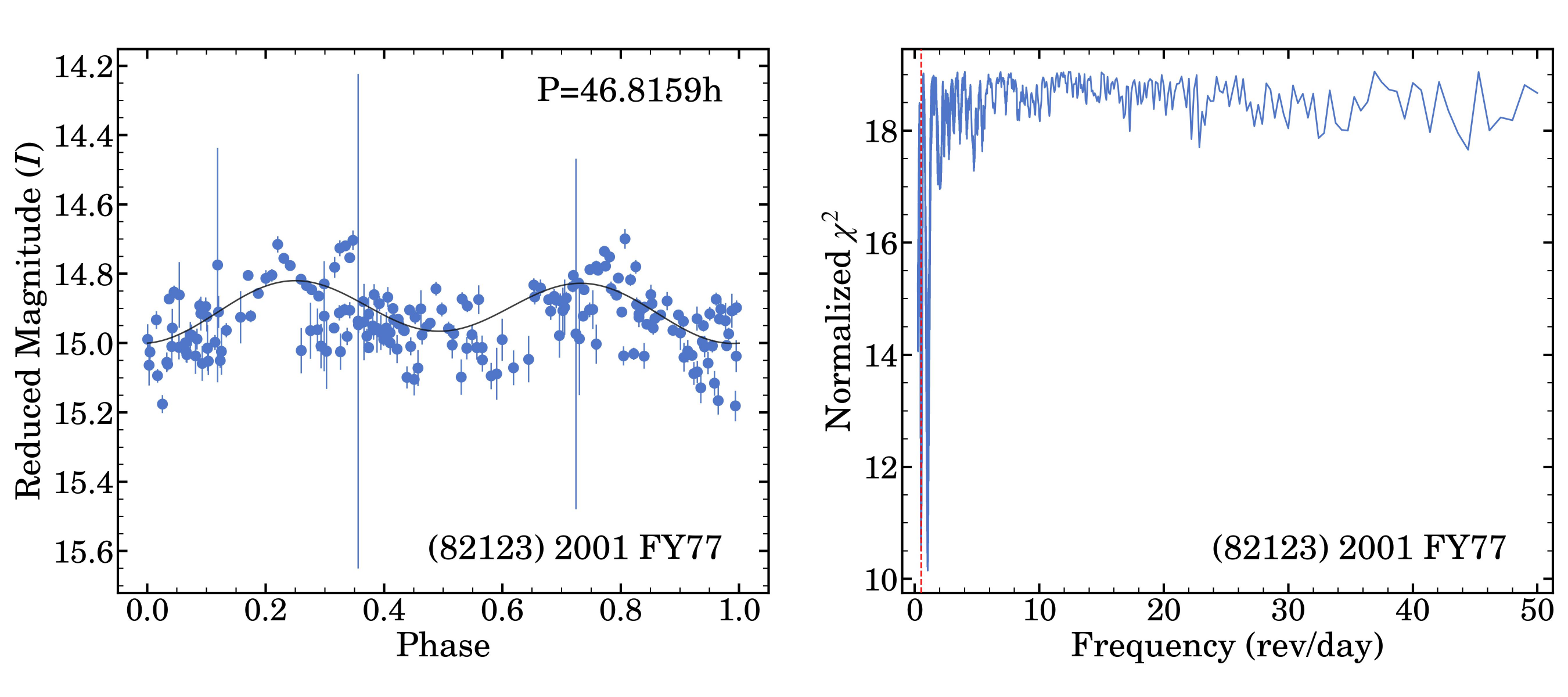}
    \includegraphics[width=0.45\linewidth]{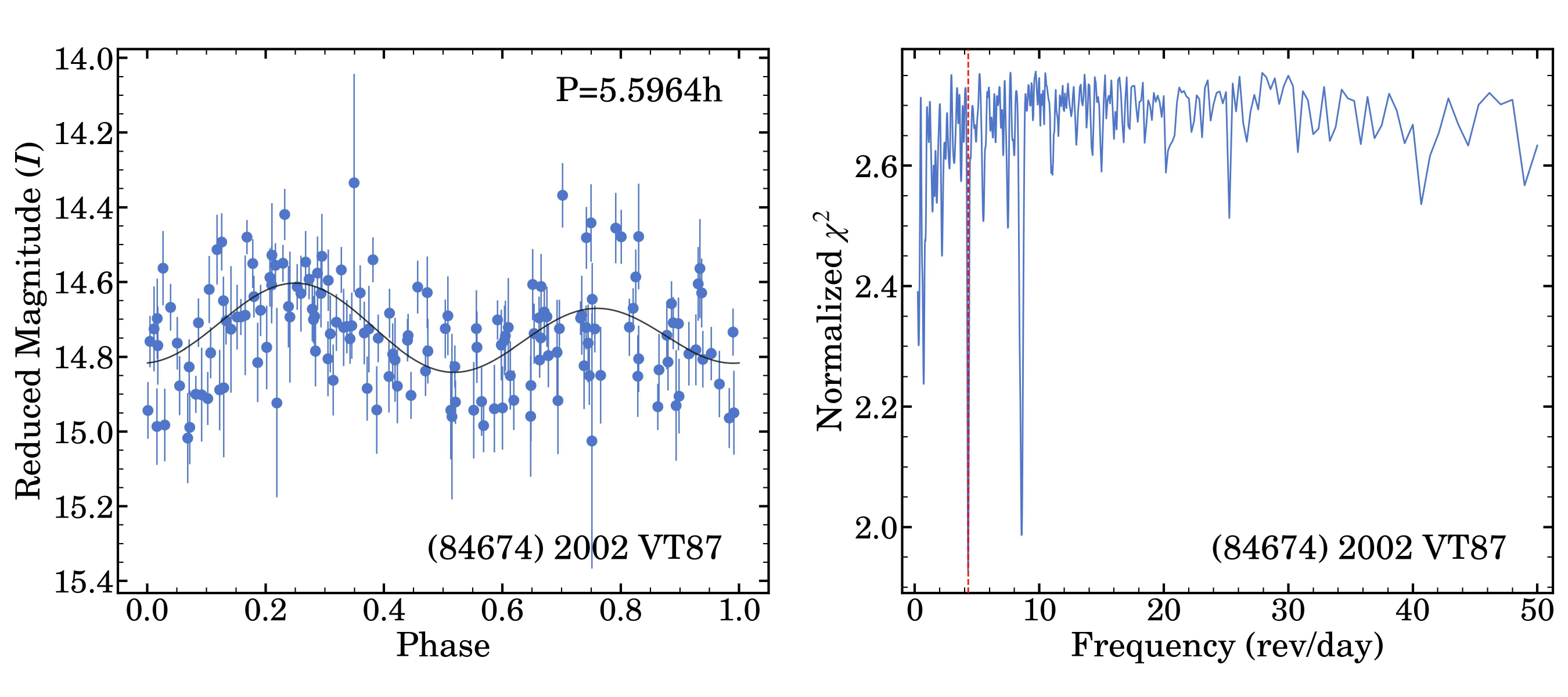}
    \includegraphics[width=0.45\linewidth]{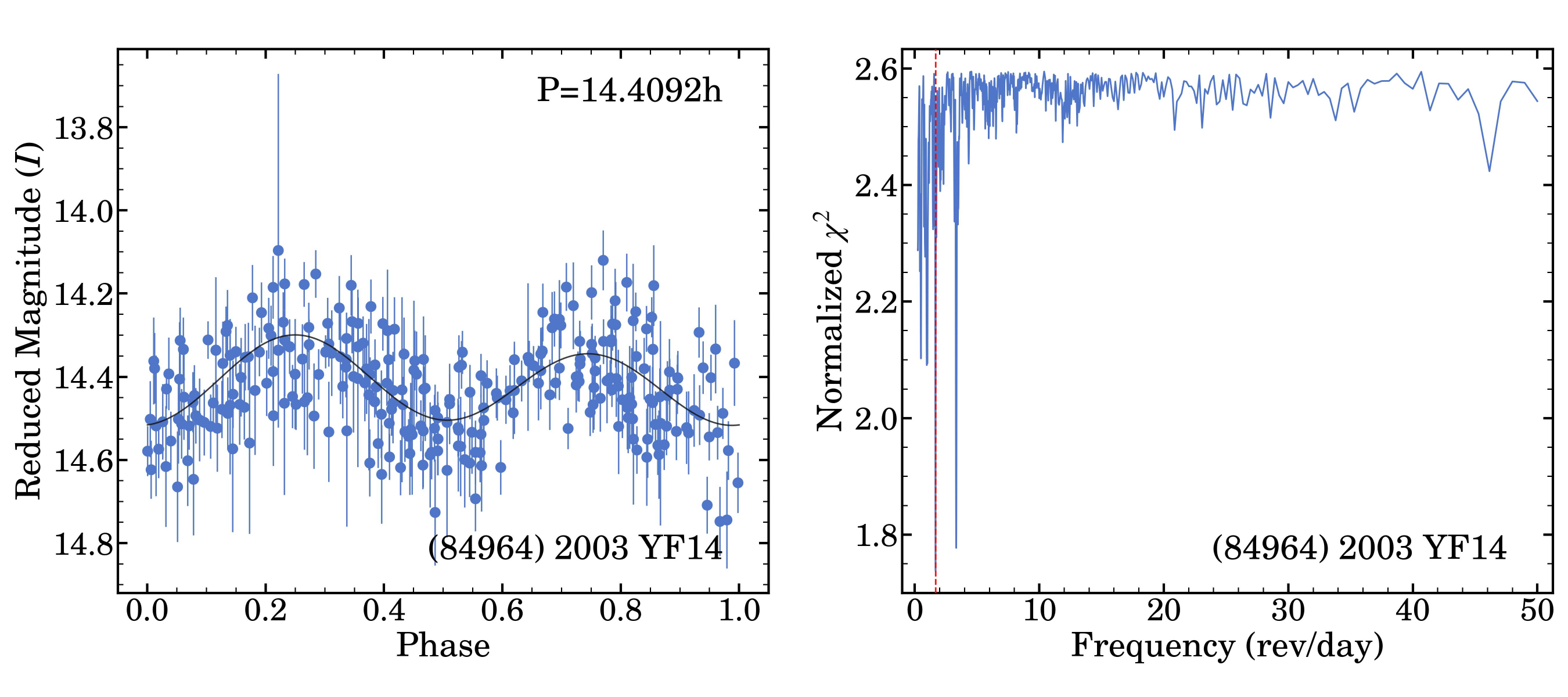}
    \includegraphics[width=0.45\linewidth]{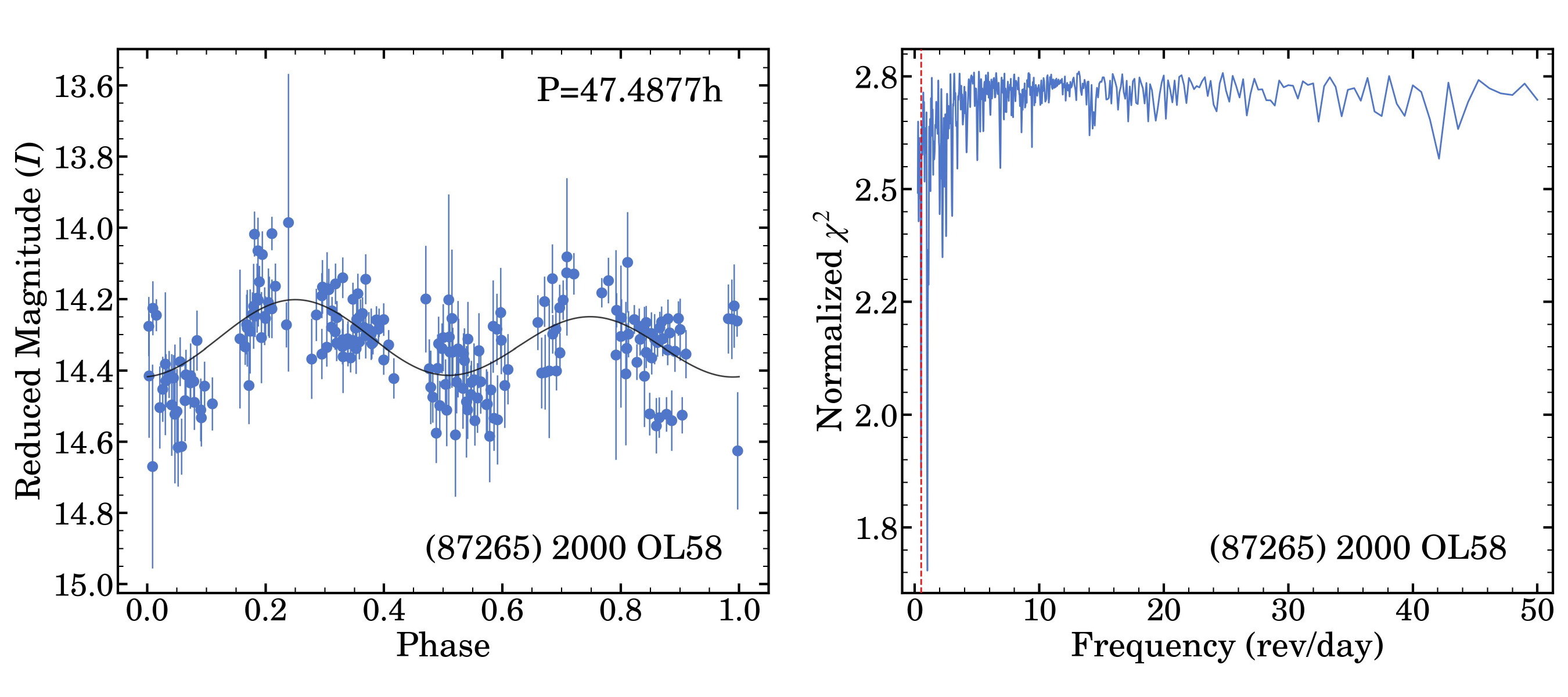}
    \includegraphics[width=0.45\linewidth]{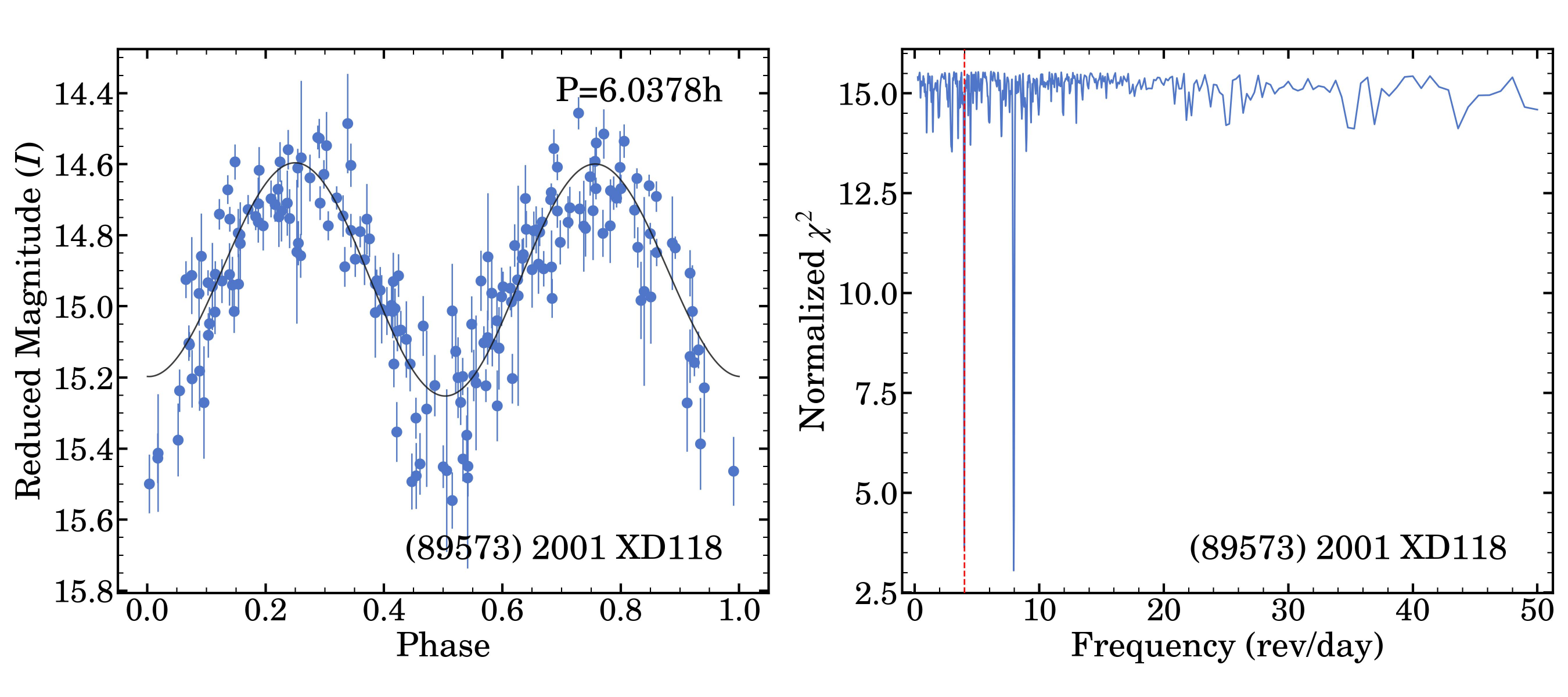}
    \caption{Continued (U = 2+).}
\end{figure*}

\begin{figure*}
    \addtocounter{figure}{-1}
    \centering
    \includegraphics[width=0.45\linewidth]{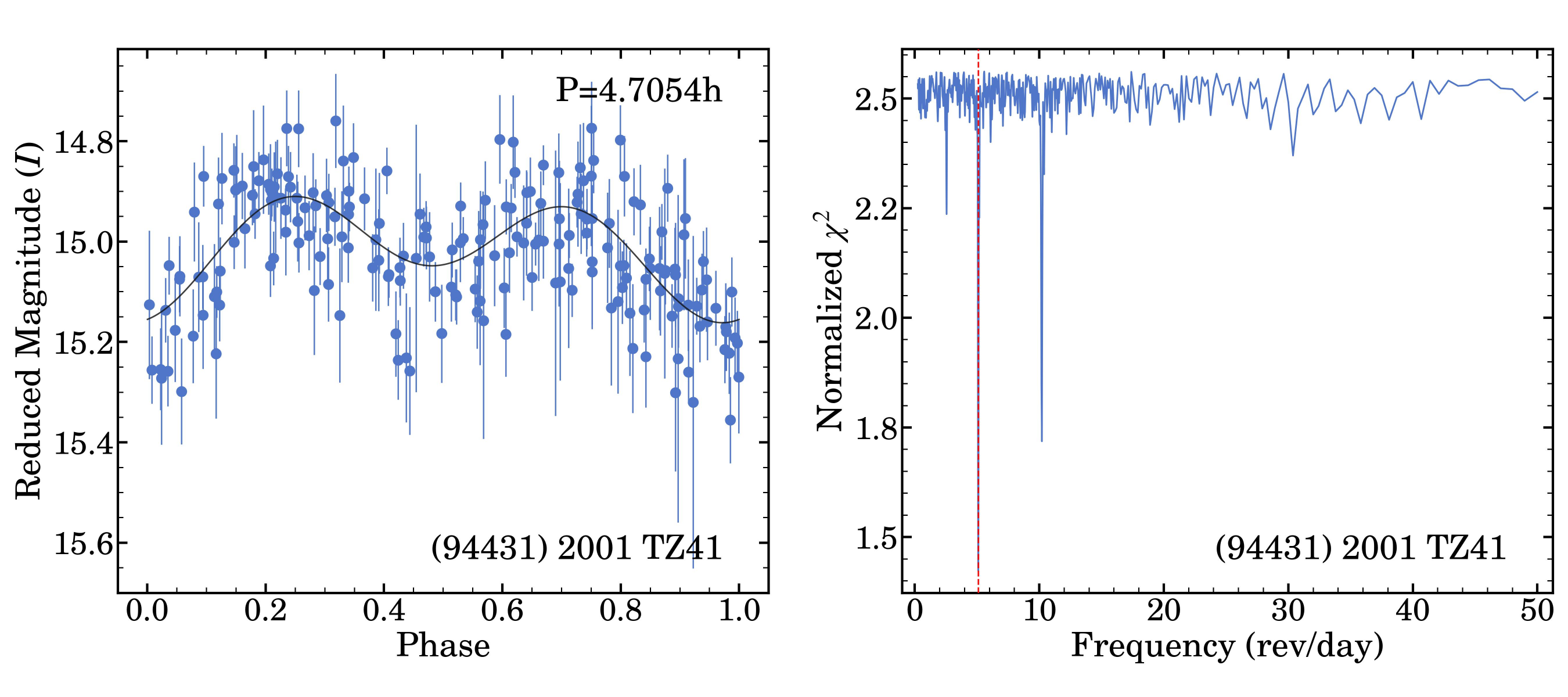}
    \includegraphics[width=0.45\linewidth]{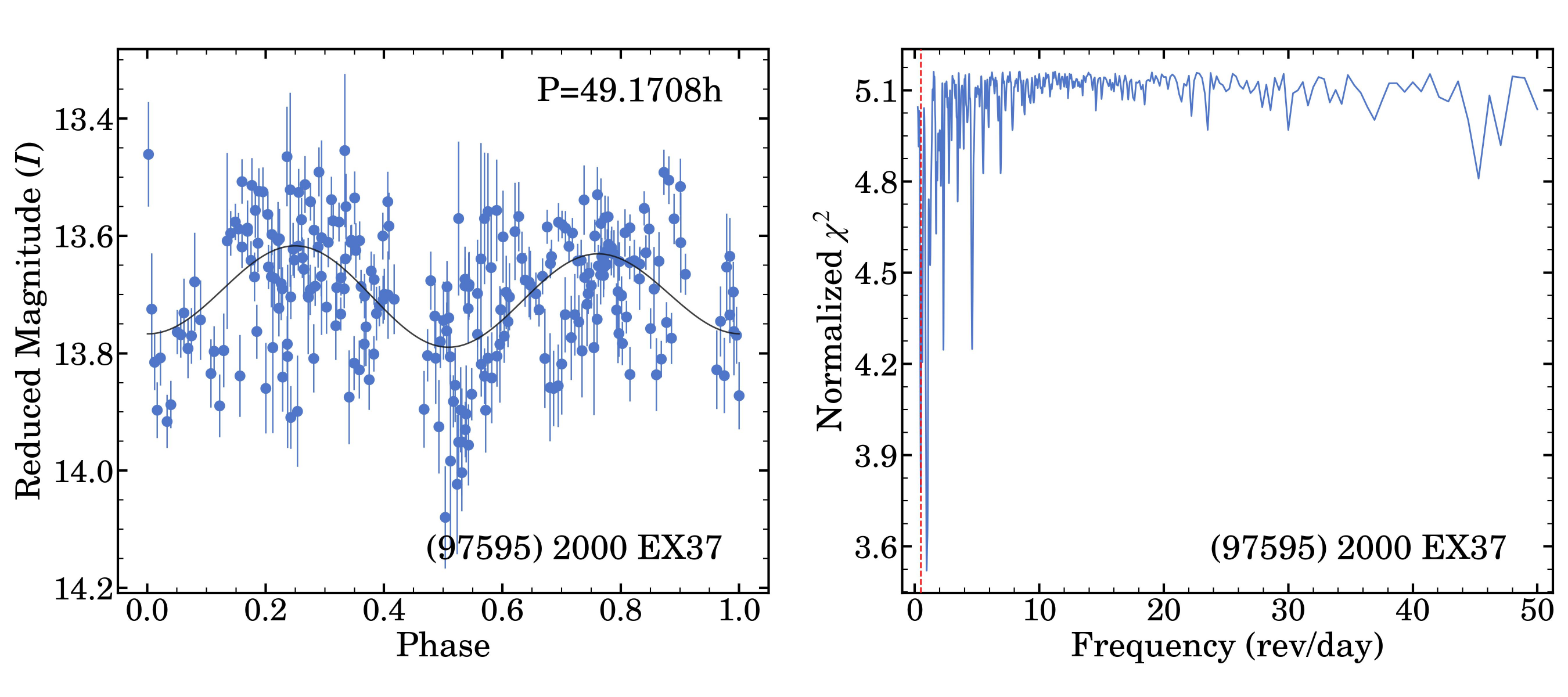}
    \includegraphics[width=0.45\linewidth]{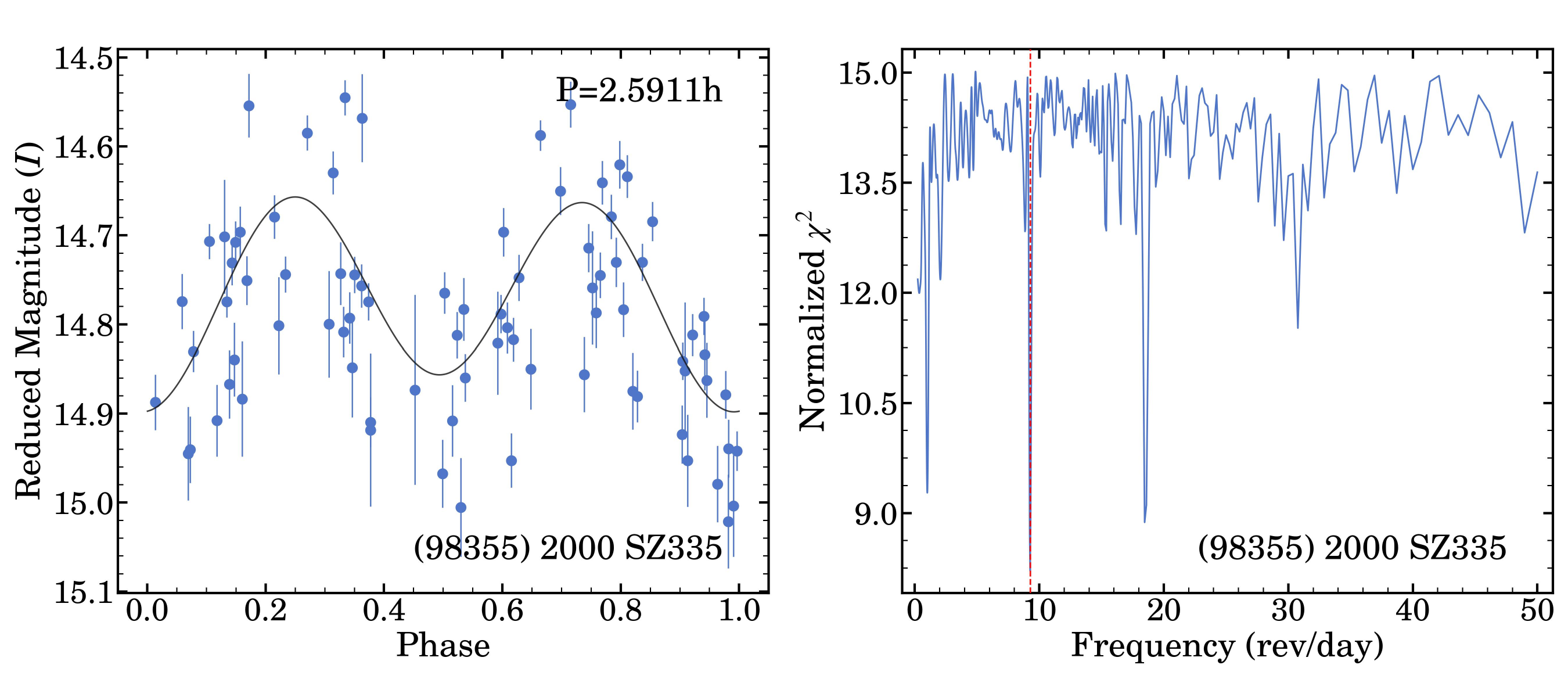}
    \includegraphics[width=0.45\linewidth]{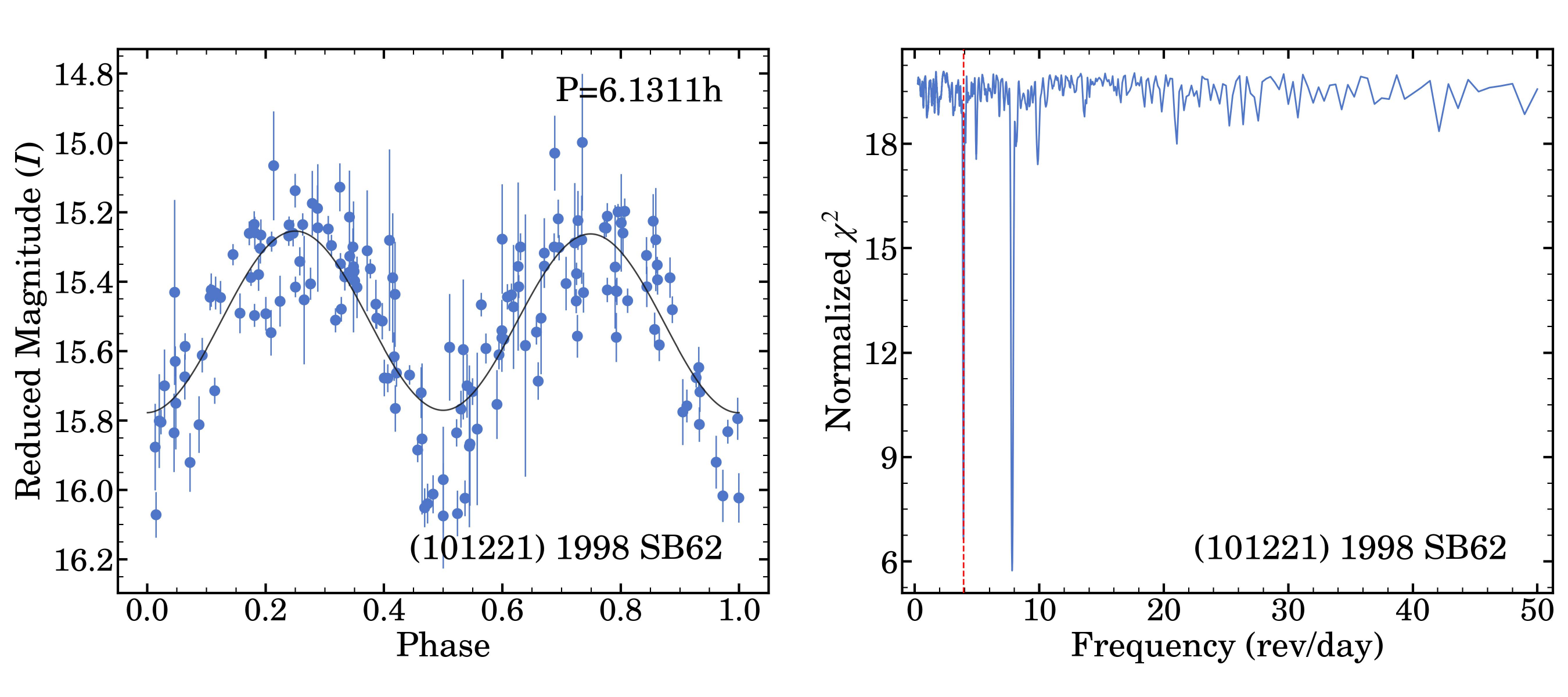}
    \includegraphics[width=0.45\linewidth]{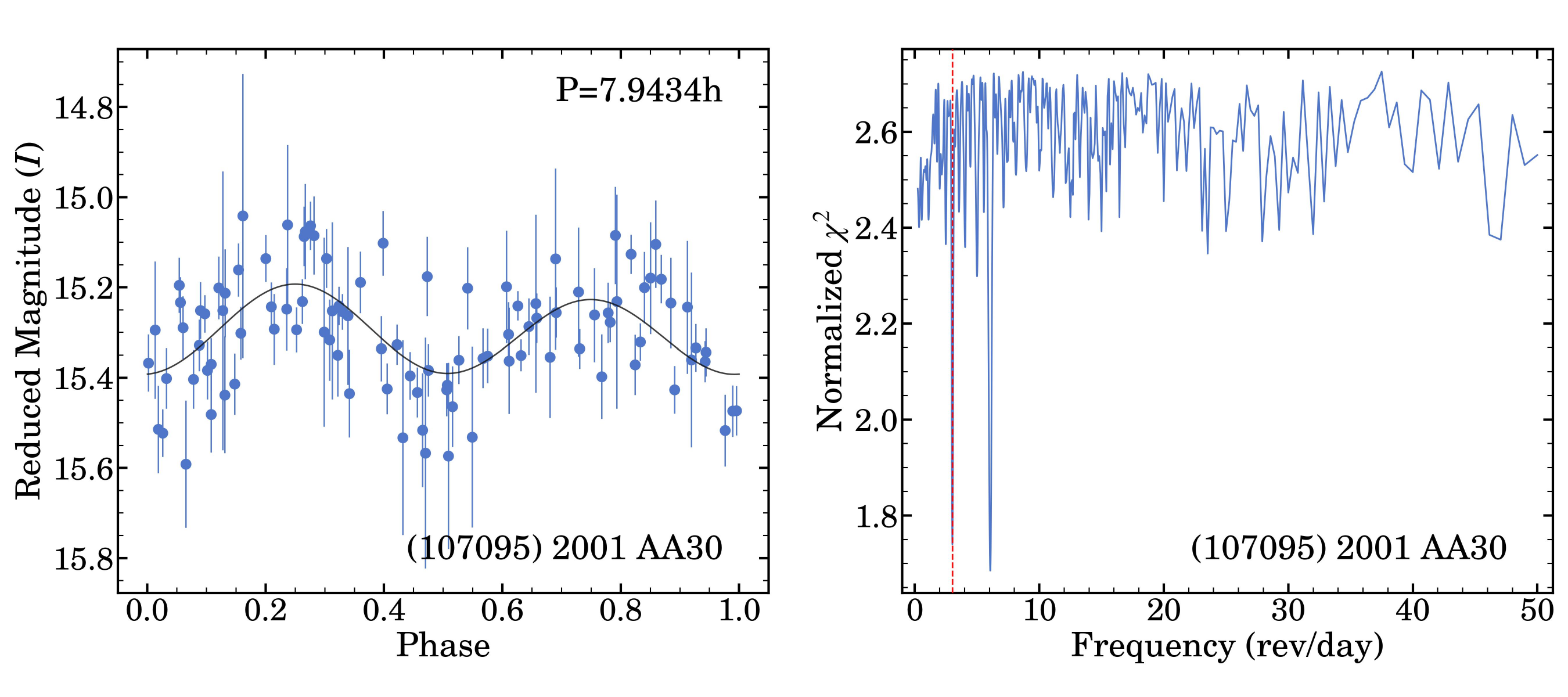}
    \includegraphics[width=0.45\linewidth]{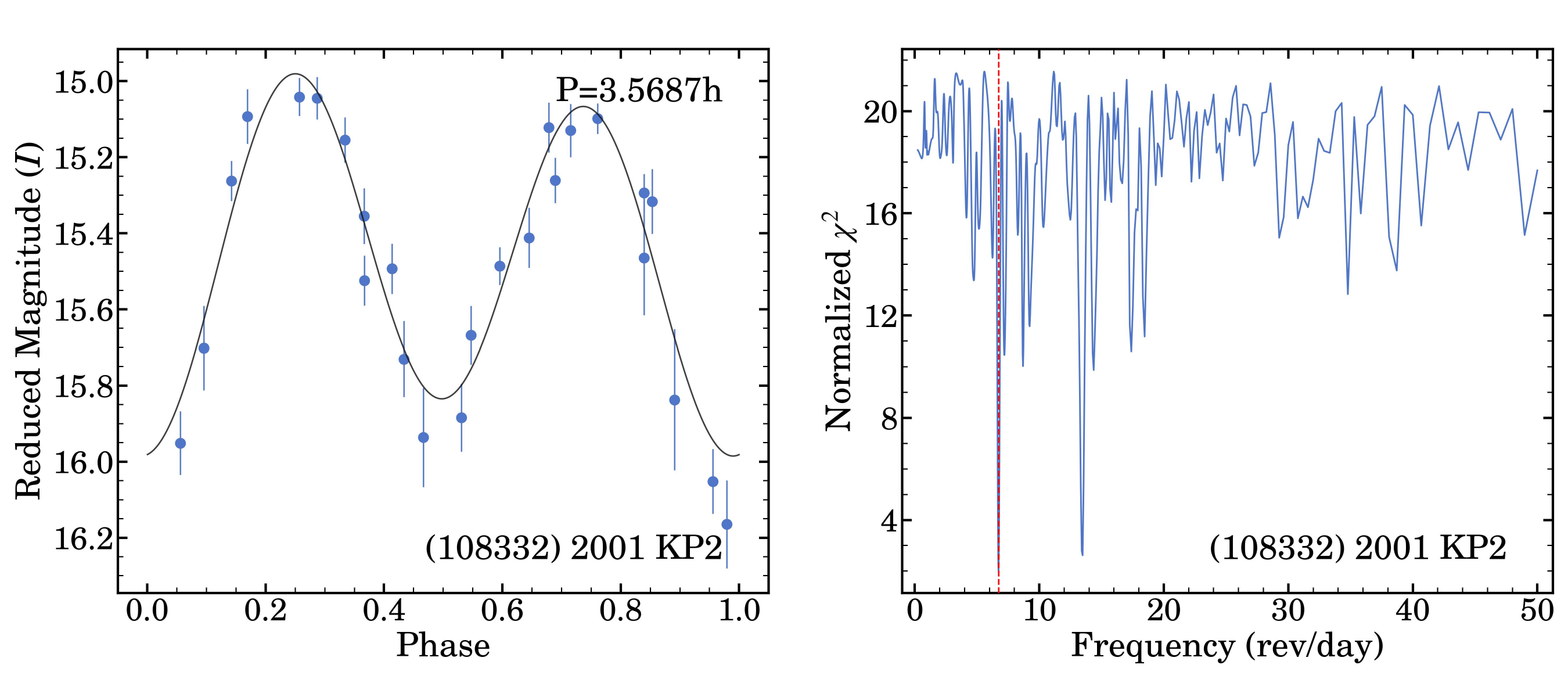}
    \includegraphics[width=0.45\linewidth]{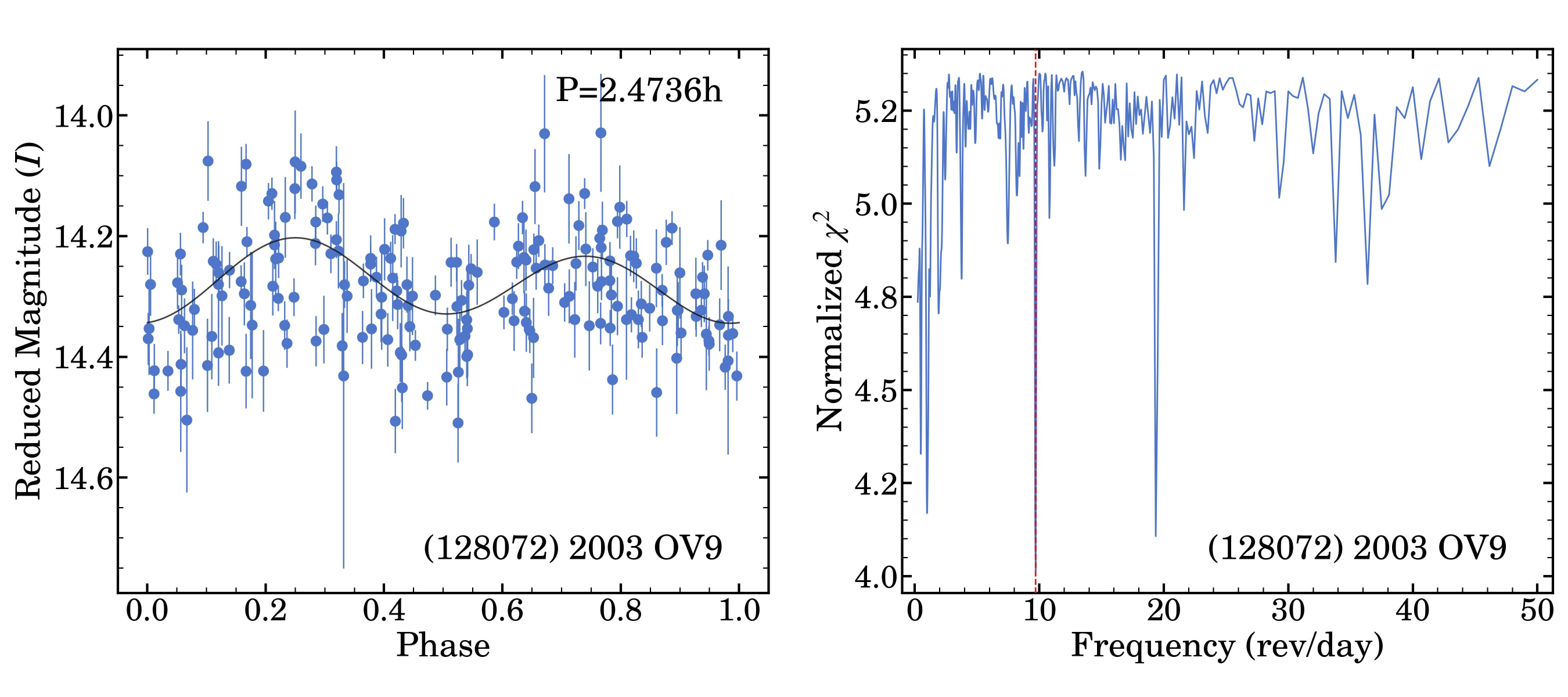}
    \includegraphics[width=0.45\linewidth]{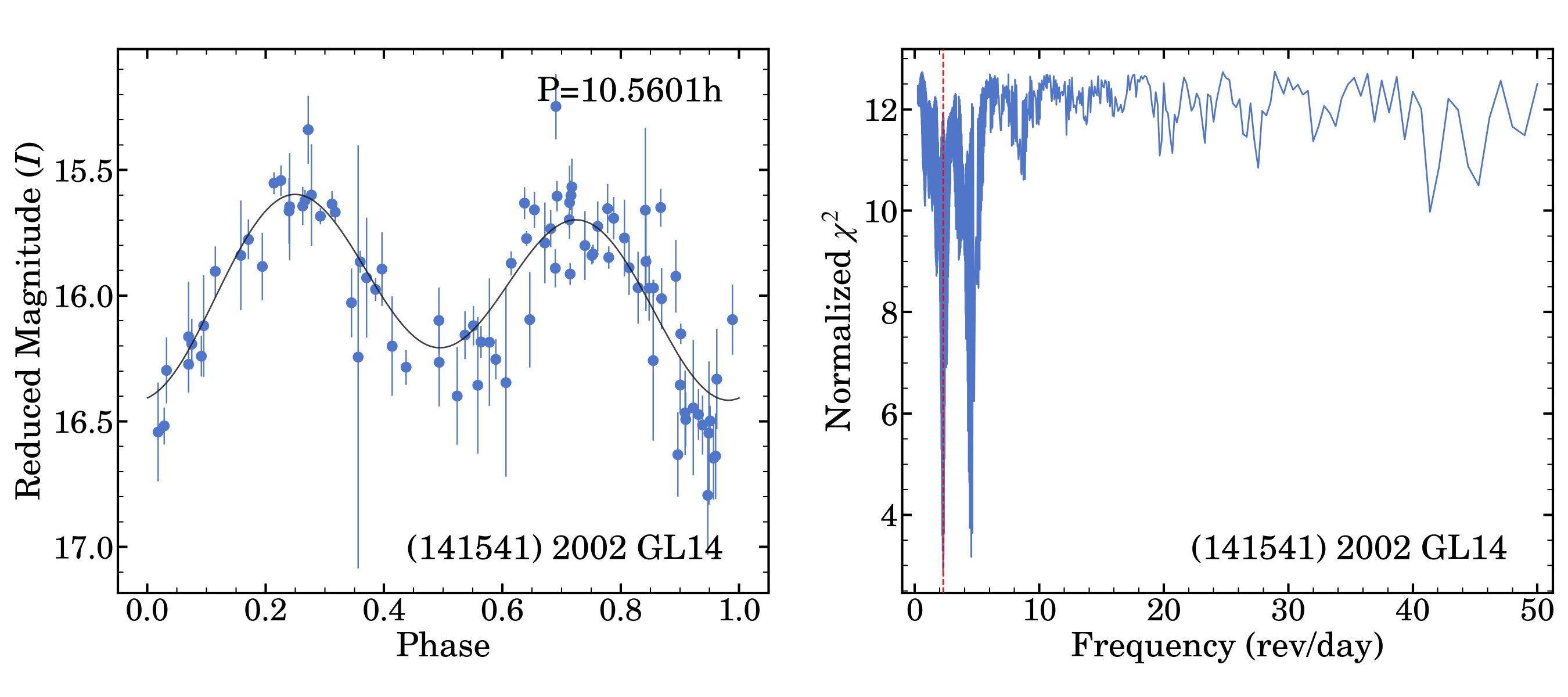}
    \includegraphics[width=0.45\linewidth]{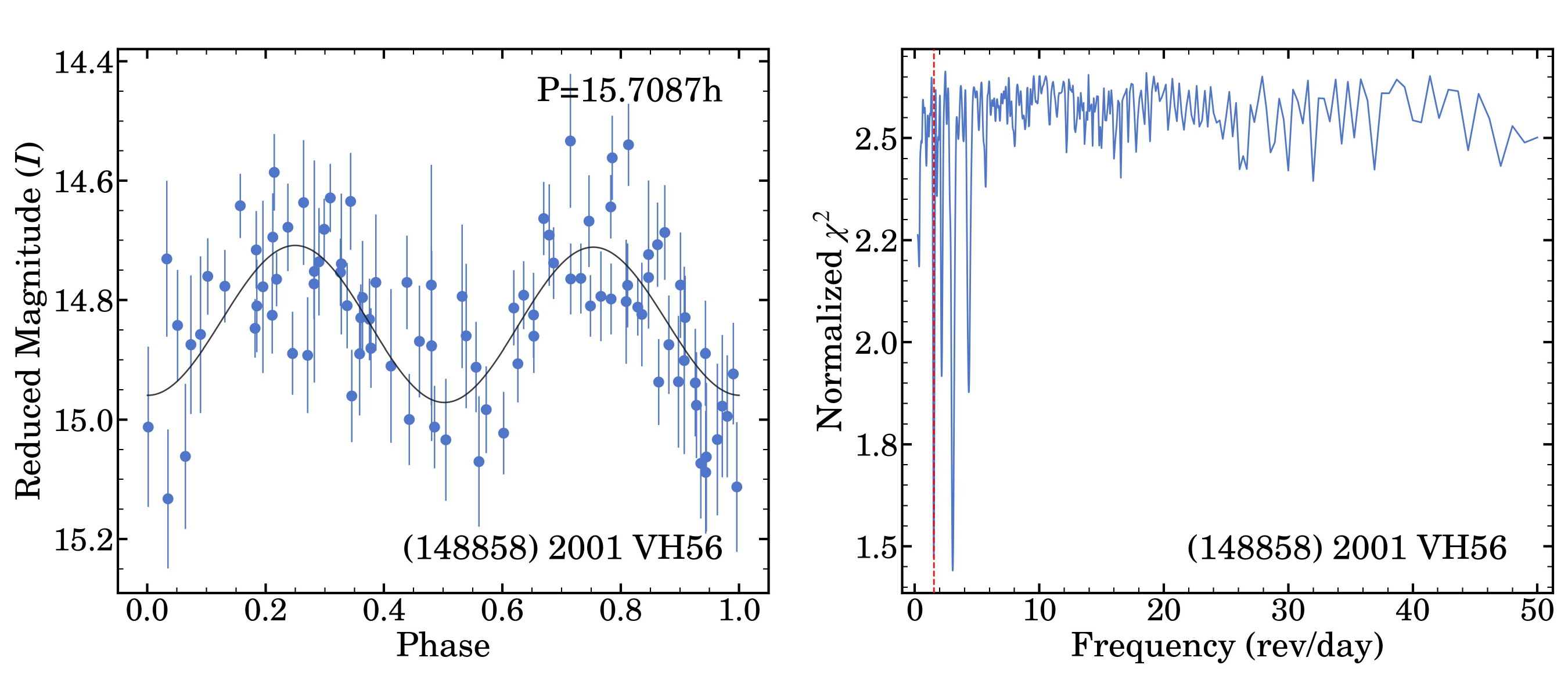}
    \includegraphics[width=0.45\linewidth]{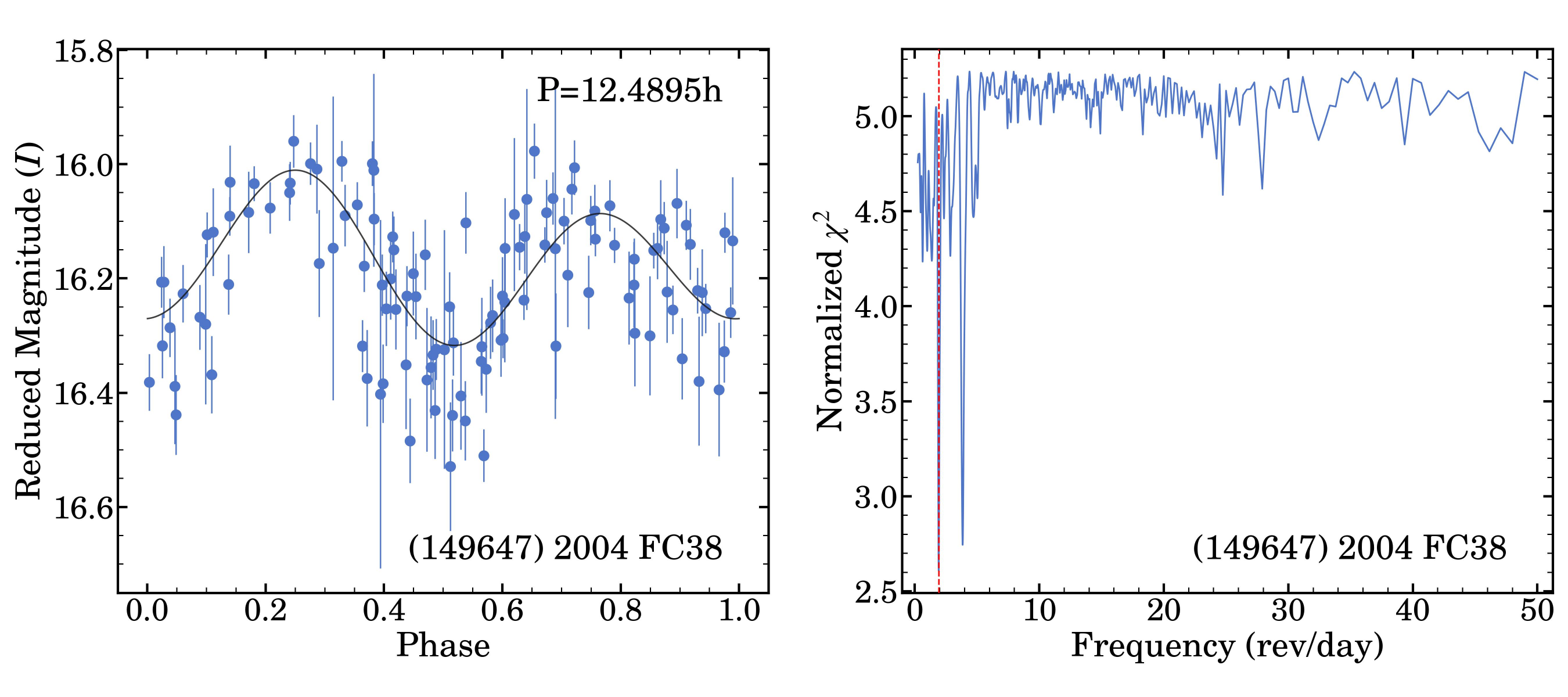}
    \includegraphics[width=0.45\linewidth]{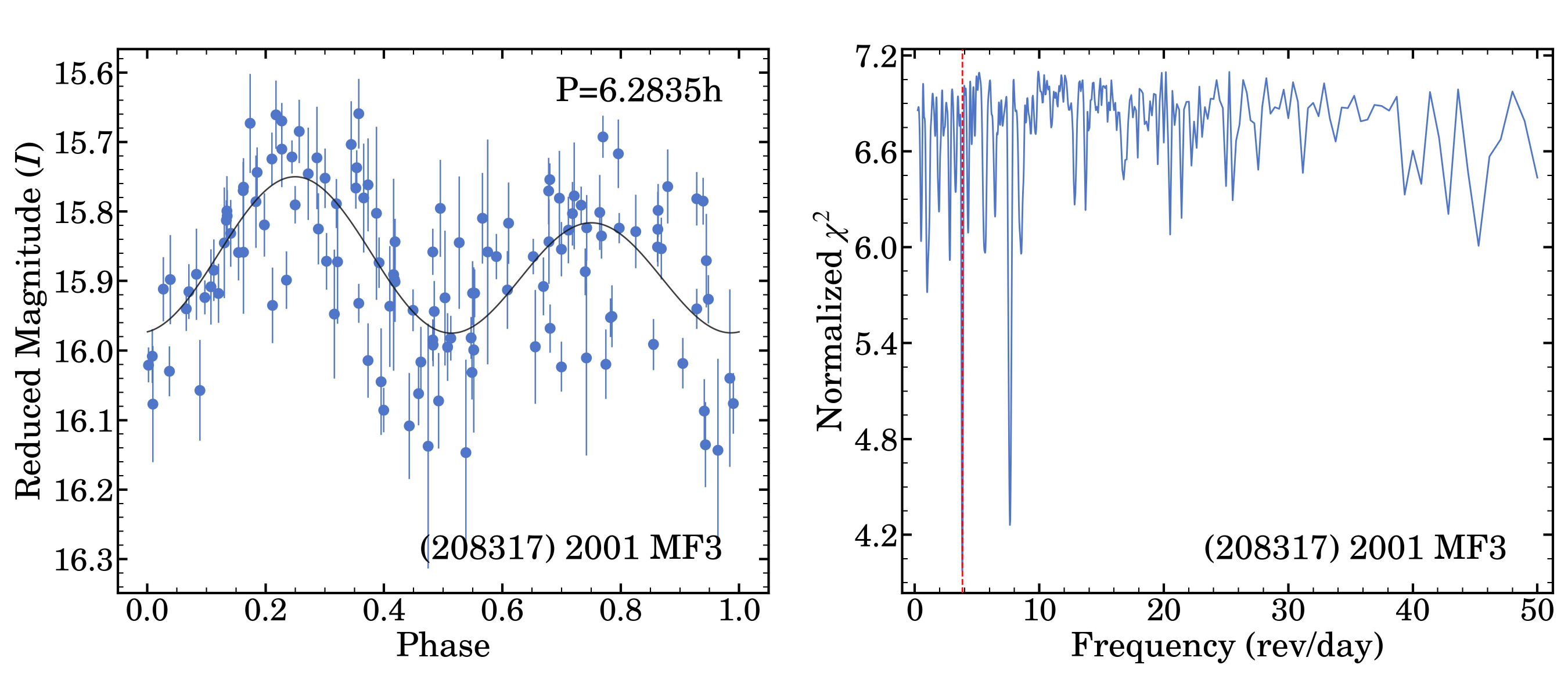}
    \includegraphics[width=0.45\linewidth]{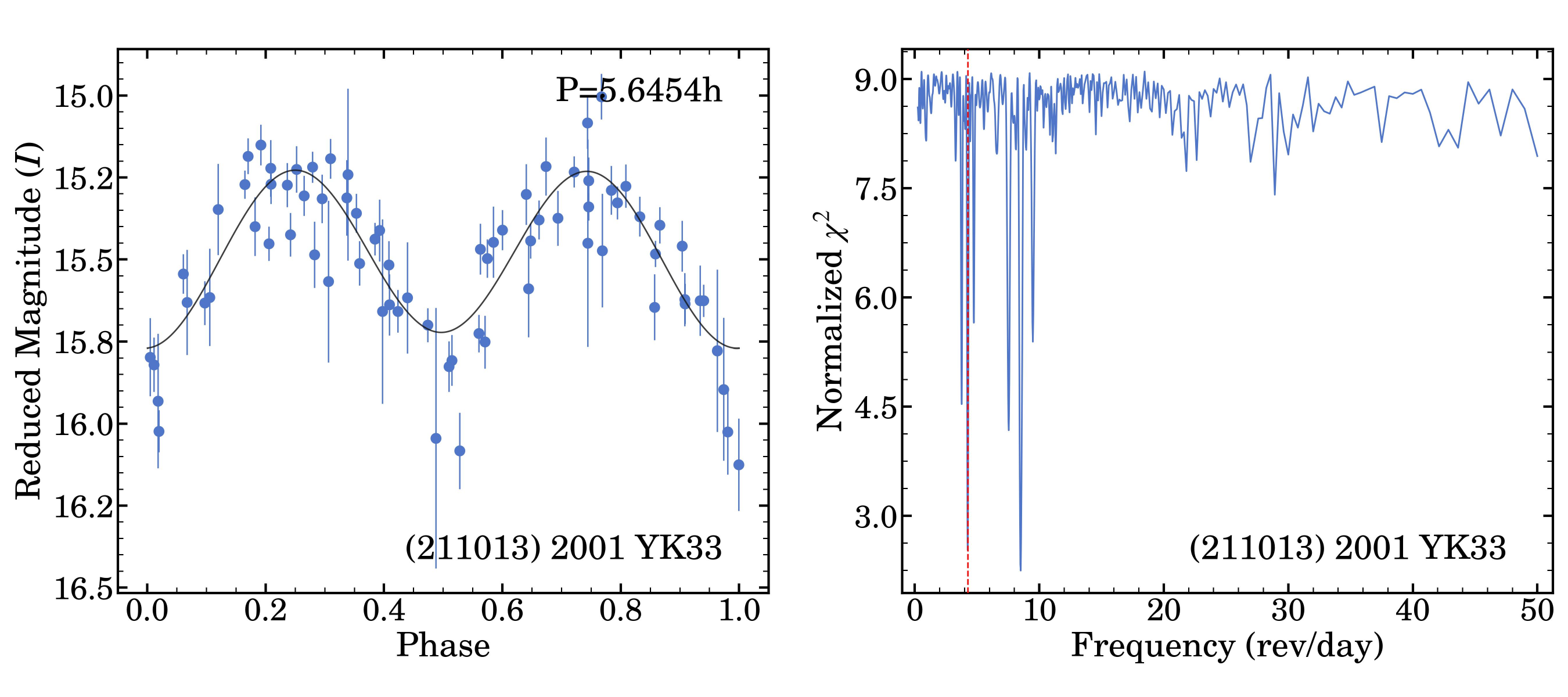}
    \caption{Continued (U = 2+).}
\end{figure*}

\begin{figure*}
    \addtocounter{figure}{-1}
    \centering
    \includegraphics[width=0.45\linewidth]{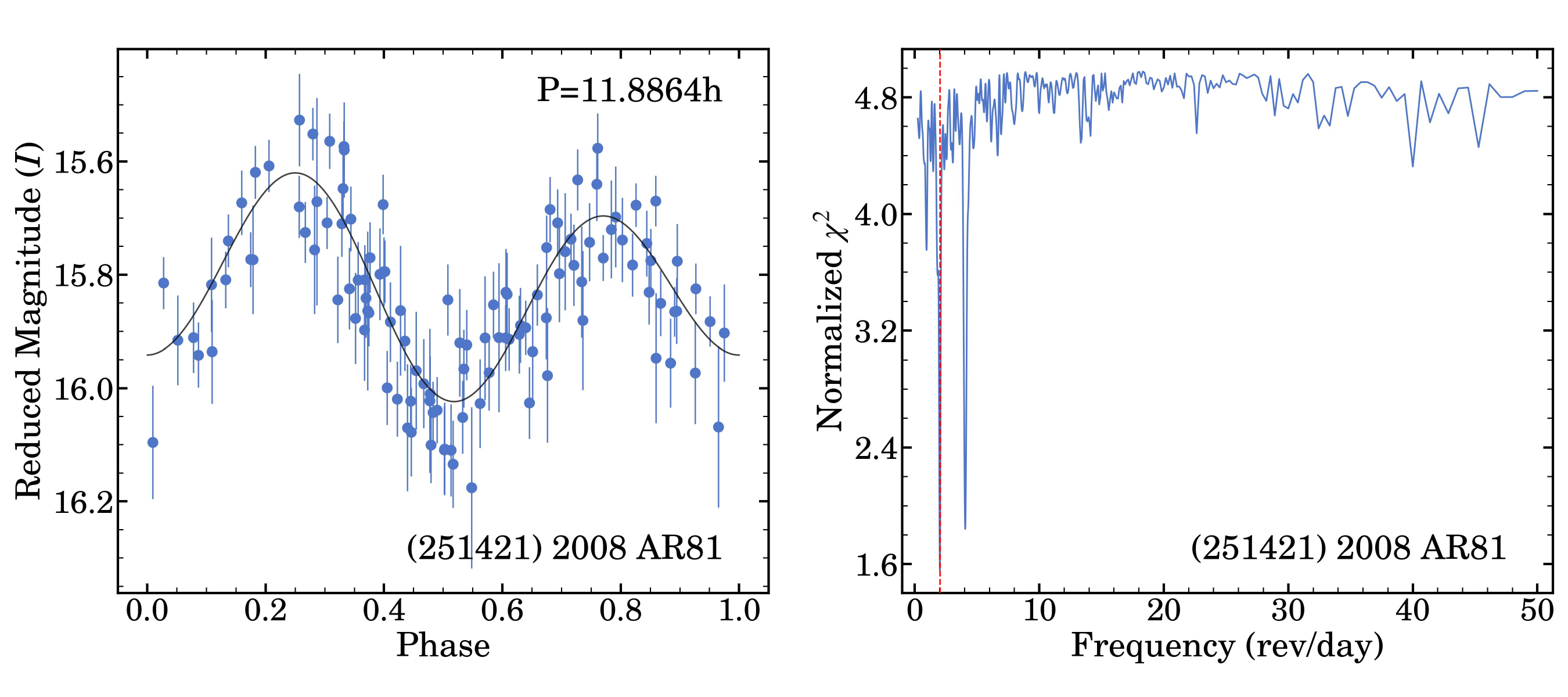}
    \includegraphics[width=0.45\linewidth]{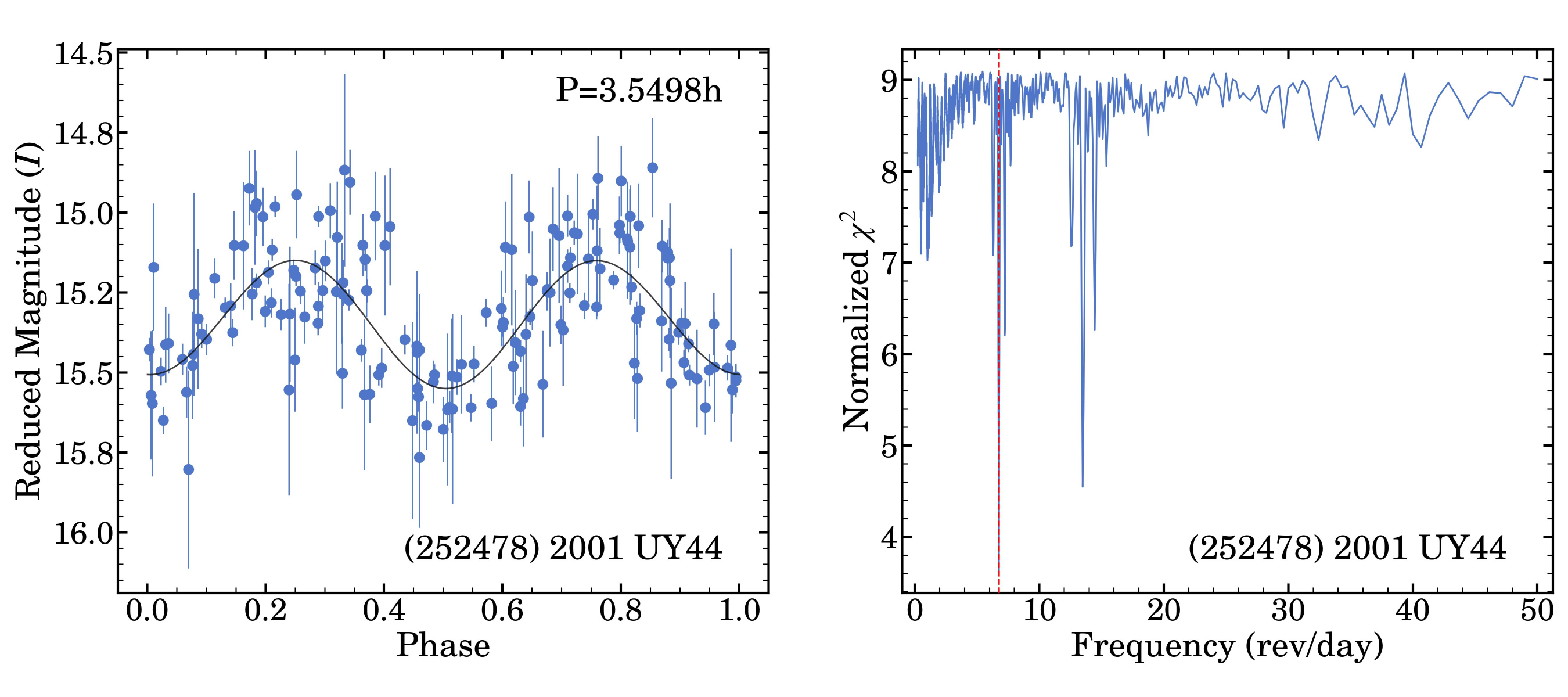}
    \includegraphics[width=0.45\linewidth]{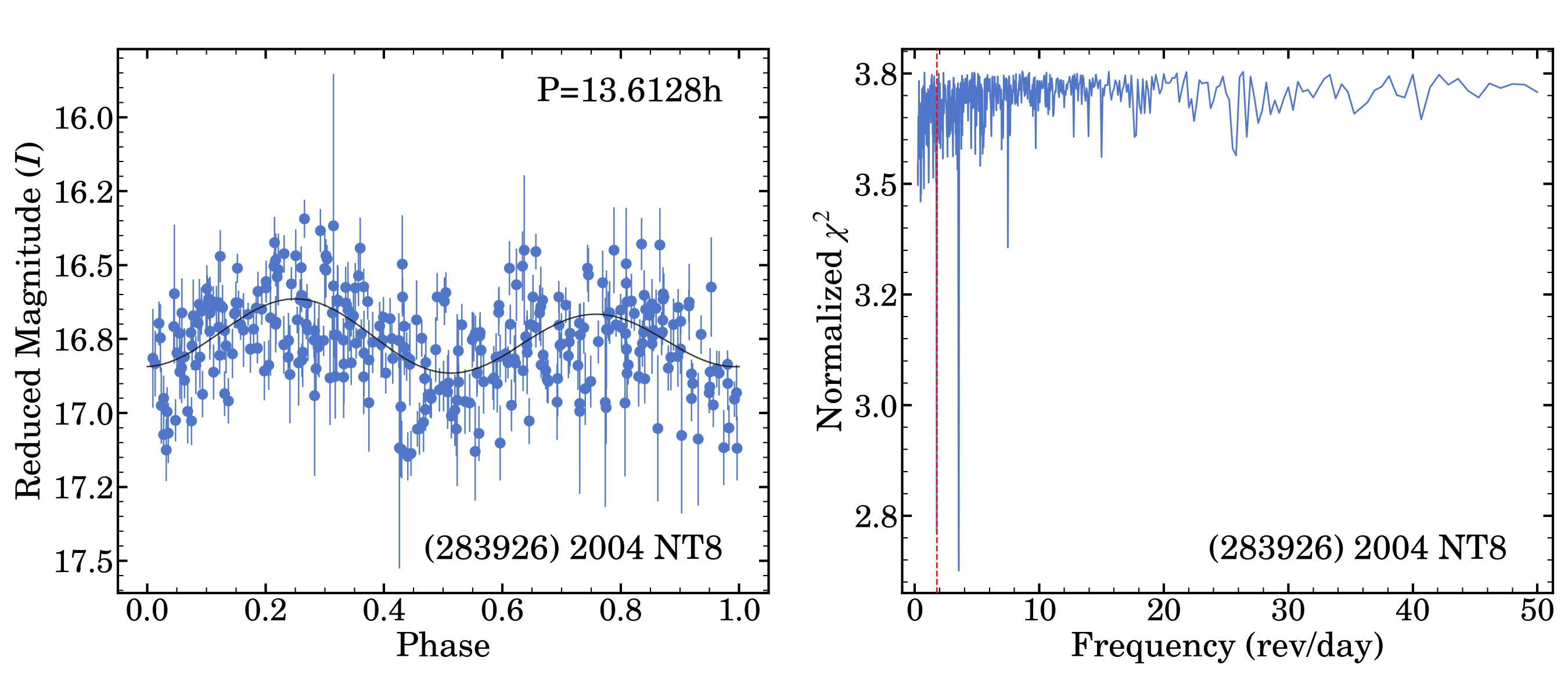}
    \includegraphics[width=0.45\linewidth]{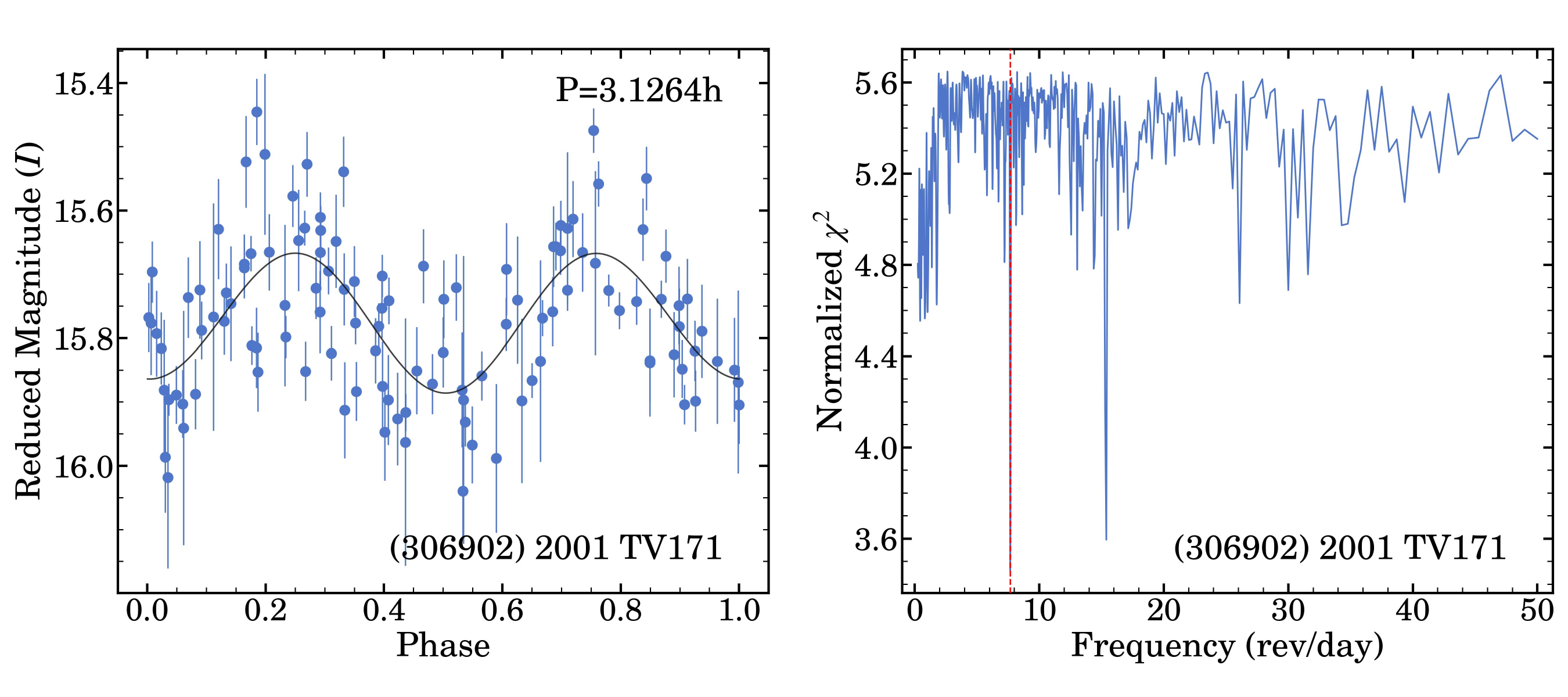}
    \includegraphics[width=0.45\linewidth]{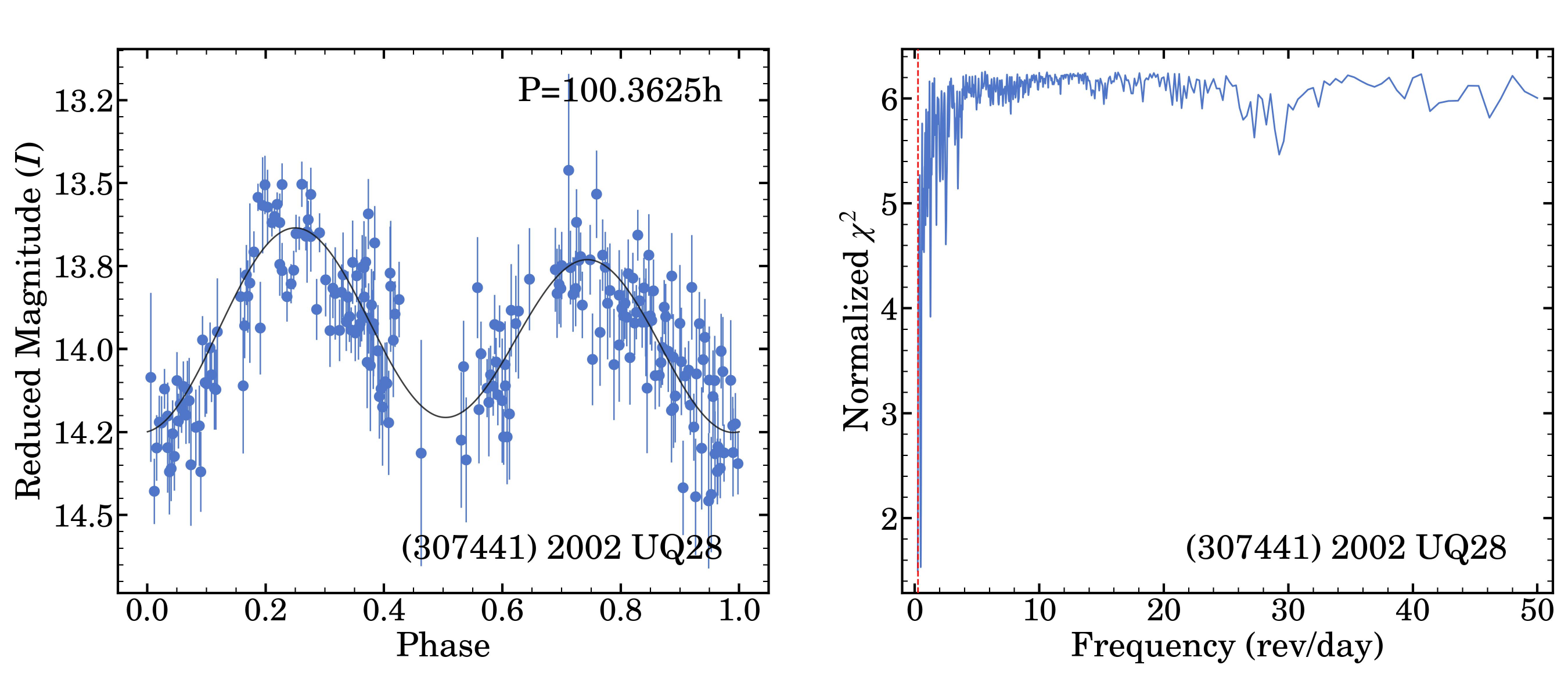}
    \includegraphics[width=0.45\linewidth]{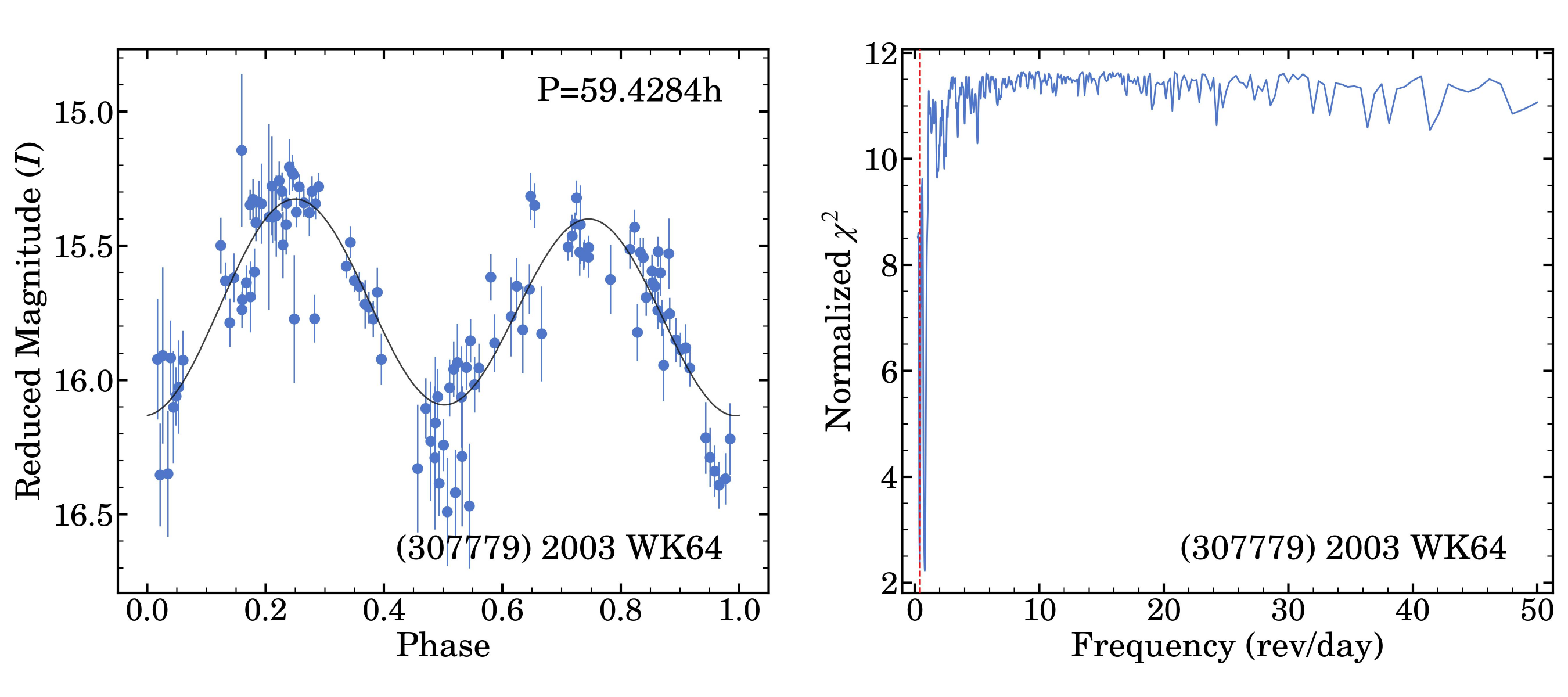}
    \includegraphics[width=0.45\linewidth]{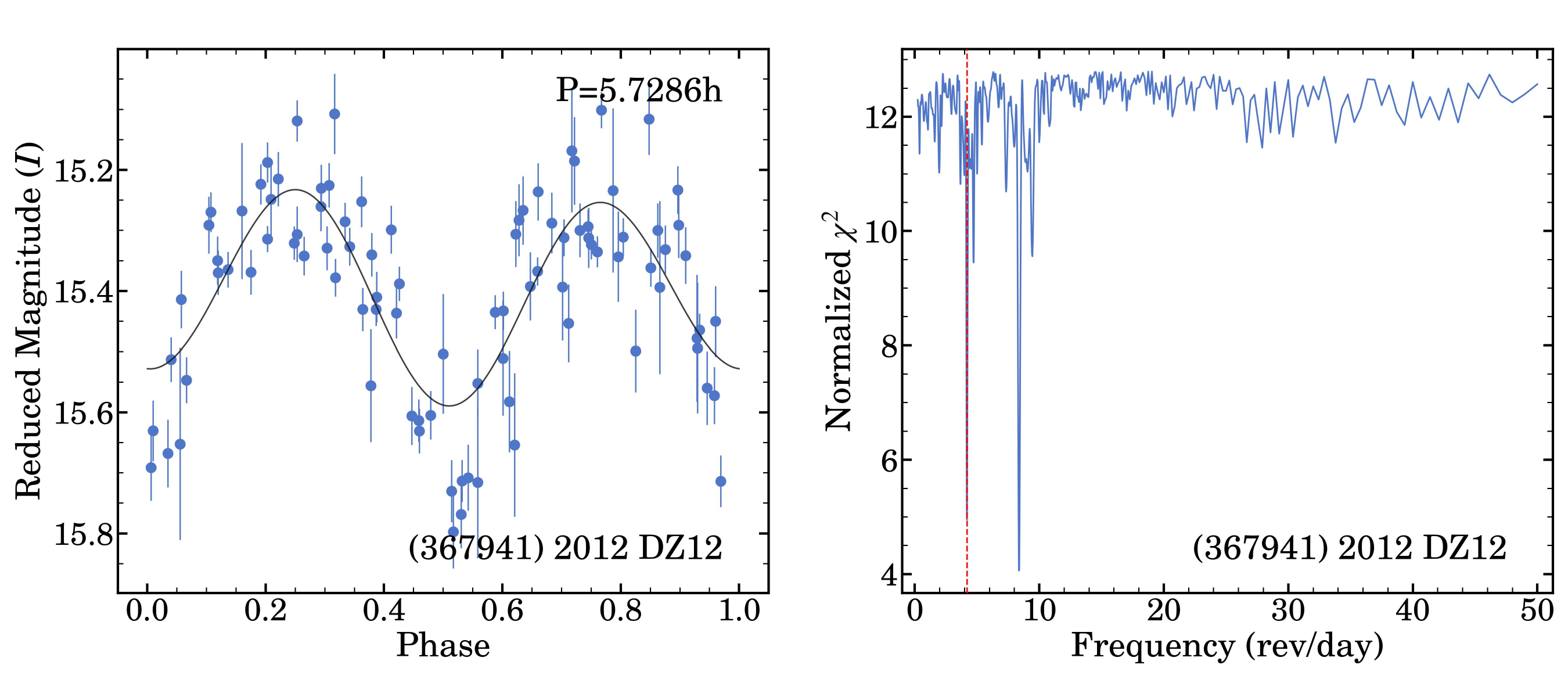}
    \includegraphics[width=0.45\linewidth]{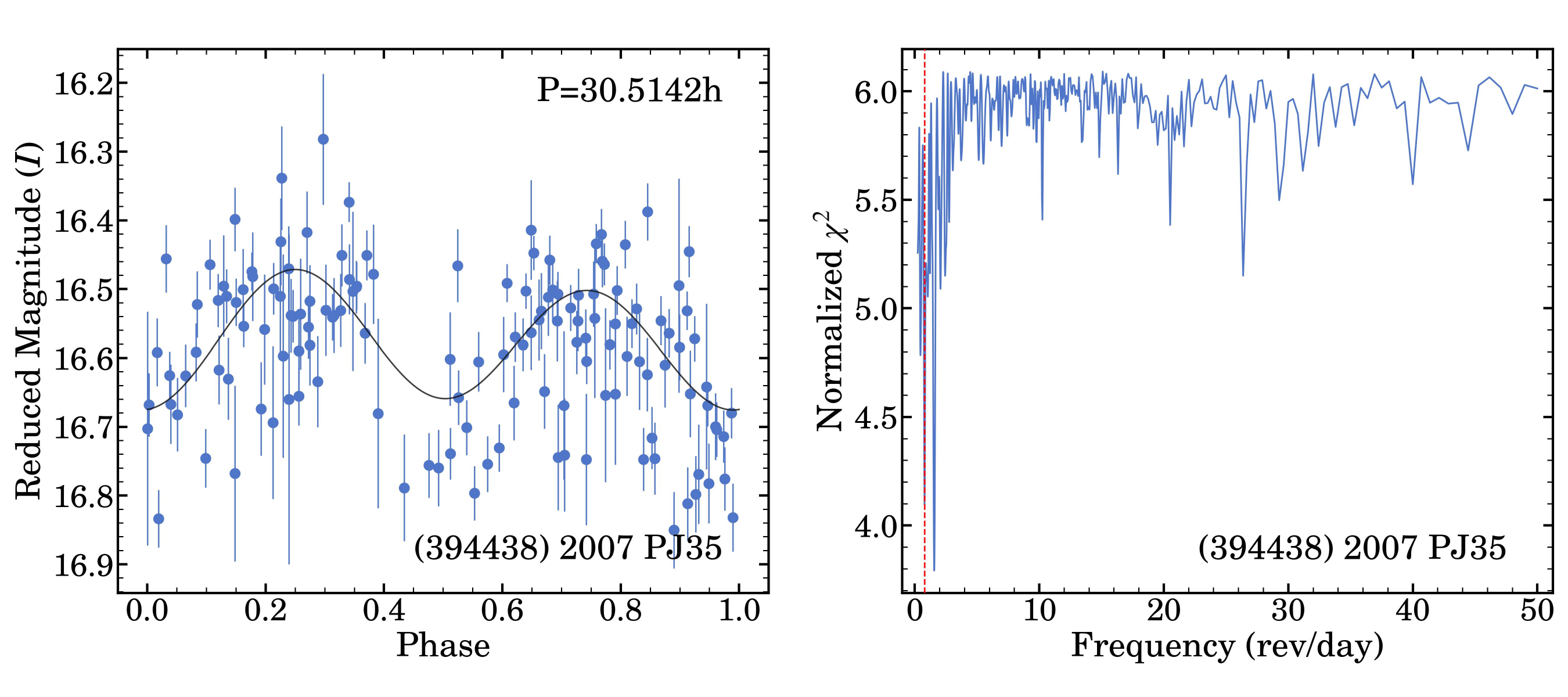}
    \includegraphics[width=0.45\linewidth]{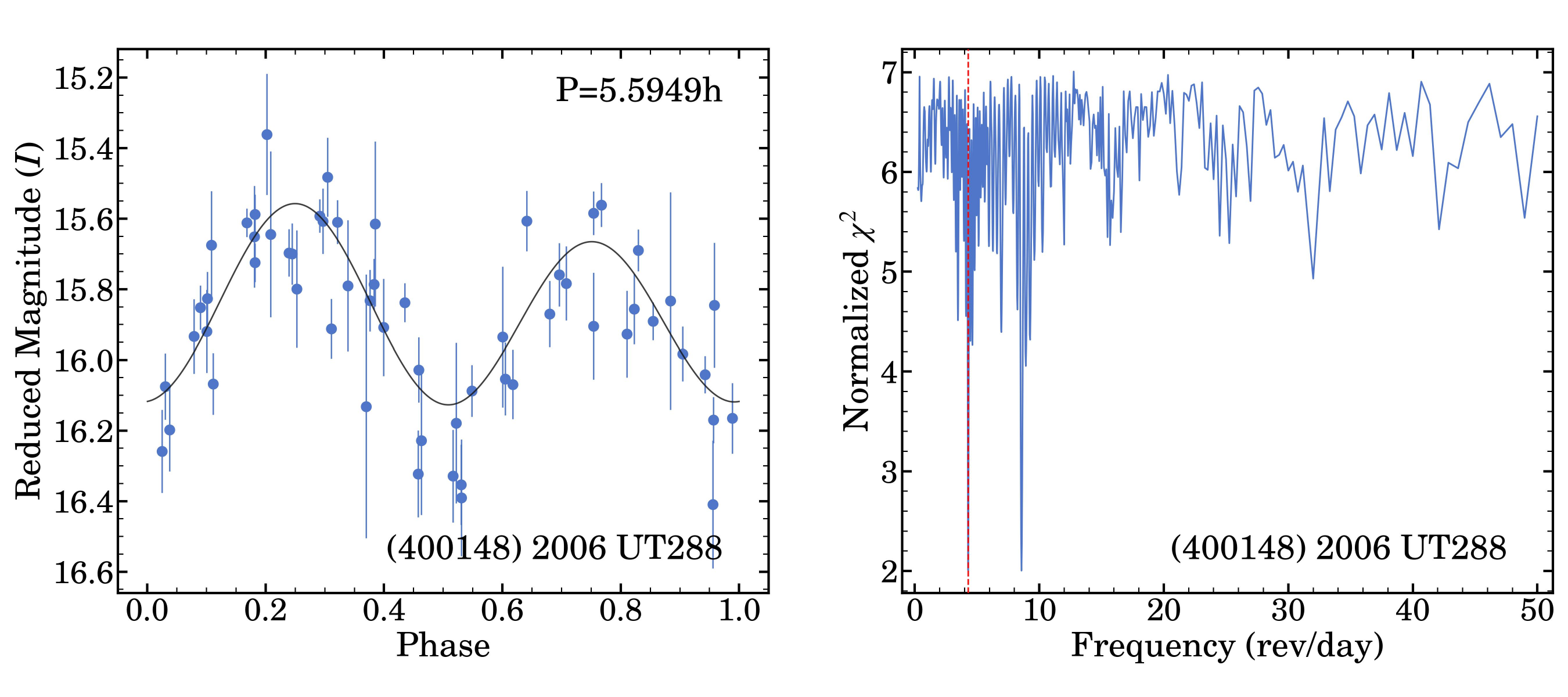}
    \includegraphics[width=0.45\linewidth]{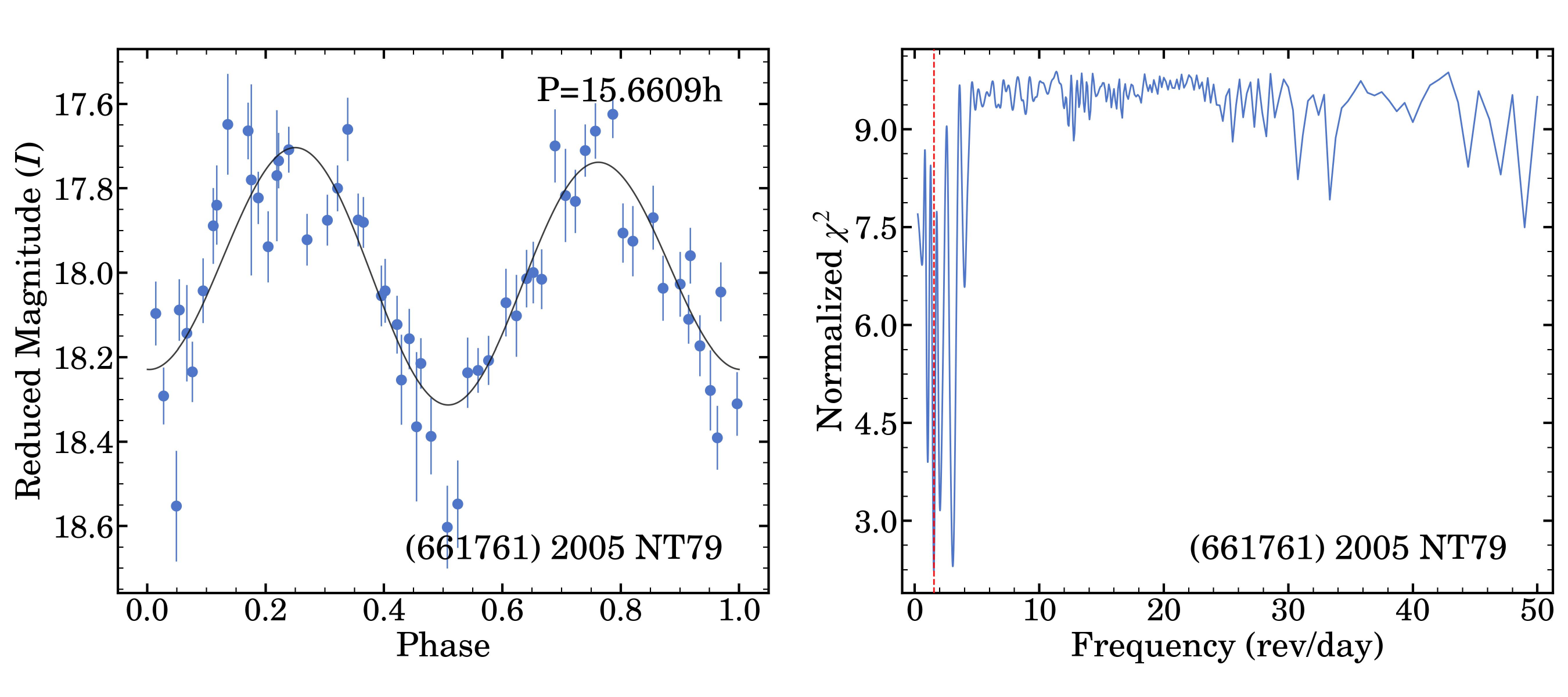}
    \includegraphics[width=0.45\linewidth]{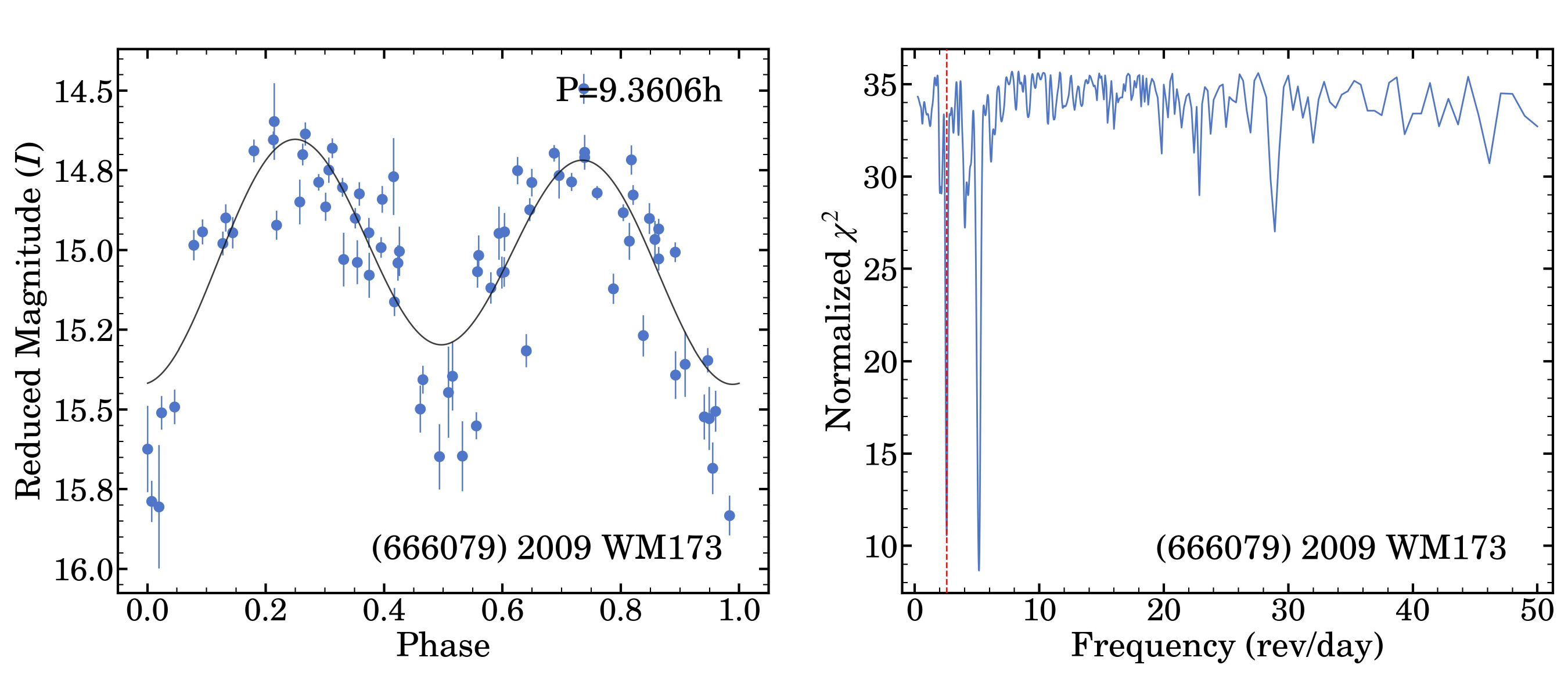}
    \includegraphics[width=0.45\linewidth]{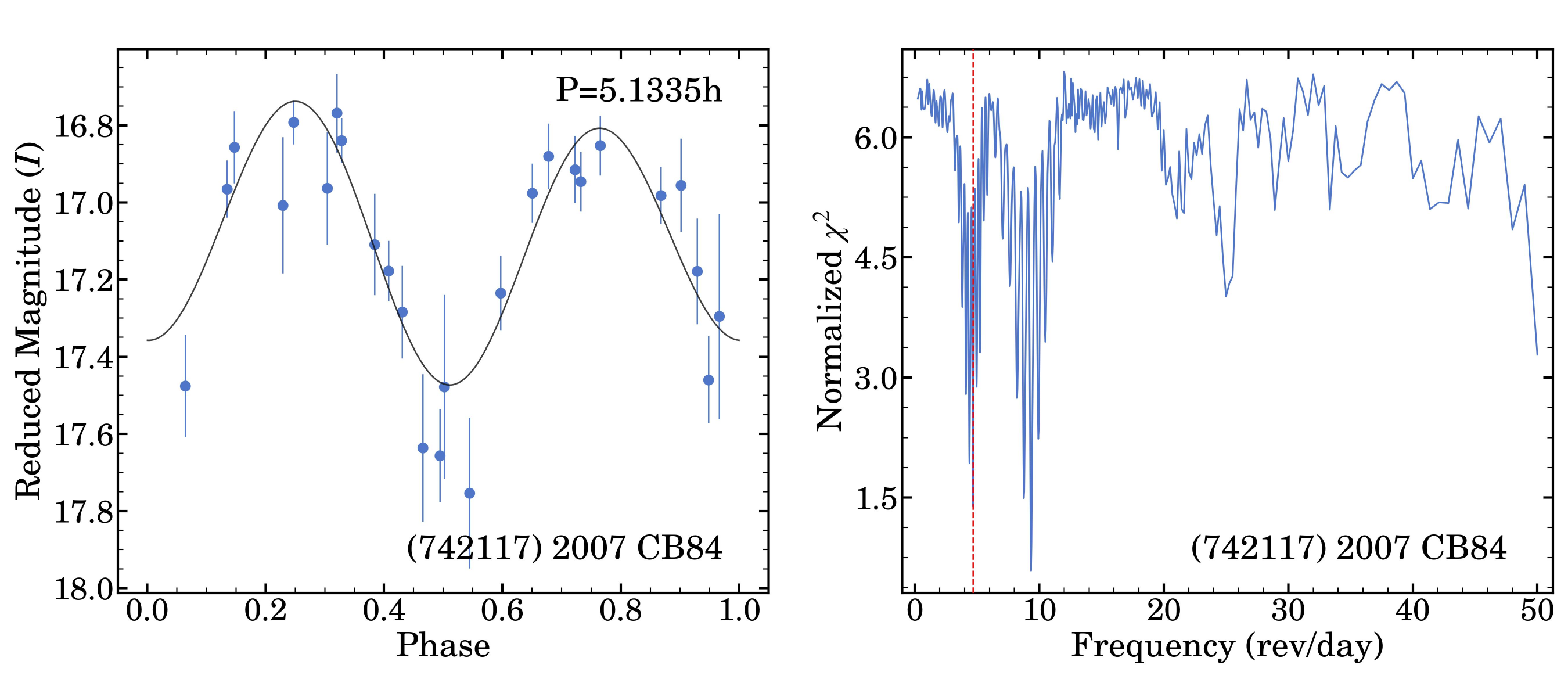}
    \caption{Continued (U = 2+).}
\end{figure*}

\end{document}